\documentclass[12pt]{article}

\usepackage{jheppub} 
\pdfoutput=1
\usepackage{xcolor}
\usepackage{hyperref}
\usepackage{subcaption}
\usepackage[utf8]{inputenc}
\usepackage{graphicx}
\usepackage{verbatim}
\usepackage{tensor}
\usepackage{setspace}
\usepackage{amsmath, amssymb, amsthm, float, graphicx, amsfonts, dsfont}
\numberwithin{equation}{section}

\def\beq{\begin{eqnarray}}\def\eeq{\end{eqnarray}}
\def\be{\begin{equation}}\def\ee{\end{equation}}
\def\g{\gamma}

\def\s{\sigma}
\def\m{\mu}

\def\a{\alpha}
\def\e{\epsilon}
\def\k{\kappa}
\def\b{\beta}
\def\d{\delta}
\def\c{\chi}
\def\f{\phi}
\def\t{\theta}
\def\D{\Delta}
\def\G{\Gamma}
\def\l{\lambda}

\def\Dphi{\Delta_{\phi}}

\def\ra{\rangle}

\def\G{\Gamma}

\def\3s{{s \choose 3}}
\def\4s{{s \choose 4}}
\def\5s{{s \choose 5}}
\def\6s{{s \choose 6}}

\def\12{\dfrac{1}{2}}

\def\2{\ell_2}

\def\pr{\partial}

\def\ra{\rightarrow}

\def\q{\quad}
\def\qq{\quad\quad}

\def\be{\begin{equation}}
\def\ee{\end{equation}}
\def\bea{\begin{eqnarray}}
\def\eea{\end{eqnarray}}
\def\ba{\begin{array}}
\def\ea{\end{array}}
\def\bec{\begin{center}}
\def\ec{\end{center}}

\def\g{\gamma}

\def\s{\sigma}
\def\m{\mu}

\def\a{\alpha}
\def\e{\epsilon}
\def\k{\kappa}
\def\b{\beta}
\def\d{\delta}
\def\c{\chi}
\def\f{\phi}
\def\t{\theta}
\def\D{\Delta}
\def\G{\Gamma}
\def\l{\lambda}

\def\Dphi{\Delta_{\phi}}

\def\ra{\rangle}

\def\G{\Gamma}

\def\3s{{s \choose 3}}
\def\4s{{s \choose 4}}
\def\5s{{s \choose 5}}
\def\6s{{s \choose 6}}

\def\12{\dfrac{1}{2}}

\def\2{\ell_2}

\def\pr{\partial}

\def\ra{\rightarrow}

\def\q{\quad}
\def\qq{\quad\quad}

\def\be{\begin{equation}}
\def\ee{\end{equation}}
\def\bea{\begin{eqnarray}}
\def\eea{\end{eqnarray}}
\def\ba{\begin{array}}
\def\ea{\end{array}}
\def\bec{\begin{center}}
\def\ec{\end{center}}

\def\a{\alpha} 
\def\b{\beta}  
\def\g{\gamma} 
\def\G{\Gamma}
\def\d{\delta} 
\def\D{\Delta}
\def\e{\varepsilon}
\def\z{\zeta}
\def\h{\eta}

\def\k{\kappa}
\def\l{\lambda}

\def\m{\mu}

\def\p{\psi}

\def\s{\sigma}

\def\t{\tau}
\def\f{\phi}

\def\c{\chi}

\def\df{\dot\varphi}

\def\cA{{\cal A}}

\def\cC{{\cal C}}

\def\cF{{\cal F}}
\def\cG{{\cal G}}

\def\cI{{\cal I}}

\def\cK{{\cal K}}

\def\cN{{\cal N}}
\def\cO{{\cal O}}

\def\cS{{\cal S}}
\def\cT{{\cal T}}

\def\Li{{\text{Li}}}
\def\df{{\Delta_{\phi}}}

\abstract{
We develop the technology for Polyakov-Mellin (PM) bootstrap in one-dimensional conformal field theories (CFT$_1$). By adding appropriate contact terms, we bootstrap various effective field theories in AdS$_2$ and analytically compute the CFT data to one loop. The computation can be extended to higher orders in perturbation theory, if we ignore mixing, for any external dimension. We develop PM bootstrap for $O(N)$ theories and derive the necessary contact terms for such theories (which also involves a new higher gradient contact term absent for $N=1$). We perform cross-checks which include considering the diagonal limit of the $2d$ Ising model in terms of the $1d$ PM blocks. As an independent check of the validity of the results obtained with PM bootstrap, we propose a suitable basis of transcendental functions, which allows to fix the four-point correlators of identical scalar primaries completely, up to a finite number of ambiguities related to the number of contact terms in the PM basis. We perform this analysis both at tree level (with and without exchanges) and at one loop. We also derive expressions for the corresponding CFT data in terms of harmonic sums. Finally, we consider the Regge limit of one-dimensional correlators and derive a precise connection between the latter and the large-twist limit of CFT data. Exploiting this result, we study the crossing equation in the three OPE limits and derive some universal constraints for the large-twist limit of CFT data in Regge-bounded theories with a finite number of exchanges.}

\title{\bf Crossing symmetry, transcendentality and the Regge behaviour of 1d CFTs}
\date{}

\author{\!\!\!\! Pietro Ferrero$^{c,}$\footnote{pietro.ferrero@maths.ox.ac.uk}, Kausik Ghosh$^{d,}$\footnote{kau.rock91@gmail.com}, Aninda Sinha$^{d,}$\footnote{asinha@iisc.ac.in} and Ahmadullah Zahed$^{d,}$\footnote{ahmadullah@iisc.ac.in}\\ ~~~~\\
\it $^{c}$Mathematical Institute, University of Oxford,\\ \it Andrew Wiles Building, Radcliffe Observatory Quarter,\\ \it Woodstock Road, Oxford, OX2 6GG, UK. \\
\it ${^d}$Centre for High Energy Physics,
\it Indian Institute of Science,\\ \it C.V. Raman Avenue, Bangalore 560012, India. }
\begin{document}

\maketitle

\section{Introduction}
Conformal symmetry puts stringent constraints on the structure of the correlators. One interesting fact about all unitary Conformal Field Theories (CFTs) is that the local operators in the theory, which are labeled by their scaling dimension ($\Delta$) and spin ($\ell$), satisfy an algebra, called the Operator Product Expansion (OPE). The structure constants of this algebra are commonly referred to as OPE coefficients, and together with the set of the quantum numbers ($\Delta,\,\ell$) of all local operators they are collectively known as CFT (or OPE) data. The CFT data contain all the dynamical information of a CFT, and therefore they characterize the theory uniquely. The conformal bootstrap is an approach to CFTs which is based on unitarity, crossing symmetry and associativity of the OPE, whose goal is to extract the OPE data non-perturbatively. After the seminal work \cite{rrtv}, there has been significant progress in constraining the solution space of CFT data, in particular putting stringent bounds on the spectrum of low dimension operators present in the theory \cite{stuff}. 

In \cite{caronhuot}, an inversion formula was derived for CFTs in $d\ge2$, which allows to extract the CFT data using only the so-called double discontinuity of the four-point function. This also puts large spin perturbation theory \cite{alday1} on a firm footing by showing that the CFT data are analytic in spin, except for a finite number of low spins. In a series of papers \cite{aldayeps} the CFT data for the Wilson-Fisher and the critical $O(N)$ model was analytically obtained using this method, in a perturbative expansion in a suitable small parameter. Crucially, the inversion formula of \cite{caronhuot} relies on a suitable Regge behaviour of the correlator, and in general it does not apply to scalar exchanges.

An alternative to the above scenario, following \cite{polya}, was proposed in \cite{KSAS, usprl,longpap, Dey:2016mcs}, where correlators are expanded in a crossing-symmetric basis. Then, demanding consistency with the OPE gives rise to constraints on the CFT data -- following \cite{gs}, we will refer to this approach as the Polyakov-Mellin (PM) bootstrap \footnote{In general Mellin space techniques have proven very effective in various different contexts \cite{Rastelli:2016nze,Penedones:2010ue,Fitzpatrick:2011ia,Costa:2012cb,Goncalves:2014rfa}.}. These constraint equations can be solved to find the OPE data of operators of all spins, including scalars. The crossing symmetric basis consists of exchange Witten diagrams and contact terms, and in order to have completeness of this basis, it is important to know which contact terms one has to add \cite{dgs,gs,KGloop}. Recently, in a series of papers \cite{Mazac1,Mazac:2018mdx,mp,Mazac:2018qmi, kpaul}, the contact terms issue was resolved in one-dimensional CFTs. There, the authors started with the crossing equations and acted with suitable functionals on these equations in order to find a sum rule, which turns out to coincide with the PM bootstrap.  In terms of the equations presented in \cite{usprl, longpap}, the key difference was to add a constant contact term \cite{gs}. 

In this paper, we consider one-dimensional CFTs dual to scalar effective field theories (EFTs) on a fixed AdS$_2$ background, via the celebrated AdS/CFT correspondence. We work on a setup which can be seen as the $1d$ analogue of \cite{PolchinskiPenedones}, where bootstrap techniques were applied to study scalar contact terms at tree level in $d=2$ and $d=4$. Hence, the CFTs we consider are toy models with only one ``single-trace'' scalar primary operator\footnote{When turning to the study of exchanges, we shall consider the existence of a second operator, $\cO_E$, which appears in the $\f \times \f$ OPE.} $\f$ of finite dimension $\df$, while all the other single-trace operators of the theory are considered to be decoupled due to their large dimension. This model has proven useful to address both conceptual and technical questions about the nature of the AdS/CFT duality, such as the emergence of bulk locality from the dual CFT \cite{PolchinskiPenedones}, or whether at loop level new constraints arise on the CFT data to guarantee that a given CFT has a bulk dual \cite{Aharony:2016dwx}. 

Specializing to one dimension allows for remarkable technical simplifications: $1d$ conformal four-point correlators are functions of one cross-ratio only ($z$), as opposed to two ($z$, $\bar{z}$) in $d>1$. In fact, $1d$ CFTs can be seen as arising in the so-called diagonal limit ($z=\bar{z}$) of higher-dimensional theories, {\it i.e.,} the kinematical configuration in which the four operators of the correlator all lie on the same line. Additionally, there is no spin in one dimension, and therefore the study of scalars essentially comprises all possible cases\footnote{One can also consider one-dimensional fermions, which however are simply Grassmann variables with no spin indices, and therefore they introduce no technical complications - for instance, the fermionic conformal blocks are the same as the bosonic ones.}. On the other hand, techniques that have proven extremely powerful in higher dimension, such as large spin perturbation theory \cite{alday1} or the inversion formula \cite{caronhuot}, do not apply to the case of $d=1$, and therefore it seems a very suitable problem for the application of PM bootstrap, as shown by recent work on the analytic functional bootstrap \cite{Mazac1,Mazac:2018mdx,mp,Mazac:2018qmi, kpaul}. Furthermore, one can wonder whether new techniques can be developed that are tailored for the one-dimensional case, but which may have potential ramifications for higher dimensions.

As an additional motivation for this study, let us stress that there are inherently one-dimensional models that one can consider, and that have been only partially explored using the conformal bootstrap. To begin with, a well-known instance in which the diagonal limit of higher-dimensional CFTs is interesting is that of line defects, such as the monodromy line defect in the 3D Ising model \cite{Billo:2013jda, Gaiotto:2013nva}, Wilson lines in four-dimensional $\cN=4$ SYM \cite{Giombi:2017cqn, Beccaria:2017rbe, Giombi:2018qox, Liendo:2018ukf, Giombi:2018hsx, Gimenez-Grau:2019hez, Beccaria:2019dws} and Wilson lines in the ABJM theory \cite{Bianchi:2017ozk, Bianchi:2018scb}. One can also consider purely one-dimensional theories, such as the SYK model \cite{Maldacena:2016hyu, Gross:2017hcz} or its conformal version \cite{Gross:2017vhb, Gross:2017aos}, whose gravitational dual is not well-understood. Both for applications to line defect theories and to the SYK model, it is also interesting to consider extensions of the setup of \cite{PolchinskiPenedones}, in which one has a multiplet of primary fields and an $O(N)$ global symmetry that rotates them. Finally, another reason of interest in $1d$ CFT's is related to recent work on the $\cI$-extremization principle in the context of black hole entropy and the AdS/CFT correspondence, for AdS$_4$ black holes with AdS$_2$ near horizon limit (see \cite{Zaffaroni:2019dhb} for a recent review). However, in the latter case, it is not clear whether the conformal bootstrap can provide new insights from the field theory side of the correspondence.

In this paper, we develop the technology for $1d$ PM bootstrap, achieving several technical simplifications in the process. For starters, since we are in one dimension, there will be only one Mellin variable. We accomplish this reduction by starting with the two variable ($s$ and $t$) expression in general dimensions, performing the $t$-integral and setting $d=1$. This approach will also be useful for future work on the diagonal limit of the PM bootstrap. While the series of papers \cite{Mazac1,Mazac:2018mdx,mp,Mazac:2018qmi} relied on unitarity, in the form of Regge boundedness of the four-point functions, we drop this assumptions and consider generic EFTs in AdS$_2$, where interactions can have an arbitrarily high number of derivatives -- see \cite{Fitzpatrick:2010zm,KGloop} for studies in higher dimensions using the AdS/CFT correspondence.  We also consider PM bootstrap for unitary scalar theories with $O(N)$ global symmetry since, as mentioned, it is interesting for some physical theories in one dimension. We shall fix the contact terms for such theories, and it turns out that in addition to the constant contact term considered in \cite{Mazac1,Mazac:2018mdx,mp,Mazac:2018qmi}, we also need to add a gradient contact term. 
As an application, we bootstrap $O(5)$ and $O(3)$, and we are able to reproduce the tree-level CFT data found in \cite{Giombi:2017cqn} for a half-BPS Wilson line in $\mathcal{N}=4$ SYM. We also perform two non-trivial consistency checks -- a) we show that the diagonal limit of the $2d$ Ising model can be expanded in terms of the $1d$ PM blocks and b) we show that the fermionic GFF correlator in one dimension (which is a bona fide CFT quantity to consider) 
can be expanded in terms of the bosonic $1d$ PM blocks. Both these consistency checks serve as non-perturbative evidence for the correctness of the $1d$ PM basis.

We also adopt an independent approach and discuss the possibility to fix the correlators only relying on simple constraints and a transcendentality principle. To be more precise, for every problem we consider, we shall provide suitable ansätze in terms of functions up to some fixed transcendentality and rational functions. Then we use crossing symmetry, combined with properties of the one-dimensional OPE, to fix the rational functions, and this allows to find correlators up to one loop with only a finite number of ambiguities, which correspond precisely to the contact terms that one needs to add to the sum over Witten exchange diagrams in order to build a complete basis of Polyakov blocks. We were also able to find closed-form expressions for the corresponding CFT data, in terms of generalized harmonic sums. Interestingly enough, a given transcendentality in the correlators directly translates into that of the CFT data, which are also found to always satisfy the reciprocity principle \cite{Dokshitzer:2005bf, Basso:2006nk, Alday:2015eya}. Transcendentality principles\footnote{There is a vast literature on transcendentality principles that appear in various contexts, and our references do not provide an exhaustive bibliography. We are only citing a few papers, and we refer the interested reader to the references therein.} were already employed in many contexts, including ${\mathcal N}=4$ super Yang-Mills \cite{Kotikov:2001sc, Bern:2005iz, Beisert:2006ez} up to seven loops \cite{Marboe:2016igj}, splitting functions in QCD (\cite{Vogt:2004mw, Moch:2004pa}) and other CFT problems \cite{aldayeps,Guha:2019ipe}, and our findings show that one-dimensional CFT’s that arise as duals of AdS$_2$ EFT’s provide a simplified setup where one can further investigate their origin. The CFT data extracted using this approach match with the PM bootstrap results exactly, and this serves as an independent check on the validity of our results as well as on the contact terms we have added to the exchange Witten diagrams basis to form the PM basis. A minor shortcoming of the transcendentality approach is the requirement of integer external operator dimensions, however, since the PM bootstrap approach agrees exactly with this, we take it as non-trivial evidence for the validity of PM bootstrap for any external operator dimension.

In the study of exact correlators, a strong connection has become apparent between the Regge (or $u$-channel OPE) limit considered in \cite{mp} for one-dimensional theories, and the large-twist limit of CFT data. First, we shall make this link more precise using the observation that the Regge limit OPE is dominated by operators with large dimension, and provide a formula that relates the expansion of correlators in the Regge limit to that of the CFT data for large twist, as an expansion in $1/\D$ ($\D=2\df+2n+\g_n$ being the physical dimension of double trace operators). Then, we shall consider the crossing equation in all the three OPE limits (namely $s$-, $t$- and $u$-channel), and observe that in two out of three the OPE is controlled by operators with large dimension ($\D$), while the third is dominated by the identity. This allows, exploiting the Regge-limit expansions previously derived, to put some constraints on the CFT data order by order in $1/\D$. Although this is not enough to completely fix the CFT data, and some of the expansions are only asymptotic, one can still learn some useful lessons from this analysis. Most notably, under the assumptions of Regge-boundedness and a finite number of exchanges, we find for the anomalous dimensions of double-trace operators a universal expansion of the type\footnote{Here we imagine that our theory has only one coupling constant, which controls all the interactions. Alternatively, if one adopts the point of view of an effective field theory (EFT), the functions $f_i$ will in general depend on many arbitrary couplings.}
\be\nonumber
\begin{aligned} \gamma(\Delta)=& \lambda\left(\frac{1}{J^{2}}+\frac{2 \Delta_{\varphi}\left(\Delta_{\varphi}-1\right)}{J^{4}}\right)+f_{1}(\lambda) \frac{1}{J^{6}}+\cdots \\ &+\log J\left(2 \lambda^{2}\left(\frac{1}{J^{6}}+\frac{2\left(3 \Delta_{\varphi}^{2}\right)-3 \Delta_{\varphi}-2}{J^{8}}\right)+f_{2}(\lambda) \frac{1}{J^{10}}+\cdots\right) \\ &+\log ^{2} J\left(24 \lambda^{3}\left(\frac{1}{J^{10}}+\frac{10 \left(\df^2-\df-2\right)}{J^{12}}\right)+f_{3}(\lambda) \frac{1}{J^{14}}+\cdots\right)+\cdots\,\,\, , \end{aligned}
\ee
where $J^2=\D(\D-1)$ is the so-called {\it conformal spin} (eigenvalue of the quadratic Casimir), and $f_i(\l)$ are arbitrary functions of the coupling constant $\l$ that we are not able to fix with this analysis.

The paper is organized as follows. In Section \ref{sec:PMhighd}, we begin by setting up the PM bootstrap equations in $1d$. This will also clarify the conventions used in our work. We will also set up the equations for the $O(N)$ case. Since the contact terms needed for the $O(N)$ case are somewhat different than for the $N=1$ case, this is not just a trivial exercise. In Section \ref{sec:impPMboot}, we will turn to implementing the bootstrap constraints. In Sections \ref{transc - tree} and \ref{sec:transc - loop}, we will implement the idea based on pure transcendentality to completely fix the four point functions perturbatively for many choices of integer external operator dimensions, {up to one loop and both with and without $O(N)$ symmetry}. The CFT data extracted from this agrees perfectly with the PM bootstrap method. In Section \ref{sec:Reggelimit}, we shall discuss the Regge limit for one-dimensional CFTs, and the implication of crossing symmetry in such regime. In Section \ref{sec:highdconclu}, we briefly address issues in higher dimensions before concluding in Section \ref{sec:Discussion}.
The appendices supplement many computational details we used in the main text.

\section{Polyakov-Mellin Bootstrap}\label{sec:PMhighd}
 We will consider four-point functions of identical scalar primary operators with conformal dimension $\Dphi$. Our conventions are the same as \cite{gs}, which we will review below. As is the usual convention in many CFT papers, we will use $h=d/2$ where $d$ is the number of spacetime dimensions in which the CFT lives. Since our approach in the case $d=1$ can be applied to the diagonal limit of CFTs in any dimension, for the moment we shall work with arbitrary $d$. In a CFT, the four-point function of scalar primaries has the form
 \begin{equation}
 \langle\phi(x_1)\phi(x_2)\phi(x_3)\phi(x_4)\rangle=\frac{1}{x_{12}^{2\Delta_{\phi}}x_{34}^{2\Delta_{\phi}}}\mathcal{A}(u,v)\,,
 \end{equation}
 where the cross ratios $u$ and $v$ are given by
 \begin{equation}
 u=\frac{x_{12}^2 x_{34}^2}{x_{13}^2 x_{24}^2}=z \bar{z}, \qquad v=\frac{x_{14}^2 x_{23}^2}{x_{13}^2 x_{24}^2}=(1-z)(1- \bar{z}).
 \end{equation}
$\mathcal{A}(u,v)$ admits a Mellin representation \cite{Mack:2009mi}\footnote{Where $ [dx]=\frac{dx}{2\pi i}$.}
\begin{equation} \label{mellinrep}
\int [ds] [dt] \mathcal{M}(s,t)\Gamma^2(\Delta_{\phi}-s)\Gamma^2(-t)\Gamma^2(s+t)\,.
\end{equation}
The reduced Mellin Amplitude $\mathcal{M}(s,t)$ can be expanded in the s-channel conformal blocks as
\begin{equation} \label{schdecomp}
\mathcal{M}(s,t)=\sum_{\Delta,\ell}c_{\Delta,\ell}B_{\Delta,\ell}(s,t),
\end{equation}
where
\begin{equation}
B_{\Delta,\ell}(s,t)=\frac{\Gamma(\frac{\Delta-\ell}{2}-s)\Gamma(\frac{2h-\Delta-\ell}{2}-s)\widehat{P}_{\Delta-h,\ell}(s,t)}{\Gamma^2(\Delta_{\phi}-s)},
\end{equation}
{where $\widehat{P}_{\Delta-h,\ell}(s,t)$ are the Mack polynomials, given in Appendix (\ref{mackpol})}. In the context of PM bootstrap, we expand $\mathcal{M}(s,t)$ {as}
\begin{equation} \label{wittenbexp}
\mathcal{M}(s,t)=\sum_{\Delta,\ell}c_{\Delta,\ell}\left(W^{(s)}_{\Delta,\ell}(s,t)+W^{(t)}_{\Delta,\ell}(s,t)+W^{(u)}_{\Delta,\ell}(s,t)\right)+c(s,t),
\end{equation}
where
\begin{equation}
\begin{split}
W^{(s)}_{\Delta,\ell}(s,t)=& \widehat{P}_{\Delta-h,\ell}(s,t)\frac{\Gamma^2(\frac{\Delta+\ell}{2}+\Delta_{\phi}-h)}{(\frac{\Delta-\ell}{2}-s)\Gamma(\Delta-h+1)}\\
& \times\, _3F_2\bigg[\frac{\Delta-\ell}{2}-s,1+\frac{\Delta-\ell}{2}-\Delta_{\phi},1+\frac{\Delta-\ell}{2}-\Delta_{\phi};1+\frac{\Delta-\ell}{2}-s,\Delta-h+1;1\bigg],
\end{split}
\end{equation}is the Witten diagram for the exchange of a primary operator of dimension $\Delta$ and spin $\ell$ in the $s$-channel, whose OPE coefficient squared is given by $c_{\Delta,\ell}$. The term $c(s,t)$ is a polynomial in $s,t$ and represents the  potential set of contact terms that one needs to add (in principle for each $\Delta,\ell$) to have a well-defined basis of PM blocks \cite{gs}.
The other channels are given by the following transformations from the $s$ channel.
\begin{equation}
t \,\text{channel} :s \rightarrow t+\Delta_{\phi},t \rightarrow s-\Delta_{\phi}; 
\qquad u\,\text{channel}:s\rightarrow \Delta_{\phi}-s-t,t\rightarrow t.
\end{equation}
The difference between this approach and the traditional approach to the conformal bootstrap is that this basis is manifestly crossing symmetric. Furthermore, $W^{(s)}_{\Delta,\ell}(s,t)$ does not have double zeros at $s=\Delta_{\phi}+n$, $n\in \mathbb{Z}$, unlike the conformal block $B^{(s)}_{\Delta,\ell}(s,t)$ . Hence, performing the integration in eq. \eqref{mellinrep}, one finds powers of $u$ which are physical, of the type $u^{\frac{\Delta-\ell}{2}+n}$, along with spurious powers $u^{\Delta_{\phi}+n}\log(u)$ and $u^{\Delta_{\phi}+n}$. Explicitly, we have\footnote{$\cA_{c}(u,v)$ is just the connected part of the correlator.}
\begin{equation} \label{fourpoinfunc}
\begin{split}
\mathcal{A}_{c}(u,v)=&\sum_{\Delta,\ell} c_{\Delta,\ell} \sum_{n}u^{\frac{\Delta-\ell}{2}+n} f_{\Delta,\ell,n}(v)\\
& +\sum_{n}u^{\Delta\phi+n}\log(u)\sum_{\Delta,\ell} c_{\Delta,\ell}  \tilde{f}_{\Delta,\ell,n}(v)+\sum_{n}u^{\Delta\phi+n}\sum_{\Delta,\ell} c_{\Delta,\ell}  \tilde{g}_{\Delta,\ell,n}(v) .
\end{split}
\end{equation}
The first line of eq. \eqref{fourpoinfunc} is the usual $s$-channel conformal blocks decomposition of the four-point function, while the second line represents unphysical spurious contribution: the requirement that the latter vanish gives the consistency conditions that are exploited in the PM bootstrap. One crucial criterion for these conditions to hold is that the summand should decay at large $\Delta$ and large $\ell$ sufficiently fast, only then there is a hope of canceling the second line of eq.(\ref{fourpoinfunc}). In the expansion of the four point function eq. (\ref{wittenbexp}), we know the explicit expressions of exchange Witten blocks, but it is not completely clear how to fix the contact terms $c(s,t)$ in an effective manner. In \cite{usprl,longpap,Dey:2016mcs}, this method was implemented successfully for the Wilson-Fisher fixed point and for the $O(N)$ model up to $O(\epsilon^3)$ and to the first non-trivial order in $1/N$. In \cite{gs,dgs,KGloop}, the issue of the  contact terms was addressed perturbatively. Following \cite{gs}, we can parametrize $c(s,t)$ as  
\begin{equation} \label{contacteq}
c(s,t)=\sum_{{mn}=0}^{m+n=L/2} a_{mn} \Bigg([s(s+t-\Delta_{\phi})(t+\Delta_{\phi})]^m[t(s+t)+s(s-\Delta_{\phi})]^n\Bigg),
\end{equation}
which also gives us the correct number of contact terms for a specfic $L$, {\it i.e.}, $\frac{(L+2)(L+4)}{8}$, as discussed in \cite{PolchinskiPenedones}. In this paper we shall make an alternative choice of contact terms, given by\footnote{This notation will be useful in the $O(N)$ case considered later and also has a nice decomposition in the continuous Hahn basis.}
\begin{equation}\label{contactterm}
c(s,t)=\sum_{m+n=0}^{L}c_{mn}\left[(-t)_n(s+t)_m+(\Delta_{\phi}-s)_m(s+t)_n+(-t)_m(\Delta_{\phi}-s)_n\right]\,,
\end{equation}
where $c_{mn}$ are constants, symmetric in their indices $c_{mn}=c_{nm}$. The functions $c(s,t)$ are arbitrary crossing-symmetric polynomials of $s$ and $t$. In practice, we find that the presence of contact terms in eq. \eqref{wittenbexp} is required in order to ensure convergence of the sum over the spectrum in. In all cases that we have encountered, when convergence of this sum fails, it does so in a similar manner for all constraint equations. To fix this problem, we have to choose the contact terms in such a way that the divergences cancel out. Then we can solve the rest of the equations completely. In \cite{gs}, it was found that this procedure leads to identical constraints to those in \cite{mp}, where the contact term in $1d$ was fixed by demanding consistency with Regge boundedness. In what follows, we will focus our attention on the $1d$ problem, deriving a new contact term for the $O(N)$ case as well as extracting analytic results from the constraint equations.

\subsection{Bootstrapping with no global symmetry}
Let us now turn our attention to the main subject of this paper, {\it i.e.} $1d$ CFTs. For a four-point function in one dimension there is only one independent cross ratio\footnote{From the point of view of higher dimensional CFTs, we can recover the case $d=1$ by placing all operators on the same line, which corresponds to the ``diagonal limit'' $z=\bar{z}$ of the cross ratios.}, and we can express a correlator of identical scalars as
\be
\left\langle\phi\left(x_{1}\right) \phi\left(x_{2}\right) \phi\left(x_{3}\right) \phi\left(x_{4}\right)\right\rangle=\frac{1}{\left|x_{12}\right|^{2 \Delta_{\phi}}\left|x_{34}\right|^{2 \Delta_{\phi}}} \mathcal{A}(z)\,.
\ee
The function $\mathcal{A}(z)$ has singularities for values of $z$ corresponding to those configurations in which two points are coincident, {\textit i.e.} $z=0,\,1$ and $\infty$. In fact, $\mathcal{A}(z)$ is not analytic on the complex $z$-plane -- rather, it reduces to three different functions in three different regions of the real $z$ line:
\begin{align}
\mathcal{A}(z)=
\begin{cases}
\mathcal{A}^{-}(z) \text{ for } z \in  (-\infty,0),\\
\mathcal{A}^{0}(z) \text{ for } z \in  (0,1),\\ \mathcal{A}^{+}(z) \text{ for } z \in  (1,+\infty).
\end{cases}
\end{align}
These functions are related to $\mathcal{A}^{0}(z)$ because of Bose symmetry of four-point function, and although they can independently be analytically continued to the complex $z$ plane, they are not analytical continuations of each other \cite{Mazac:2018qmi}. Physically, this is related to the fact that in one dimension one cannot move operators around each other. $\mathcal{A}^{0}(z)$ also admits an expansion in conformal blocks\footnote{From now onwards when we write $A(z)$, we actually mean $A^{0}(z)$. } 
\be\label{eq:Amp}
\mathcal{A}(z)=\sum_{\Delta} C_{\Delta} G_{\Delta}(z),
\ee
with 
\be
G_{\D}(z)=z^\D {}_2F_1(\D,\D;2\D;z)\,.
\ee
To get the constraint equations for this case, it is enough set $d=1$ in our previous equations, as those results hold for general dimension. Furthermore, one needs to set $\ell=0$ as there are only scalars in one dimension. We also set $z=\bar{z}$ and find 
\begin{equation}
\mathcal{A}_c(z)=\int[ds][dt] z^{2s}(1-z)^{2t} \mathcal{M}(s,t)\Gamma^2(\Delta_{\phi}-s)\Gamma^2(-t)\Gamma^2(s+t)\,,
\end{equation}
where\footnote{Here $c_{\Delta}$ are the squared OPE coefficients which come from $1d$ conformal blocks decomposition,  \textit{{\it i.e.}} $c_\D=C_\D N_{\D,0}$ (or more generally  $c_{\D,\ell}=C_{\D,\ell} N_{\D,\ell}$ ) and $N_{\D,0}$ is given below in eq.(\ref{f1(s)}) and $C_\Delta$ are the standard OPE coefficients--see \cite{usprl}.}
\begin{equation}
\mathcal{M}(s,t)=\sum_{\Delta}c_{\Delta}(W^{(s)}_{\Delta,0}(s,t)+W^{(t)}_{\Delta,0}(s,t)+W^{(u)}_{\Delta,0}(s,t))+c(s,t).
\end{equation}
Let us study the above equation more carefully, without the contact terms $c(s,t)$. This will be instructive since unlike the discussions in the literature so far, which focus on higher-dimensional cases, we will perform the $t$-integral before writing down the consistency conditions. There exist many definitions of Polyakov blocks in the literature, depending on how one modifies the exchange Witten diagrams by adding contact terms. For our purposes, we define Polyakov blocks to be simply sum of exchange Witten diagrams in all three channels. In Mellin space, the Polyakov block $\mathcal{PB}_{\D,\ell}(u,v)$ in general dimension can be written as  
\be
\begin{split}
\mathcal{PB}_{\D,\ell}(u,v)=&\int_{-i \infty}^{i \infty}[ds]~ [dt]~u^s v^t \G^2(\D_\phi-s)\G^2(s+t)\G^2(-t)\\
&\left[ \sum_{\ell'=0}^{\infty}\left(q^{(s)}_{\D,\ell'|\ell}(s)+q^{(t)}_{\D,\ell'|\ell}(s)+q^{(u)}_{\D,\ell'|\ell}(s)\right)Q^{2s+\ell'}_{\ell',0}(t)\right]\,,
\end{split}
\ee
where we have decomposed the exchange Witten block in orthogonal continuous Hahn polynomials (see \ref{ap:cHahn}):
\begin{equation}
W^{(i)}_{\Delta,\ell}(s,t)=\sum_{\ell'} q^{(i)}_{\Delta,\ell'|\ell}(s) Q^{2s+\ell'}_{\ell',0}(t),
\end{equation}
and $i$ represents $s$, $t$ or $u$ channel. Since $(-1)^{\ell'} q^{(t)}_{\D,\ell'|\ell}(s)=q^{(u)}_{\D,\ell'|\ell}(s)$, we have
\be\label{gendPM}
\begin{split}
\mathcal{PB}_{\D}(z)=&\int_{-i \infty}^{i \infty}[ds]~ [dt]~z^{2s} (1-z)^{2t} \G^2(\D_\phi-s)\G^2(s+t)\G^2(-t)\\
&\left[ \sum_{\ell'=0}^{\infty}\left(q^{(s)}_{\D,\ell'|0}(s)\d_{\ell',0}+\left(1+(-1)^{\ell'}\right)q^{(t)}_{\D,\ell'|0}(s)\right)Q^{2s+\ell'}_{\ell',0}(t)\right]\,.
\end{split}
\ee
In order to perform the $t$-integral, we write $(1-z)^{2t}=\sum_{r=0}^{\infty}(-1)^r ~{}^{2t}C_{r}z^r$, and expand ${}^{2t}C_{r}$ in a basis of continuous Hahn polynomial $Q^{2s+\ell''}_{\ell'',0}(t)$ (see appendices \ref{ap:cHahn} and \ref{ap:dderi}),
\be\label{delldef}
{}^{2t}C_{r}=\sum_{\ell''=0}^{r} d _{r,\ell''}(s)Q^{2s+\ell''}_{\ell'',0}(t)\,,
\ee
where $d_{r,\ell}(s)$ is given in eq. (\ref{dell}). We can now perform the integral over $t$ using orthogonality of the continuous Hahn polynomials, and we get
\be
\begin{split}
\mathcal{PB}_{\D}(z)=&\int_{-i \infty}^{i \infty}[ds]~~z^{2s} \left(\sum_{r=0}^{\infty}\sum_{\ell''=0}^{r}(-z)^r  ~~ d _{r,\ell''}(s)\right)\G^2(\D_\phi-s)\\
&\left[ \sum_{\ell'=0}^{\infty}\left(q^{(s)}_{\D,\ell'|0}(s)\d_{\ell',0}+\left(1+(-1)^{\ell'}\right)q^{(t)}_{\D,\ell'|0}(s)\right)\k_{\ell'}(s)\d_{\ell',\ell''}\right]\,,
\end{split}
\ee
where $k_{\ell}(s)$ is given in eq. (\ref{khahn}). Making a change of variable $s\to s-r/2$, then performing the sum over $\ell'$ and relabelling $\ell'' \to \ell'$, we can express the final result as
\be\label{main}
\begin{split}
\mathcal{PB}_{\D}(z)=&\int_{-i \infty}^{i \infty}[ds]~~z^{2s}\G^2(\D_\phi-s) \Bigg[\left(\sum_{r=0}^{\infty}\sum_{\ell'=0}^{r}(-1)^r(\Dphi-s)^2_{\frac{r}{2}}  ~~ d _{r,\ell'}(s-\frac{r}{2})\right)\\
&\left\lbrace q^{(s)}_{\D,\ell'|0}(s-\frac{r}{2})\d_{\ell',0}+\left(1+(-1)^{\ell'}\right)q^{(t)}_{\D,\ell'|0}(s-\frac{r}{2})\right\rbrace\k_{\ell'}(s-\frac{r}{2})\Bigg]\,.
\end{split}
\ee
Notice that the shift $s \to s-r/2$ moves some poles inside the contour of integration, even thought these were not giving any contribution before the shift. Therefore, we do not include the contribution from these poles in the calculation, as it would contradict the initial result. We can think of this procedure as a modification of the contour, where any spurious poles from the point of view of the original contour before the shift are excised.

The term within the square bracket in eq. \eqref{main} can be taken as the crossing symmetric PM block in $1d$. Then, the consistency conditions come from demanding that the contributions of the spurious poles, coming from $\G^2(\D_\phi-s)$, are zero. From this requirement, we get the constraints
\be\label{eqnprev}
\sum_{\D}C_{\D}N_{\D,0}f_{\D}(\D_\phi+n)=0~;~n= 0,1,2,3\dots;
\ee
\be\label{2eqnprev}
\sum_{\D} C_{\D}N_{\D,0}f_{\D}'(\D_\phi+n)+ q_{dis}'(\D_\phi+n)=0~;~n= 0,1,2,3\dots ;
\ee
where
\be\label{f1(s)}
\begin{split}
q_{dis}(s)=&\left(\frac{1}{s\G^2(\Dphi-s)}+\frac{1}{(s-\Dphi)\G^2(\Dphi-s)}-\frac{2\Gamma \left(1-2 \Delta _{\phi }\right) \Gamma \left(2 \Delta _{\phi }-2 s\right)}{\Gamma (1-2 s)\G^2(\Dphi-s)}\right)\\
N_{\Delta,0}=&\frac{ (\D-1)\G(\D-1)\G(\D+\frac{1}{2})}{\G^4(\frac{\D}{2})\G^2(\Dphi-\frac{\D}{2})\G^2(\Dphi-\frac{1-\D}{2})}\,,\\
f_{\D}(s)=& \Bigg[\left(\sum_{r=0}^{\infty}\sum_{\ell'=0}^{r}(-1)^r(\Dphi-s)^2_{\frac{r}{2}}  ~~ d _{r,\ell'}(s-\frac{r}{2})\right)\\
&\left\lbrace q^{(s)}_{\D,\ell'|0}(s-\frac{r}{2})\d_{\ell',0}+\left(1+(-1)^{\ell'}\right)q^{(t)}_{\D,\ell'|0}(s-\frac{r}{2})\right\rbrace\k_{\ell'}(s-\frac{r}{2})\Bigg]\,.
\end{split}
\ee
Detailed calculations for $q_{dis}(s)$ are given in appendix \ref{identity}. $q_{dis}(s)$ represents the contribution of the identity exchange, which is the sum of the disconnected contributions in all three channels. It can be seen that the identity exchange only has single poles at $s=\Delta_{\phi}+n$, where $n$ is a non-negative integer, and therefore it only appears in equation (\ref{2eqnprev}).  We have expanded the four-point function in exchange Witten blocks and we pick the residue at $s=\frac{\Delta}{2}$, which contributes to the four point function, and this residue is given by $C_{\Delta} (N_{\Delta,0})^{-1}$, $C_{\Delta}$ being the square of the OPE coefficient. So we multiply our equations with  $N_{\D,0}$ to have an agreement of OPE coefficients with what appear in the conventional conformal blocks expansion (also see \cite[eq 5.3]{gs}). Eq. (\ref{eqnprev}) is the coefficient of $z^{2\Dphi+2n}\log{z}$ and eq. (\ref{2eqnprev}) is the coefficient of $z^{2\Dphi+2n}$. 

We now check if the $\Delta$ sum of equation (\ref{eqnprev}) is convergent when one replaces the GFF OPE coefficients. We get the following asymptotics in the large $\Delta$ limit for the coefficient of $z^2\log(z)$, \textit{{\it i.e.},} the case $n=0$ of eq. (\ref{eqnprev}) for $\Dphi=1$:
\begin{equation}
C_{\D}N_{\D,0}f_{\D}(\D_\phi)\sim-\frac{24 \Delta }{\pi ^2}+\frac{12}{\pi ^2}-\frac{64}{\pi ^2 \Delta ^3}.
\end{equation}  
The coefficient of $z^{4}\log(z)$, \textit{{\it i.e.}} the case $n=1$ of eq.(\ref{eqnprev}) for $\Dphi=1$ is
 \begin{equation}
C_{\D}N_{\D,0}f_{\D}(\D_\phi+1)\sim-\frac{24 \Delta }{\pi ^2}+\frac{12}{\pi ^2}-\frac{288}{\pi ^2 \Delta ^3}.
\end{equation}  
It is clear that the individual expressions  grow with $\Delta$ but if we subtract one from the other, the result falls off as $\frac{1}{\Delta^3}$. We find that for any arbitrary $\Delta_{\phi}$ this is the general picture: with only one subtraction we can make the sum over $\Delta$ convergent, so that the constraint equations are well-defined. This is equivalent to choosing a contact term $c(s,t)=c_{00}$ (see equation (\ref{contactterm})), hence sacrificing the $n=0$ equation. An important point to note is the following. We could just try to remove the divergent piece from individual equations by suitably choosing $c(s,t)$, without losing the $n=0$ condition. However, we would still have the freedom to add an arbitrary finite piece, e.g., a constant to $c(s,t)$. Then all the OPE data would be expressed in terms of this unknown constant. This is equivalent to expressing  all the anomalous dimension $\gamma_n$ in terms of one undetermined quantity (say $\gamma_0$)--this is easy to see since taking pairwise differences of the constraint equations would get rid of the unknown constant. Although in the main text of this paper we choose the $n=0$ equation to get rid of the divergences, it must be stressed that this choice is completely arbitrary. This procedure is equivalent to a redefinitions of the Polyakov blocks. More precisely, we defined the Polyakov blocks as the crossing symmetric sum of exchange Witten diagrams, without any contact term. The procedure outlined above is equivalent to redefining the blocks, adding a contact term (\textit{e.g.} $\phi^4$) with an arbitrary coefficient, say $c_{00}$. The introduction of this arbitrary coefficient is equivalent, in practice, to the idea of ``sacrificing'' the equation for $n=0$, which we subtract from all the other equations in order to have convergent sums\footnote{In our examples we subtract the equation with $n=0$, but note that this is an arbitrary choice, and in principle any equation would serve the purpose.}. Note that it is non-trivial that such a simple modification of our Polyakov blocks, \textit{i.e.} the addition of a contact term, is enough to guarantee convergence of our sums: this is due to the detailed structure of the contact terms as functions of $z$. An interesting observation is that our motivation for the addition of contact terms to the original basis of Polyakov blocks was to guarantee convergence of the sums over the spectrum ($\Delta$), and therefore we required every term to decay at least with $\Delta^{-2}$ for large $\Delta$. However, we could in principle add two contact terms, with two undetermined coefficients, and require an even stronger decay of the summand\footnote{Interestingly, we find that the conditions that are necessary for convergence also kill the $\Delta^{-2}$ terms, and the terms in the sum decay with $\Delta^{-3}$. Then, if we add another contact term, and require (in principle) a decay with $\Delta^{-4}$, other terms automatically cancel, and the behaviour is $\Delta^{-7}$. The pattern goes on, and with the addition of $k$ contact terms we find a decay with $\Delta^{1-4k}$.}. This is equivalent to the subtraction of two equations, say for $n=0$ and $n=1$. In the language of the analytic functionals, this corresponds to demanding a softer Regge behaviour. Finally, the consistency conditions take the following form
\be\label{eq} 
\sum_{\D}C_{\D}N_{\D,0}f_{\D}(\D_\phi+n)=0~;~n= 0,1,2,3\dots;
\ee
\be\label{2eq}
\sum_{\D}\left(C_{\D}N_{\D,0}f_{\D}'(\D_\phi+n)\right)+ q_{dis}'(\D_\phi+n)=0~;~n= 0,1,2,3\dots ;
\ee
where
\be\label{f(s)}
\begin{split}
f_{\D}(s)=& \Bigg[\left(\sum_{r=0}^{\infty}\sum_{\ell'=0}^{r}(-1)^r(\Dphi-s)^2_{\frac{r}{2}}  ~~ d _{r,\ell'}(s-\frac{r}{2})\right)\\
&\left\lbrace q^{(s)}_{\D,\ell'|0}(s-\frac{r}{2})\d_{\ell',0}+c_{00}~\d_{\ell',0}+\left(1+(-1)^{\ell'}\right)q^{(t)}_{\D,\ell'|0}(s-\frac{r}{2})\right\rbrace\k_{\ell'}(s-\frac{r}{2})\Bigg]\,,
\end{split}
\ee
where $c_{00}$ is a constant. In the language of an AdS$_2$ effective field theory this simply corresponds a $\phi^4$ contact interaction, with no derivatives. An important observation is that, as shown in \cite{gs}, the $n=0$ equation does not converge. We will use the equation for $n=0$ to determine $c_{00}$, and this leads to the same results of \cite{Mazac:2018mdx}. More explicitly, looking at the case $n=0$, we have
\be
\sum_{\D}C_{\D}N_{\D,0}\left(q^{(s)}_{\D,0|0}(\D_\phi)+2q^{(t)}_{\D,0|0}(\D_\phi)+c_{00}\right)\k_{0}(\D_\phi)=0\,.
\ee
This will give,
\be\label{eq0}
c_{00}=-\left(q^{(s)}_{\D,0|0}(\D_\phi)+2q^{(t)}_{\D,0|0}(\D_\phi)\right)\,.
\ee

For a generic contact term of the type in eq. (\ref{contactterm}) (corresponding to quartic derivative interactions in an AdS$_2$ effective field theory), $f_{\D}(s)$ gets slightly modified to
\be\label{f(s)gen}
\begin{split}
f_{\D}(s)=& \Bigg[\left(\sum_{r=0}^{\infty}\sum_{\ell'=0}^{r}(-1)^r(\Dphi-s)^2_{\frac{r}{2}}  ~~ d _{r,\ell'}(s-\frac{r}{2})\right)\\
&\left\lbrace q^{(s)}_{\D,\ell'|0}(s-\frac{r}{2})\d_{\ell',0}+a_{\ell'}(s-\frac{r}{2})+\left(1+(-1)^{\ell'}\right)q^{(t)}_{\D,\ell'|0}(s-\frac{r}{2})\right\rbrace\k_{\ell'}(s-\frac{r}{2})\Bigg]\,.
\end{split}
\ee
where $a_\ell(s)$ is given in  appendix (\ref{ap:aell}). Note that in the consistency conditions of eqs. (\ref{eq}) and (\ref{2eq}), only even $r$ contributes.



\subsection{Bootstrapping $O(N)$ global symmetry}
Now we consider the case of a $1d$ CFT with $O(N)$ global symmetry. The four-point function of scalar fields $\phi_{i}$ can be expanded as a sum over three irreducible representations of $O(N)$
\begin{equation}
\langle\phi_i \phi_j \phi_k \phi_l\rangle=\delta_{i j}\delta_{k l}\,\mathcal{G}_{S} +\left(\frac{\delta_{ik}\delta_{j l}+\delta_{i l} \delta_{j k}}{2}-\frac{1}{N}\delta_{ij}\delta_{kl}\right)\, \mathcal{G}_{T}+ \frac{(\delta_{i k}\delta_{j l}-\delta_{il}\delta_{j k})}{2}\,\mathcal{G}_{A},
\end{equation}
where $\mathcal{G}_S$ corresponds to the singlet, $\mathcal{G}_T$ to the symmetric traceless and $\mathcal{G}_A$ the anti-symmetric representation.

The sum of exchange Witten diagrams in a certain channel can be decomposed accordingly into three channels, as
\begin{equation}
\begin{split}
\sum_{\Delta,\ell} c_{\Delta,\ell}W^{(i)}(u,v)=& \int [ds] [dt] u^s v^t \Gamma^2(\Delta_{\phi}-s)\Gamma^2(-t) \Gamma^2(s+t) \\
& \Bigg( \delta_{ij}\delta_{kl} M^{S,(i)}(s,t)+(\frac{\delta_{ik}\delta_{j l}+\delta_{i l} \delta_{j k}}{2}-\frac{1}{N}\delta_{ij}\delta_{kl})M^{T,(i)}(s,t)\\
& +\frac{(\delta_{ik}\delta_{j l}-\delta_{i l}\delta_{j k})}{2}M^{A,(i)}\Bigg),
\end{split}
\end{equation}
where $i$ can stand for either of $s,t,u$ channels, and the total crossing symmetric amplitude is be given by
\begin{equation}
\begin{split}
&\mathcal{A}(u,v)=\int [ds] [dt] u^s v^t \Gamma^2(\Delta_{\phi}-s)\Gamma^2(-t) \Gamma^2(s+t) \\
& \times  \Bigg[\delta_{ij}\delta_{kl}\left(M^{S,(s)}(s,t)-\frac{1}{N}M^{T,(s)}(s,t)+\frac{1}{2}(M^{T,(t)}(s,t)+M^{T,(u)}(s,t)+M^{A,(t)}(s,t)\right.\\
&  -M^{A,(u)}(s,t))\Bigg)+\delta_{il}\delta_{jk} \left(M^{S,(t)}(s,t)-\frac{1}{N}M^{T,(t)}(s,t)+\frac{1}{2}(M^{T,(s)}(s,t)+M^{T,(u)}(s,t)\right.\\
&  +M^{A,(s)}(s,t) +M^{A,(u)}(s,t))\Bigg) + \delta_{ik}\delta_{jl}\left(M^{S,(u)}(s,t)-\frac{1}{N}M^{T,(u)}(s,t)+\frac{1}{2}(M^{T,(s)}(s,t)\right.\\
& +M^{T,(t)}(s,t)-M^{A,(s)}(s,t) -M^{A,(t)}(s,t)) \Bigg)\Bigg].
\end{split}
\end{equation}
Again, we decompose the above equation in three irreducible sectors and perform the integral over $t$. The constraint equations for each sector are
\be\label{eqnglobal}
\sum_{\Delta} f^{(i)}_{\D}(\Dphi+n)=0,
\ee
\be\label{eqnglobal2}
\sum_{\Delta} f^{(i)~\prime}_{\D}(\Dphi+n)-q_{dis}^{(i)~\prime}(\Dphi+n)=0,
\ee
where $i$ stands for $S, T, A$ and the corresponding modified blocks take the form
\be\label{eq:S_no_conatct}
\begin{split}
f_{\D}^{(S)}(s)=& \sum_{r=0}\sum_{\ell'=0}^{r}(-1)^r(\Delta_{\phi}-s)_{\frac{r}{2}}^2 d_{r,\ell'}(s-\frac{r}{2})\kappa_{\ell'}(s-\frac{r}{2})\Bigg[ c^{(S)}_{\Delta} q^{(s)}_{\Delta,\ell'|0}(s-\frac{r}{2})\d_{\ell',0}\\
&+\frac{(1+(-1)^{\ell'})}{N}c^{(S)}_{\Delta} q^{(t)}_{\Delta,\ell'|0}(s-\frac{r}{2})
+\frac{(1+(-1)^{\ell'})}{2}(1+\frac{1}{N}-\frac{2}{N^2})c^{(T)}_{\Delta} q^{(t)}_{\Delta.\ell'|0}(s-\frac{r}{2})\\
&-(1-\frac{1}{N})\frac{(1+(-1)^{\ell'})}{2}c^{(A)}_{\Delta}q^{(t)}_{\Delta,\ell'|1}(s-\frac{r}{2})\Bigg]\,,
\end{split}
\ee

\be\label{eq:T_no_conatct}
\begin{split}
 f^{(T)}_{\D}(s)=&\sum_{r=0}\sum_{\ell'=0}^{r}(-1)^r(\Delta_{\phi}-s)_{\frac{r}{2}}^2 d_{r,\ell'}(s-\frac{r}{2})\kappa_{\ell'}(s-\frac{r}{2})\Bigg[(1+(-1)^{\ell'}) c^{(S)}_{\Delta} q^{(t)}_{\Delta,\ell'|0}(s-\frac{r}{2})\\
 &+ c^{(T)}_{\Delta}q^{(s)}_{\Delta,\ell'|0}(s-\frac{r}{2})\d_{\ell',0}+\frac{(1+(-1)^{\ell'})}{2}(1-\frac{2}{N})c^{(T)}_{\Delta} q^{(t)}_{\Delta,\ell'|0}(s-\frac{r}{2})\\
&+c^{(A)}_{\Delta}\frac{(1+(-1^{\ell'})}{2}q^{(t)}_{\Delta,\ell'|1}(s-\frac{r}{2})\Bigg]\,,
\end{split}
\ee

\be\label{eq:A_no_conatct}
\begin{split}
f^{(A)}_{\D}(s)=& \sum_{r=0}\sum_{\ell'=0}^{r}(-1)^r(\Delta_{\phi}-s)_{\frac{r}{2}}^2 d_{r,\ell'}(s-\frac{r}{2})\kappa_{\ell'}(s-\frac{r}{2})\Bigg[-(1-(-1)^{\ell'})c^{(S)}_{\Delta}q^{(t)}_{\Delta,\ell'|0}(s-\frac{r}{2})\\ 
&+\frac{(1-(-1)^{\ell'})}{2}(1+\frac{2}{N})c^{(T)}_{\Delta} q^{(t)}_{\Delta,\ell'|0}(s-\frac{r}{2})+ c^{(A)}_{\Delta}q^{(s)}_{\D,\ell'|1}(s-\frac{r}{2})\delta_{\ell',1}\\
& +\frac{(1-(-1)^{\ell'})}{2}c^{(A)}_{\Delta}q^{(t)}_{\D,\ell'|1}(s-\frac{r}{2})\Bigg]\,.
\end{split}
\ee
The functions $q_{dis}^{(S)}(s)$, $q_{dis}^{(T)}(s)$ and $q_{dis}^{(A)}(s)$ are given  in appendix \ref{ap:ONqdis}.

Now let us look at the form of large $\Delta$ expansion of the consistency condition arising from the singlet sector inserting the GFF OPE coefficients as before. The   coefficient of $z^2\log(z)$ (for $\Dphi=1$) gives the behaviour
\begin{equation} \label{eq1}
f_{\D}^{(S)}(\Delta_{\phi})\sim-\frac{8 \Delta  (N+2) (2 N-1)}{\pi ^2 N^2}+\frac{4 (N+2) (2 N-1)}{\pi ^2 N^2}-\frac{16 (N-1) (2 N-1)}{\pi ^2 \Delta  N^2},
\end{equation}
and from the coefficient of $z^4\log(z)$ we get
\begin{equation} \label{eq2}
f_{\D}^{(S)}(\Delta_{\phi}+1)\sim-\frac{8 \Delta  (N+2) (2 N-1)}{\pi ^2 N^2}+\frac{4 (N+2) (2 N-1)}{\pi ^2 N^2}-\frac{24 (N-1) (2 N-1)}{\pi ^2 \Delta  N^2}.
\end{equation}

These equations have terms proportional to $\Delta$ as well as $\frac{1}{\Delta}$, unlike in the case when there was no $O(N)$ global symmetry, where there was no $\frac{1}{\Delta}$ term. Therefore, with one subtraction we cannot make the terms in the series fall off as $\frac{1}{\Delta^2}$, as required for convergence. So let us look at the summand arising from the cancellation of coefficient of $z^6 \log(z)$ given by
\begin{equation}\label{eq3}
-\frac{8 \Delta  (N+2) (2 N-1)}{\pi ^2 N^2}+\frac{4 (N+2) (2 N-1)}{\pi ^2 N^2}-\frac{40 (N-1) (2 N-1)}{\pi ^2 \Delta  N^2}.
\end{equation}

Now we can take linear combinations of equations \eqref{eq1}, (\ref{eq2}) and (\ref{eq3}) to cancel the divergent term in the expansion. This turns out to be the feature for general $\Delta_{\phi}$. Since we need two subtractions \footnote{Again,this cancellation introduces undetermined parameters, and in this case their number is two. This is equivalent to removing divergences from all equations with two contact terms, which multiply just the diverging pieces. However, such a procedure is again ambiguous upto addition of two constants.}, therefore our $c(s,t)$ turns out to be 
\be
\begin{split}
c(s,t)&=\delta_{ij}\delta_{k l}\sum_{m+n=0}^{1}c_{mn}(-t)_m (s+t)_n+\delta_{il}\delta_{j k}\sum_{m+n=0}^{1}c_{mn}(-s+\Delta_{\phi})_m (s+t)_n\\
&+\delta_{ik}\delta_{j l}\sum_{m+n=0}^{1}c_{mn}(-t)_m (\Delta_{\phi}-s)_n\,.
\end{split}
\ee
We decompose these contact terms in the irreducible sectors and our new equation becomes
\be\label{eqnglobal}
\sum_{\Delta}f^{(i)}_{\D}(\Dphi+n)=0,
\ee

\be\label{eqnglobal2}
\sum_{\Delta}f^{(i)~\prime}_{\D}(\Dphi+n)+q_{dis}^{(i)~\prime}(\Dphi+n)=0,
\ee
and the corresponding modified blocks would take the following modified form compared to eqs.(\ref{eq:S_no_conatct}, \ref{eq:T_no_conatct}, \ref{eq:A_no_conatct}):
\be
\begin{split}
&f_{\D}^{(i)}(s)\rightarrow f_{\D}^{(i)}(s)+\sum_{r=0}\sum_{\ell'=0}^{r}(-1)^r(\Delta_{\phi}-s)_{\frac{r}{2}}^2 d_{r,\ell'}(s-\frac{r}{2})\kappa_{\ell'}(s-\frac{r}{2})\big[~a^{(i)}_{\ell'}(s-\frac{r}{2})~\big],
\end{split}
\ee
where $i$ stands for $S, T, A$ respectively (see appendix \ref{ap:aell}).

\subsection*{Expanding Generalized free fermion as a check}

As a sanity check, we consider the $O(N)$ generalized free fermion model and show that the bootstrap equations are satisfied by putting the spectrum of this model. Since we have constructed the PM basis for  the $O(N)$ model where the double zeros are at the location of $\Delta_n=2\Delta_{\phi}+2n$ the fermionic model serves as a non perturbative example as the dimensions are far away from $\Delta_n$. The four-point function of the $O(N)$ generalized free fermion model is given by
\begin{equation}
\langle\psi(x_1)\psi(x_2)\psi(x_3)\psi(x_4)\rangle=\delta_{ij}\delta_{kl} \mathcal{A}^{S}(z)+(\frac{\delta_{ik}\delta_{jl}+\delta_{il}\delta_{jk}}{2}-\frac{1}{N}\delta_{ij}\delta_{kl} )\mathcal{A}^{T}(z)+\frac{(\delta_{ik}\delta_{jl}-\delta_{il}\delta_{jk})}{2}\mathcal{A}^{A}(z),
\end{equation}
where
\begin{equation}
\begin{split}
& z^{-2\Delta_{\phi}} \mathcal{A}^{S}(z)=\frac{1}{N} (-1+(1-z)^{-2\Delta_{\phi}}+N z^{-2\Delta_{\phi}}),\\
&z^{-2\Delta_{\phi}} \mathcal{A}^{T}(z)= (-1+(1-z)^{-2\Delta_{\phi}}),\\
&z^{-2\Delta_{\phi}} \mathcal{A}^{A}(z)= (-1-(1-z)^{-2\Delta_{\phi}}).
\end{split}
\end{equation}
We can decompose this in $1d$ conformal blocks as\footnote{The spin-1 block is related to this $1d$ scalar block by a factor of $ (-\frac{1}{2})$. The antisymmetric block is derived setting $d=1$ and $\ell=1$, then integrating out $t$ from the higher dimensional blocks. In the second half of the paper we find the OPE coefficients of all channels using the scalar block. Therefore, we have multiplied the antisymmetric block by $(-\frac{1}{2})$ to make the normalizations same.},
\begin{equation}
 \mathcal{A}^{i}(z)=\sum_{\Delta}C^{MFT}_{\Delta} z^{\Delta-2\Delta_{\phi}}\,_2F_1(\Delta,\Delta,2\Delta,z),
\end{equation}
and $\Delta=2\Delta_{\phi}+2n+1$ for singlet and traceless symmetric part and $\Delta=2\Delta_{\phi}+2n$
for the antisymmetric sector. The OPE coefficients are given by,
\begin{equation}
\begin{split}
& C^{T}_{n}=\frac{2 \Gamma^2 (2 n+2 \Delta_\phi +1) \Gamma (2 n+4 \Delta_\phi )}{\Gamma^2 (2 \Delta_\phi ) \Gamma (2 n+2) \Gamma (2 (2 n+2 \Delta_\phi +1)-1)},\\
& C^{S}_{n}=\frac{1}{N}C^{T}_{n},\\
& C^{A}_{n}=-\frac{2 \Gamma \left(2 n+2 \Delta _{\phi }\right){}^2 \Gamma \left(2 n+4 \Delta _{\phi }-1\right)}{\Gamma (2 n+1) \Gamma \left(2 \Delta _{\phi }\right){}^2 \Gamma \left(2 \left(2 n+2 \Delta _{\phi }\right)-1\right)},
\end{split}
\end{equation}
Here we explicitly show how to solve for the constraint equations coming from double poles in $s$. The same procedure will be followed to find the results of section \ref{theoryo(n)global}.  Let us start with eq. (\ref{eqnglobal}) for the singlet exchange in the s-channel at $s=\Delta_{\phi}$ and for illustration purpose we set $\Delta_{\phi}=1$,
\begin{equation} \label{seqdph0}
\begin{split}
& c^{S}_{\Delta}  4\sqrt{\pi} \left(\frac{1}{-\Delta ^2+\Delta +2}+\frac{\psi ^{(1)}\left(\frac{\Delta +1}{2}\right)-\psi ^{(1)}\left(\frac{\Delta }{2}\right)}{N}\right) +c^{T}_{\Delta} \sqrt{\pi}  \left(-\frac{4}{N^2}+\frac{2}{N}+2\right) \Bigg(\psi ^{(1)}\left(\frac{\Delta +1}{2}\right)\\
& -\psi ^{(1)}\left(\frac{\Delta }{2}\right)\Bigg) -c^{A}_{\Delta} \left(1-\frac{1}{N}\right) \frac{\sqrt{\pi } \left((\Delta -1)^2 \Delta ^2 \left(\psi ^{(1)}\left(\frac{\Delta }{2}\right)-\psi ^{(1)}\left(\frac{\Delta +1}{2}\right)\right)-2 (\Delta -1) \Delta -4\right)}{8 (\Delta -1) \Delta }\\
& -\frac{2 (c_{00} (N+2)+c_{01} (N+1))}{N},
\end{split}
\end{equation}
where the last line is coming from the contact term
\begin{equation}
\sum_{r=0}\sum_{\ell'=0}^{r}(-1)^r(\Delta_{\phi}-s)_{\frac{r}{2}}^2 d_{r,\ell'}(s-\frac{r}{2})\kappa_{\ell'}(s-\frac{r}{2})\big[~a^{(s)}_{\ell'}(s-\frac{r}{2})~\big]|_{s=\Delta_{\phi}},
\end{equation}
$a^{(s)}_{\ell'}(s)$ is given by\footnote{For details of contact terms and explicit  expression of $\Omega_{\ell}^{(m,n)}(s)$ we refer the readers to equation (\ref{eq:Omega}). }
\begin{equation}
a_{\ell}^{(S)}(s)=\sum_{m+n=0}^{1}c_{mn}\bigg(\Omega_{\ell}^{(m,n)}(s)+\frac{1}{N}((\Delta_{\phi}-s)_m  \Omega_{\ell}^{(0,n)}(s)+(\Dphi-s)_n  \Omega_{\ell}^{(m,0)}(s))\bigg),
\end{equation}
and at  $s=\Delta_{\phi}+1$,
\begin{equation}\label{seqdph1}
\begin{split}
& \frac{2}{5\sqrt{\pi}}\sum_{\Delta}c^{S}_{\Delta}\Bigg\{\pi  \left(\frac{9}{-\Delta ^2+\Delta +2}+\frac{1}{-\Delta ^2+\Delta +12}+\frac{5 ((\Delta -1) \Delta +3)}{N}\right)+\frac{5 \pi  ((\Delta -2) \Delta +3)}{2 N}\\
&\times  \left(\Delta ^2+2\right)\left( \psi ^{(1)}\left(\frac{\Delta +1}{2}\right)-\psi ^{(1)}\left(\frac{\Delta }{2}\right)\right)\Bigg\}+\frac{\sqrt{\pi}  \sum_{\Delta}c^{T}_{\Delta} \left(N^2+N-2\right)}{2 N^2} \Bigg\{ 6+2\Delta(\Delta-1)\\
& + ((\Delta -2) \Delta +3) \left(\Delta ^2+2\right) \left(\psi ^{(1)}\left(\frac{\Delta +1}{2}\right)-\psi ^{(1)}\left(\frac{\Delta }{2}\right)\right)\Bigg\}-\sum_{\Delta}c^{A}_{\Delta}\left(1-\frac{1}{N}\right)\Bigg\{\frac{\sqrt{\pi }}{32} (\Delta -1)\\
& \times \left.  \Delta  ((\Delta -2) \Delta +3) \left(\Delta ^2+2\right) \left(\psi ^{(1)}\left(\frac{\Delta }{2}\right)-\psi ^{(1)}\left(\frac{\Delta +1}{2}\right)\right)-\frac{\sqrt{\pi } ((\Delta -1) \Delta +3) }{16 (\Delta -1) \Delta }\right.\\
& \times \left((\Delta -1)^2 \Delta ^2+4\right)\Bigg\}-\frac{6 c_{00} (N+2)+c_{01} (7 N+5)}{3 N}.
\end{split}
\end{equation}
The constraint equation at $s=\Delta_{\phi}+2$ is
\begin{equation}\label{seqdph2}
\begin{split}
&\sum_{\Delta}-\frac{\sqrt{\pi}  c^{T}_{\Delta} \left(N^2+N-2\right)}{288 N^2}\Bigg(-2 ((\Delta -1) \Delta  ((\Delta -1) \Delta  ((\Delta -1) \Delta +19)+490)+1300)\\
& +((\Delta -1) \Delta  ((\Delta -1) \Delta +4) ((\Delta -1) \Delta  ((\Delta -1) \Delta +16)+444)+1440) \left(\psi ^{(1)}\left(\frac{\Delta }{2}\right)\right.\\
&\left. -\psi ^{(1)}\left(\frac{\Delta +1}{2}\right)\right)\Bigg)+\sum_{\Delta}c^{S}_{\Delta}  \frac{\sqrt{\pi}}{144} \Bigg(\frac{2 ((\Delta -1) \Delta  ((\Delta -1) \Delta  ((\Delta -1) \Delta +19)+490)+1300)}{N}\\
& -\frac{576 ((\Delta -1) \Delta  ((\Delta -1) \Delta -39)+274)}{(\Delta -6) (\Delta -4) (\Delta -2) (\Delta +1) (\Delta +3) (\Delta +5)}+\frac{1}{N}\left(\psi ^{(1)}\left(\frac{\Delta +1}{2}\right)-\psi ^{(1)}\left(\frac{\Delta }{2}\right)\right)\\
& \times (\Delta -1) \Delta  ((\Delta -1) \Delta +4) ((\Delta -1) \Delta  ((\Delta -1) \Delta +16)+444)+1440 \Bigg)-\sum_{\Delta}c^{A}_{\Delta}(1-\frac{1}{N})\\
& \frac{ \sqrt{\pi }}{4608 (\Delta -1) \Delta }\Bigg(-2 ((\Delta -1) \Delta  ((\Delta -1) \Delta  ((\Delta -1) \Delta  ((\Delta -1) \Delta  ((\Delta -1) \Delta +19)+490)+1300)\\
& +576)+2880)\left.+(\Delta -1)^2 \Delta ^2 ((\Delta -1) \Delta  ((\Delta -1) \Delta +4) ((\Delta -1) \Delta  ((\Delta -1) \Delta +16)+444)\right.\\
& +1440)  \left.  \left(\psi ^{(1)}\left(\frac{\Delta }{2}\right)-\psi ^{(1)}\left(\frac{\Delta +1}{2}\right)\right)\right)-\frac{2 c_{00} (N+2)+c_{01} (3 N+1)}{N}\,.
\end{split}
\end{equation}
Now we use eq. (\ref{seqdph0}) and (\ref{seqdph1}) to fix the two unknown parameters $c_{00}$, $c_{01}$ which come from two contact terms. Unlike the $N=1$ case, here we lose two equations.  Although here we have used constraint equations arising from $(s=\Delta_{\phi})$ and $(s=\Delta_{\phi}+1)$, one can use any other two equations. Now we substitute $c_{00}$, $c_{01}$  in eq. (\ref{seqdph2}),
\begin{equation}
\begin{split}
&\sum_{n} \Bigg[\frac{1}{18 \pi ^2 N^2}(n+1)\Bigg((N-1) N (2 n+1) (4 n+3) \left(-n (2 n+3) \left(n (2 n+3) \left(4 n^2+6 n+25\right)+73\right)\right.\\
&\left. -51\right) + 2 (2 n+3) (4 n+5)\Bigg( -\frac{36 N}{8 n^6+60 n^5+98 n^4-135 n^3-313 n^2+30 n+72}+1+(n+1)\\
& \times  (2 n+3)(n (2 n+5) (2 n (2 n+5)+31)+104)\Bigg)-(2 n+3) (4 n+5)(-313-n (2 n+5)(197+n \\
& (2 n+5)(2 n (2 n+5)+37)))(N^2+N-2)\Bigg)-\frac{(N+1) (2 n+3)^2 (4 n+5) (n (2 n+5)+5)}{9 \pi ^2 N}\\
& \times (n (2 n+5) (n (2 n+5)+14)+36)+\frac{2(1+n)^2}{9\pi^2 N} \Bigg(2 n (n+2) (n (n+2) (8 n (n+2) (22 n (n+2)\\
&+193) +4395)+5022)+3996+N (n+1) (n (n+2) (16 n (n+2) (n (n+2) (8 n (n+2)+97)\\
& +366)+8469) +4104)\Bigg)  \left(\psi ^{(1)}(n+1)-\psi ^{(1)}\left(n+\frac{3}{2}\right)\right)\Bigg]
\end{split}
\end{equation}
Quite remarkably, 
this sum identically vanishes\footnote{We can use these formulas and derivatives of these to perform the sums \be \begin{split} & \sum_{n=0}^{\infty}\psi ^{(1)}(n+\frac{3}{2})e^{\epsilon n }=\frac{1}{2-2 e^{\epsilon }}\left(\frac{4 \text{Li}_2\left(-\sqrt{e^{\epsilon }}\right)}{\sqrt{e^{\epsilon }}}-\frac{4 \text{Li}_2\left(\sqrt{e^{\epsilon }}\right)}{\sqrt{e^{\epsilon }}}+\pi ^2\right),\\
&  \sum_{n=0}^{\infty}\psi ^{(1)}(n+1)e^{\epsilon n }=\frac{1}{6-6 e^{\epsilon }}\Bigg(\pi ^2-6 \text{Li}_2\left(e^{\epsilon }\right)\Bigg)\,.
\end{split}
\ee} and thereby shows the correctness of our basis. Now we can also look at the equation for antisymmetric sector at $s=\Delta_{\phi}$:
\begin{equation}
\begin{split}
& c_{01}+\sum_{\Delta} \frac{3 \sqrt{\pi } c^{A}_{\Delta}}{16} \Bigg(\frac{4}{-\Delta ^2+\Delta +6}-\frac{2 \left((\Delta -1)^2 \Delta ^2+2\right)}{(\Delta -1) \Delta }+(\Delta -1) \Delta  ((\Delta -1) \Delta +1)\\
&\times  \Bigg(\psi ^{(1)}\left(\frac{\Delta }{2}\right)-\psi ^{(1)}\left(\frac{\Delta +1}{2}\right)\Bigg)\Bigg)+6 \sqrt{\pi }\sum_{\Delta} \left(c^{T}_{\Delta} \left(\frac{1}{N}+\frac{1}{2}\right)-c^{S}_{\Delta}\right)\Bigg(2+(1+\Delta -1) \Delta )\\
& \times \left.\psi ^{(1)}\left(\frac{\Delta +1}{2}\right)-\psi ^{(1)}\left(\frac{\Delta }{2}\right)\right).
\end{split}
\end{equation}
Again, the sum can be shown to vanish.

\section{Implementing Bootstrap}\label{sec:impPMboot}
In this section, we will derive solutions for the CFT data for a variety of cases. First, we shall focus on deformation of generalised free field (GFF) theories, corresponding to contact interactions in an effective field theory (EFT) on AdS$_2$. As we solve for the CFT data at higher orders in perturbation theory, we shall find new divergences in our equations, which are completely analogous to the need for new counterterms in the perturbative expansion of non-renormalizable theories. Later we shall consider the case of exchanged operators, and observe the resonance-like behaviour of anomalous dimensions, when the dimension of the double trace operators becomes comparable with that of the exchanged operator. A similar behaviour was observed in higher dimensions in \cite{Fitzpatrick:2010zm}.  

\subsection{Deforming away from GFF}
We begin by writing the OPE coefficients as 
\be\label{OPEfrom}
 C_{\D}=C_{n}=C^{(0)}_n+C^{(1)}_n g+C^{(2)}_n g^2+\dots\,\,,
\ee
and the deformations away from the GFF dimensions 
\be\label{ANMform}
\D=\D_{n}=2\D_\phi+2n+\g^{(1)}_n g+\g^{(2)}_n g^2+\dots\,\,.
\ee 
In what follows we will ignore operator mixing. 

\subsubsection{Contact term without derivatives}\label{sec:withoutderivative}

Throughout the discussion in this section, we take $\D_0=2\Dphi+g$ as definition of the coupling $g$. First of all, we aim to solve for the leading order $O(g^0)$ of eq. (\ref{2eq}), which will give us the OPE coefficient at leading order, \textit{i.e.} $C^{(0)}_n$. For the case $n=0$, eq. (\ref{2eq}) reads
\be \label{op0}
-2+k_0(\Dphi)~\partial_s\left[q^{(s)}_{\D,0|0}(\D_\phi)+2q^{(t)}_{\D,0|0}(\D_\phi)\right]_{\D_\phi}=0\,.
\ee
We expand eq. (\ref{op0}) in powers of $g$ and the leading order $O(g^0)$ in  eq. (\ref{op0}) is 
\be
-2+C^{(0)}_0=0\,,
\ee
which enables us to solve for the OPE coefficient of the operator $\D_0$ at leading order:
\be
C^{(0)}_0=2\,.
\ee
Similarly we now consider $n=1$ case in eq. (\ref{2eq}), and  at leading order  $O(g^0)$ we have 
\be
\frac{\Delta_\phi  (2 \Delta_\phi +1) (2 \Delta_\phi  (C^{(0)}_0-4)+C^{(0)}_0-2)}{8 \Delta_\phi +2}+C^{(0)}_1=0\,,
\ee
which gives 
\be
C^{(0)}_1=\frac{2 \Delta_\phi ^2 (2 \Delta_\phi +1)}{4 \Delta_\phi +1}\,.
\ee
One can solve eq. (\ref{2eq}) for any $n$ at leading order $O(g^0)$. The OPE coefficient of the operator $\D_n$ at leading order reads
\be\label{ope0}
C^{(0)}_n=\frac{2 \Gamma^2 \left(2 n+2 \Delta _{\phi }\right) \Gamma \left(2 n+4 \Delta _{\phi }-1\right)}{\Gamma (2 n+1) \Gamma^2\left(2 \Delta _{\phi }\right) \Gamma \left(4 n+4 \Delta _{\phi }-1\right)}\,.
\ee

After knowing the leading order OPE coefficient \textit{i.e} $C^{(0)}_n$, we now proceed towards solving for the leading order anomalous dimension $\g_{n}^{(1)}$. For example, we first consider the $n=1$ case of eq. (\ref{eq})
\be\label{eq1-0}
\begin{split}
\sum_{\D}C_{\D}N_{\D,0}&\bigg(\left[\left(q^{(s)}_{\D,0|0}(\D_\phi+1)+2q^{(t)}_{\D,0|0}(\D_\phi+1)\right)-\left(q^{(s)}_{\D,0|0}(\D_\phi)+2q^{(t)}_{\D,0|0}(\D_\phi)\right)\right]\k_{0}(\D_\phi+1)+\\
&q^{(t)}_{\D,2|0}(\D_\phi)\k_{2}(\D_\phi)\bigg)=0\,,
\end{split}
\ee
which will give the anomalous dimension of the operator $\D_1$. Now if we use eq. (\ref{ope0}) to replace in  eq. (\ref{eq1-0}), and expand in power of $g$, at $O(g)$ we find 
\be
\frac{g~ 2^{2 \Delta _{\phi }-1} \left(2 \g^{(1)}_{1} \left(2 \Delta _{\phi }+1\right){}^2-4 \Delta _{\phi }^2+\Delta _{\phi }\right) \Gamma \left(\Delta _{\phi }+\frac{1}{2}\right)}{\sqrt{\pi } \Delta _{\phi } \left(4 \Delta _{\phi }+1\right) \Gamma^3 \left(\Delta _{\phi }\right)}=0\,,
\ee
which immediately gives the anomalous dimension at $O(g)$ of the operator $\D_1$
\be
\g^{(1)}_{1}=\frac{\Delta _{\phi } \left(4 \Delta _{\phi }-1\right)}{2 \left(2 \Delta _{\phi }+1\right){}^2}\,.
\ee
One can solve eq. (\ref{eq}) for any $n$ at leading order $O(g)$ with the help of eq. (\ref{ope0}) which will give the anomalous dimension of the operator $\D_n$ at order $O(g)$. We get a general formula 

\be\label{anm1}
\g^{(1)}_{n}=\frac{\left(\frac{1}{2}\right)_n \left(\left(\Delta _{\phi }\right)_n\right){}^2 \left(2 \Delta _{\phi }-\frac{1}{2}\right)_n}{(1)_n \left(2 \Delta _{\phi }\right)_n \left(\left(\Delta _{\phi }+\frac{1}{2}\right)_n\right){}^2}\,.
\ee

Now we are in a position to solve for the first perturbative correction to the OPE coefficients, namely $C_{n}^{(1)}$. To this end, we have to expand eq. (\ref{2eq}) in powers of $g$ with the help of eqs. (\ref{ope0}) and (\ref{anm1}). For the operator $\D_n$, there is an enhancement in $n$-th case of eq. (\ref{eq})  which comes from the  $s$-channel Witten block.
One can solve eq. (\ref{2eq}) for any $n$ at $O(g)$. We give the results for general $n$ 
\be\label{ope1}
C^{(1)}_n=\frac{1}{2}\frac{\partial}{\partial n}\left(C^{(0)}_n \gamma^{(1)}_n\right)
\ee

\subsubsection*{ $O(g^2)$ calculations }\label{g2}
\paragraph{}

We now outline the computation of anomalous dimensions at order $O(g^2)$. To this end, we use eqs. (\ref{ope0}, \ref{anm1}, \ref{ope1}) to replace in (\ref{eq}). Again, the operator $\D_n$ will cause an enhancement in $n$-th case of eq. (\ref{eq}). This allows to find the anomalous dimension at $O(g^2)$, which we call $\g_{n}^{(2)}$, in terms of sum of  specific combinations of $\g_j^{(1)}$, $C_j^{(0)}$ and $C_j^{(1)}$. Let us explain this procedure looking, for simplicity, at the case $n=1$. At $O(g^2)$, $\gamma^{(2)}_j$ of all operators don't appear in eq. (\ref{eq1-0}), except  there is an enhancement for $\D_1$ because of a pole in $s-$ channel exchange coefficient, {\it i.e.} $\g^{(2)}_{1}$ only appear in the equation in terms of sum of  specific combinations of $\g_j^{(1)}$, $C_j^{(0)}$ and $C_j^{(1)}$\footnote{these quantities had already been determined in the previous order.}at $O(g^2)$. One can find $\g^{(2)}_{1}$, provided that the sum over all operator contributions in eq. (\ref{eq1-0}) is of  $O(g^2)$. For illustration purpose, we show the key steps for $\Dphi=1$. From equation (\ref{eq1-0})we find at order $O(g^2)$,
\be
\frac{1}{300} \left(10331-1050 \pi ^2\right) g^2+\frac{\left(22680 \gamma _1^{(2)}-712950 \pi ^2+7036049\right) g^2}{37800}+g^2\sum_{j=2}^{\infty}S_j=0~~;\D_\phi=1\,,
\ee
where we have used eq. (\ref{ope0}, \ref{anm1}, \ref{ope1}). For general $\Dphi$, $S_j$ has a complicated expression in terms of ${}_7F_6$, but for $\Dphi=1$ it simplifies and  is given by 
\be
\begin{split}
S_j&=\frac{4 j+3}{4 (j-1)^2 j^2 (j+1) (2 j+1) (2 j+3) (2 j+5)}\\
&\Bigg(-5 (j (j (2 j (j (2 j (4 j (j (2 j+11)+22)+85)+79)+24)+105)+105)+42)\\
&+2 (j-1)^2 j^2 (2 j+3) (2 j+5) (2 j (2 j+3) (j (2 j+3)+4)+7) \left(\psi ^{(1)}(j-1)-\psi ^{(1)}\left(j+\frac{3}{2}\right)\right)\Bigg)\,.
\end{split}
\ee
It is possible to evaluate the sum $\sum_j S_j$ exactly. In order to get a 5 decimal place accuracy, \footnote{\label{foot:accuracy} It is sufficient to take only 4 operators in the sum for 4 decimal place accuracy, since after adding one more \textit{{\it i.e.},} including 5 operators the effect is on $5^{th}$ decimal place and again after 5 more operators, \textit{{\it i.e.}} total 10 operators the effect is on $6^{th}$ decimal place.} we truncate the sum over $j$ at $j=15$, \textit{{\it i.e.}} including 15 operators, we find $\sum_{j=2}^{15} S_j= 0.001015$, which gives the anomalous dimension of $\D_1$ at order $O(g^2)$
\be
\g^{(2)}_{1}=0.19796~~;\quad \D_\phi=1\,.
\ee
Proceeding in a completely analogous way, one can find $\g^{(2)}_n$ for any $\Dphi$.  We tabulate   numerical calculations taking only 15 operators in the sum, up to 5 decimal place accuracy (for $\g^{(2)}_2,\g^{(2)}_3$ the conclusion is same as described in footnote \ref{foot:accuracy}) in the sum over operators
\begin{table}[hbt!]
\centering
\begin{tabular}{|c|c|c|c|c|c|c|c|c|c|c||}
\hline
& $\D_\phi=\frac{3}{4}$ & $\D_\phi=1$ & $\D_\phi=\frac{15}{10}$ & $\D_\phi=2$ & $\D_\phi=\frac{25}{10}$ & $\D_\phi=3$ & $\D_\phi=\frac{35}{10}$ & $\D_\phi=4$ & $\D_\phi=\frac{45}{10}$ \\
& & & & & & & & &  \\\hline
$\g^{(2)}_1$ & 0.13299& 0.19796 & 0.24196 & 0.25656 & 0.26362 & 0.26795 & 0.27103 & 0.27341 & 0.27534  \\
& & & & & & & & &  \\\hline
$\g^{(2)}_2$ & 0.04740& 0.08273 & 0.11792 & 0.13774 & 0.15128 & 0.16138 & 0.16931  & 0.17576 & 0.18112  \\
& & & & & & & & &  \\\hline
$\g^{(2)}_3$& 0.02383  & 0.04471 & 0.06877 & 0.08483 & 0.09709 & 0.10696 & 0.11515 & 0.12208 & 0.12806  \\
& & & & & & & & &\\\hline
\end{tabular}
\end{table}

Using a similar method, solving  $O(g^2)$ of eq. (\ref{2eq}) up to 3 decimal place accuracy\footnote{It is sufficient to take only 3 operators in the sum over operators for $C_{0}^{(2)}, \gamma_{1}^{(3)}$ and 4 operators for $C_{1}^{(2)}$ for the same reason we described in footnote \ref{foot:accuracy}.} we find
\be
\begin{array}{|c|c|c|c|c|c|}\hline & {\Delta_{\phi}=1} & {\Delta_{\phi}=\frac{15}{10}} & {\Delta_{\phi}=2} & {\Delta_{\phi}=\frac{25}{10}} & {\Delta_{\phi}=3} \\ \hline C_{0}^{(2)} & {4.186} & {1.955} & {1.419} & {1.375} & {1.518}  \\ \hline C_{1}^{(2)} & {-0.249} & {-0.443} & {-0.518} & {-0.4563} & {-0.215} \\ \hline\end{array}
\ee
In principle, one can go to higher orders but one will need to work harder by retaining more operators. Here we tabulate the $O(g^3)$ anomalous dimension \textit{{\it i.e.}} $\g^{(3)}_1$ for various $\Dphi$ up to 3 decimal place accuracy.
\be
\begin{array}{|c|c|c|c|c|c|}\hline & {\Delta_{\phi}=1} & {\Delta_{\phi}=\frac{15}{10}} & {\Delta_{\phi}=2} & {\Delta_{\phi}=\frac{25}{10}} & {\Delta_{\phi}=3} \\ \hline  \gamma_{1}^{(3)} & {0.257} & {0.193} & {0.182} & {0.179} & {0.178} \\ \hline\end{array}
\ee

Our results agree with \cite{mp} for  $\Delta_{\phi}=1$  and for other integer values of $\Dphi$ say $\Dphi=2,3,4\dots$ we verified our results with the transcendentality method which described below in Sections \ref{transc - tree} and \ref{sec:transc - loop}.

\subsubsection*{Non-perturbative bound}
We can also give non-perturbative bounds on the dimension of the leading operator numerically. In order to do so, we plot the r.h.s. of eq. (\ref{eq}) for the case $n=1$, \textit{{\it i.e.}} $f_{\D}(\Dphi+1)$, as a function of $\D$ for various values of $\Dphi$. From figure (\ref{fn}), it is clear that  beyond $\D=2\Dphi+2$ the function $f_{\D}(\Dphi+1)$ is always positive. Hence, in order to satisfy eq. (\ref{eq}) the leading operator should have dimension between $2\Dphi$ and $2\Dphi+2$, where we have  assumed that the leading operator has the dimension of the form $\D= 2\Dphi+g $ with $g>0$. One immediate conclusion is $g<2$.

\begin{figure}[hbt!]
\centering
    \includegraphics[width=0.7\linewidth]{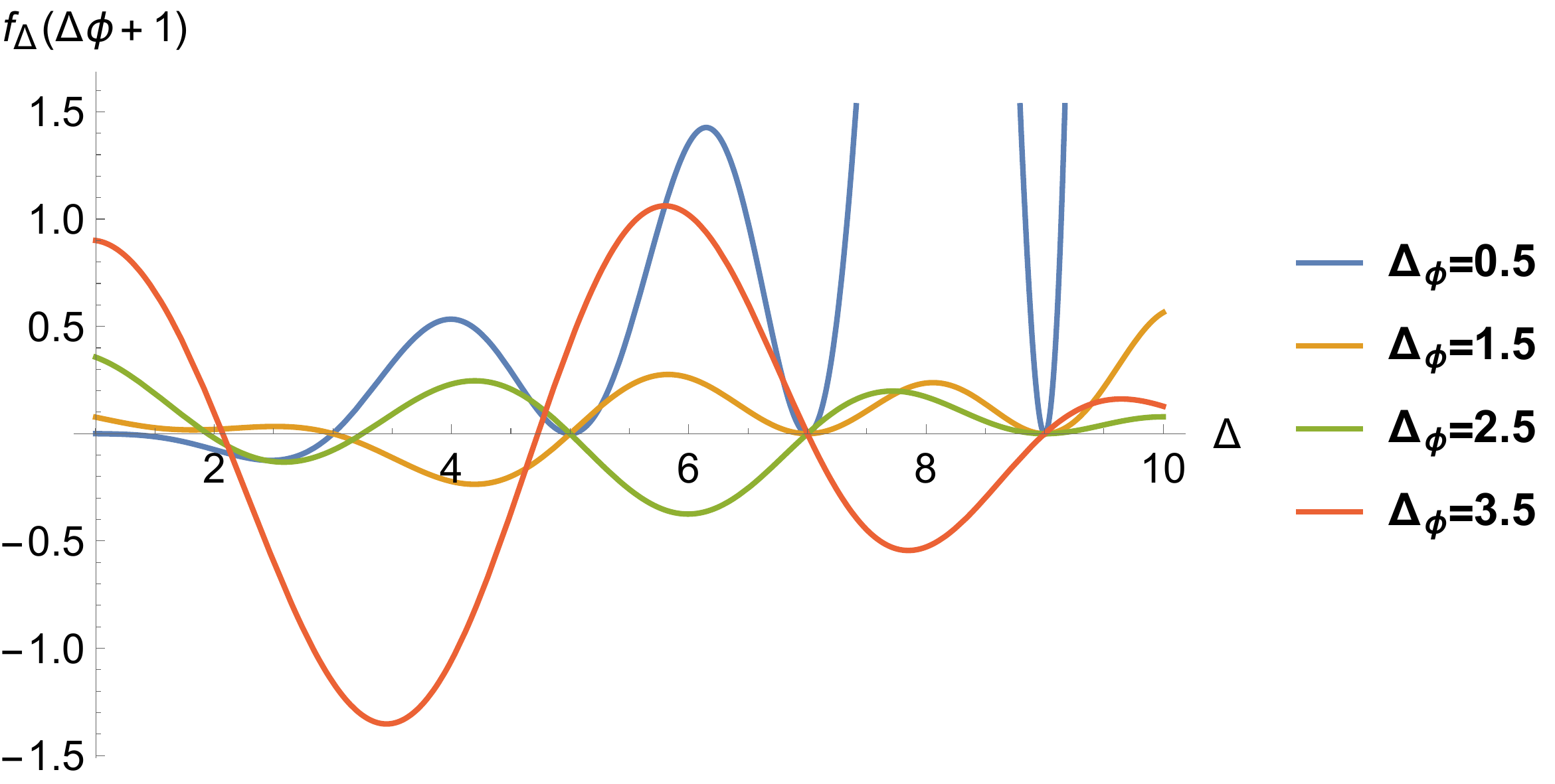}
    \caption{$f_{\D}(\Dphi+1)~vs ~ \D$}
   \label{fn}
\end{figure}

\subsubsection{Generic contact interactions}\label{sec:genericcontact}
\subsection*{Four-derivatives contact term}
We now consider an AdS EFT with interaction $(\nabla\phi)^4$. Since each power of $s$ and $t$ accounts for two derivatives, the contact term in this particular case will be given by eq. (\ref{contactterm}) with $L=2$. Hence, we have to solve eq. (\ref{eq}) and eq. (\ref{2eq}) with four unknowns $c_{00},~c_{01},~c_{11},~c_{02}$. Although we have four unknowns, it turns out that only two of them are independent and fixed in terms of $\D_0=2\Dphi+\g^{(1)}_0 g$, $\D_1=2\Dphi+2+\g^{(1)}_1 g$. In other words, we can solve the equations $n=0,1$ and these equations will fix the unknown coefficients coming from the contact term. Then, we can evaluate $n=2$ equation which we can solve to get anomalous dimension of $\D_2=6+\gamma^{(1)}_{2} \,g+\gamma^{(2)}_2 g^2+\dots$ up to $O(g^2)$. 

If we chose the normalization $\g^{(1)}_0=0$ and $\g^{(1)}_1=1$, then at $O(g)$ eq. (\ref{eq}) for $n=0$ reads
\be
\frac{\sqrt{\pi } 2^{-2 \Delta _{\phi }-1} \left(\Delta _{\phi } \left(\Delta _{\phi } \left(\left(c_{11}+4 c_{02}\right) \Delta _{\phi }+c_{11}+8 \left(c_{01}+c_{02}\right)\right)+4 \left(3 c_{00}+c_{01}+c_{02}\right)\right)+6 c_{00}\right) \Gamma^3 \left(\Delta _{\phi }\right)}{\Gamma \left(\Delta _{\phi }+\frac{3}{2}\right)}=0\,,
\ee
while for $n=1$ it reads
\be
\begin{split}
&\frac{1}{4} \Delta _{\phi } \Bigg(-\frac{8 g \Delta _{\phi } \left(2 \Delta _{\phi }+1\right)}{4 \Delta _{\phi }+1} -\frac{\sqrt{\pi } 4^{-\Delta _{\phi }} \Gamma^3 \left(\Delta _{\phi }\right)}{\Gamma \left(\Delta _{\phi }+\frac{5}{2}\right)}\bigg\{6 c_{00} \left(2 \Delta _{\phi }+3\right) \left(\Delta _{\phi } \left(3 \Delta _{\phi }+2\right)+1\right)\\
&+\Delta _{\phi } \bigg[4 c_{01} \left(2 \Delta _{\phi }+3\right) \left(\Delta _{\phi } \left(3 \Delta _{\phi }+2\right)+1\right)\\
&+\left(\Delta _{\phi }+1\right) \left(4 c_{02} \left(\Delta _{\phi } \left(\Delta _{\phi } \left(3 \Delta _{\phi }+10\right)+11\right)+3\right)+3 c_{11} \Delta _{\phi } \left(\Delta _{\phi }^2+\Delta _{\phi }-1\right)\right)\bigg]\bigg\}\Bigg)=0\,.
\end{split}
\ee
From these two equations we find 
\be
\begin{split}
c_{11}&= \frac{g 2^{2 \Delta _{\phi }+3} \left(\Delta _{\phi }+1\right) \left(2 \Delta _{\phi }+1\right) \Gamma \left(\Delta _{\phi }+\frac{5}{2}\right)}{\sqrt{\pi } \left(4 \Delta _{\phi }+1\right) \left(4 \Delta _{\phi }+3\right) \Gamma \left(\Delta _{\phi }\right){}^2 \Gamma \left(\Delta _{\phi }+3\right)}-\frac{6 c_{00}+4 c_{01} \Delta _{\phi }}{\Delta _{\phi }^2+\Delta _{\phi }}\,,\\
c_{02}&= -\frac{1}{2 \left(\Delta _{\phi }+1\right)}\left(\frac{3 c_{00}}{\Delta _{\phi }}+2 c_{01}+\frac{g 4^{\Delta _{\phi }+1} \left(2 \Delta _{\phi }+1\right) \Gamma \left(\Delta _{\phi }+\frac{5}{2}\right)}{\sqrt{\pi } \left(\Delta _{\phi }+2\right) \left(4 \Delta _{\phi }+1\right) \left(4 \Delta _{\phi }+3\right) \Gamma \left(\Delta _{\phi }\right){}^3}\right)\,.
\end{split}
\ee
We can then replace the values of $c_{11}$ and $c_{02}$ in eqs. (\ref{eq}, \ref{f(s)gen}), and following a similar method to that of section \ref{sec:withoutderivative}, we can solve for the anomalous dimensions. At $O(g)$ we find 
\be
\begin{split}
& \g^{(1)}_2=\frac{3 (4 \Delta_\phi +3) (\Delta_\phi  (\Delta_\phi  (4 \Delta_\phi +21)+29)+10)}{(2 \Delta_\phi +1) (2 \Delta_\phi +3) (2 \Delta_\phi +5) (4 \Delta_\phi +1)}\,.
\end{split}
\ee
One can in general solve for any $n$, for example we give here general formula for $\Dphi=1$ 
\begin{equation}\label{gammatreeq=$1d$=1}
\begin{split}
\g^{(1)}_n&=\frac{6 n (2 n+3) (2n^2+3n+2)}{35 (n+1) (2 n+1)}~~;~\Dphi=1\,.
\end{split}
\end{equation}

Similarly, at $O(g^2)$ one has to solve for $c_{11}$ and $c_{02}$. However, there are now infinite sums in both direct and crossed channel. The final expression for $\Delta_{\phi}=1$ is
\begin{equation}
\begin{split}
\g_2^{(2)} &=\frac{1}{1039500}\bigg(-780516 (\gamma _0^{(1)})^2-160380 \gamma _0^{(1)}-123234734766 (\gamma _1^{(1)})^2-73920 \gamma _1^{(1)}-374220 \gamma_0^{(2)}\\
&+2661120\gamma_1^{(2)} +12486474000 \pi ^2 (\gamma _1^{(1)})^2+1148755608000 \pi ^2 (\gamma _2^{(1)})^2-11337763363365 (\gamma _2^{(1)})^2\\
&+1230020 \gamma _2^{(1)} +4365900 \tilde{S}\bigg)\,,
\end{split}
\end{equation}
where $\tilde{S}$ is given by
\begin{small}
\begin{equation}
\begin{split}
\tilde{S}&=\sum_{n=3}^{\infty}\frac{-(n+1) (2 n+1) (4 n+3)}{252 (n-2) (n-1) n (2 n+3) (2 n+5) (2 n+7)}\\
& 756-n (2 n+3) \left(n (2 n+3) \left(n (2 n+3) \left(n (2 n+3) \left(7 n (2 n+3) \left(4 n^2+6 n-13\right)-1978\right)+5274\right)+25747\right)\right.\\
&\left.-174\right)  +2 (n-2) (n-1) n^2 (2 n+3)^2 (2 n+5) (2 n+7)n (2 n+3) (7 n (2 n+3) (n (2 n+3)+14)+313)+258\\
&\left(\psi ^{(1)}(n+1)-\psi ^{(1)}(n+\frac{3}{2})\right)(\hat{\gamma}_n^{(1)})^2\\
& =\frac{7128000 \zeta (3)-733149980362800 \pi ^2+7235899768502069}{424462500}\,.
\end{split}
\end{equation}
\end{small}
Assuming there is no correction to $\D_0$ and $\D_1$, or in other words that $\g_0^{(2)}=0=\g_1^{(2)}$, we find
\begin{equation} \label{gamma2d1q1_2}
\g_2^{(2)}=\frac{6019}{10500}+\frac{432 \zeta (3)}{6125}.
\end{equation}
We can solve other equations as well to get the loop corrections to dimensions of operators with higher n values, e.g. solving $n=3$ equation will give 
\be \label{gamma2d1q1_3}
\g_3^{(2)}=\frac{575916557}{240100000}+\frac{10692 \zeta (3)}{60025}\,.
\ee
These expressions match with eq. (\ref{gq$1d$1comp}) which we derived using the transcendentality method that we shall describe later.

\subsection*{Eight-derivatives contact term}
Now we turn to a theory with interaction $(\nabla^2\phi)^4$. We will have $L=4$ in (\ref{contactterm}). In this case it turns out that there are only 3 independent $c_{mn}$, therefore we use the $n=0,1,2$ equations to fix these unknown parameters. Then normalizing $\Delta_0=2\Dphi+\g^{(1)}_0 g$, $\Delta_1=2\Dphi+2+\g^{(1)}_1 g$, $\Delta_2=2\Dphi+4+\g^{(1)}_2 g$ to be $\g^{(1)}_0=0,~\g^{(1)}_1=0,~\g^{(1)}_2=1$, we find
\begin{equation}
\begin{split}
& \g^{(1)}_3=\frac{5 (4 \Delta_\phi +7) (\Delta_\phi  (\Delta_\phi  (4 \Delta \phi +43)+126)+108)}{2 (\Delta_\phi +1) (2 \Delta_\phi +5) (2 \Delta_\phi +9) (4 \Delta_\phi +3)}\,.
\end{split}
\end{equation}
One can solve for any $n$, for example
\begin{equation}\label{gammatreeq=2d=1}
\g^{(1)}_n=\frac{5 (n-1) n (2 n+3) (2 n+5) \left(7 n (2 n+3) \left(2 n^2+3 n+11\right)+124\right)}{108108 (n+1) (2 n+1)}~~;~\Dphi=1\,.
\end{equation}

Now let us consider the equation for $n=3$ with $\Delta_{\phi}=1$, and expand it to $O(g^2)$. It turns out that at this order the infinite sum is divergent. Then we have to use add a multiple, of say, the $n=4$ equation, in order for the sum to converge. Since by doing so we lose the $n=4$ equation, we have another undetermined parameter. If we evaluate loop corrections to the anomalous dimension of $\Delta_4$, normalizing the loop corrections to $\Delta_0$, $\Delta_1$, $\Delta_2$, $\Delta_3$ to be 0, we find
\begin{equation} \label{gamma2d1q2_4}
\g^{(2)}_4=\frac{502604844863939}{17816700021120}+\frac{1700 \zeta (3)}{91091}\,.
\end{equation} 
The appearance of another undetermined parameter at loop level is a feature of effective field theory and can be understood in terms of the usual perturbative renormalisation. It is indeed equivalent to the necessity of adding a new counter-term at this order, since we started with a non-renormalisable interaction.
\subsection{Theories with $O(N)$ global symmetry} \label{theoryo(n)global}

In this subsection, we shall study the bootstrap problem for a theory of $N$ scalars, with $O(N)$ global symmetry.

\subsubsection{With contact term degree 1 in $s$ and $t$ }
We now consider the addition of a contact term which is a polynomial of degree 1 in $s$ and $t$. There are two independent unknowns, which we fix using the equations for $n=0,1$ in terms of unknowns $\g_0^{(1)},\,\g_1^{(1)}$ where $\D_0=2\Dphi+\g_0^{(1)} g,\,\D_1=2\Dphi+2+\g_1^{(1)}g$. For illustration purposes, we solve here for singlet sector of the consistency conditions at tree level. A similar method is applicable for the symmetric traceless and the anti-symmetric sector. From the identity contribution, we immediately find the OPE at leading order (similarly to what discussed in section (\ref{sec:withoutderivative})):
\be \label{ope0ON_$1d$egree}
\begin{split}
C_{n}^{S(0)}=&\frac{\sqrt{\pi } 2^{-4 \Delta _{\phi }-4 n+3} \Gamma \left(2 \left(n+\Delta _{\phi }\right)\right) \Gamma \left(2 n+4 \Delta _{\phi }-1\right)}{N \Gamma (2 n+1) \Gamma^2 \left(2 \Delta _{\phi }\right) \Gamma \left(2 n+2 \Delta _{\phi }-\frac{1}{2}\right)}\,,\\
C_{n}^{T(0)}=&\frac{\sqrt{\pi } 2^{-4 \Delta _{\phi }-4 n+3} \Gamma \left(2 \left(n+\Delta _{\phi }\right)\right) \Gamma \left(2 n+4 \Delta _{\phi }-1\right)}{\Gamma (2 n+1) \Gamma^2 \left(2 \Delta _{\phi }\right) \Gamma \left(2 n+2 \Delta _{\phi }-\frac{1}{2}\right)}\,,\\
C_{n}^{A(0)}=&-\frac{\sqrt{\pi } 2^{-4 \Delta _{\phi }-4 n+1} \Gamma \left(2 n+4 \Delta _{\phi }\right) \Gamma \left(2 \left(n+\Delta _{\phi }\right)+1\right)}{\Gamma (2 n+2) \Gamma^2 \left(2 \Delta _{\phi }\right) \Gamma \left(2 n+2 \Delta _{\phi }+\frac{1}{2}\right)}\,.
\end{split}
\ee 
From $n=0,1$ of eq.\eqref{eqnglobal} at $O(g)$, one can solve for $c_{00},~c_{01}$. Then replacing $c_{00},\,c_{01}$ in the case $n=2$ of eq.\eqref{eqnglobal} and solving for $\g_{2}^{(1)}$, we find (in the singlet sector)
\be
\g_2^{S(1)}=\frac{3 \left(\Delta _{\phi }+1\right){}^2 \left(4 \left(2 \Delta _{\phi }+1\right){}^2 \left(4 \Delta _{\phi }+3\right) \gamma _1^{\text{S(1)}}-\Delta _{\phi } \left(4 \Delta _{\phi }-1\right) \left(4 \Delta _{\phi }+5\right) \gamma _0^{\text{S(1)}}\right)}{4 \left(2 \Delta _{\phi }+1\right){}^3 \left(2 \Delta _{\phi }+3\right){}^2}\,.
\ee
Similarly, replacing $c_{00},~c_{01}$ in the case $n=0$ of eq.\eqref{eqnglobal2}, and solving for $C_0^{(1)}$, we find (in the singlet sector)
\be
\begin{split}
C_0^{S(1)}&=\frac{\Delta _{\phi } \gamma _0^{\text{S(1)}}+2 \left(2 \Delta _{\phi }+1\right){}^2 \gamma _1^{\text{S(1)}}}{N \Delta _{\phi } \left(4 \Delta _{\phi }+1\right)}\\
&\times\left(-4 (1+\log (4)) \Delta _{\phi }+3 \left(4 \Delta _{\phi }+1\right) \left(\psi ^{(0)}\left(\Delta _{\phi }\right)+\gamma \right)+\left(-4 \Delta _{\phi }-1\right) H_{\Delta _{\phi }-\frac{1}{2}}+1-\log (4)\right)
\end{split}
\ee
With this method, one can solve the equations \eqref{eqnglobal} and eq.\eqref{eqnglobal2} for general $n$ and $\Dphi$, and the results are found to agree with those of section \ref{sec:O(N)tree}, where we compute the same quantities using the transcendentality method.

\subsubsection*{$O(g^2)$ calculations}
Let us now consider the following perturbative order. We start again with eq. (\ref{eqnglobal}), and consider the singlet sector, using the OPE data derived in the previous section. Once again, we have two contact terms; therefore, we can use the first three equations to get\footnote{Here all the infinite sums can be evaluated as before, since they also involve PolyGamma functions.},
\begin{equation}
\gamma^{S(2)}_2=\frac{-3600 \gamma^{S(2)}_0+33600 \gamma^{S(2)}_1+9056 N^2-55927 N+33362}{30000}\,,
\end{equation}
while in the traceless symmetric operator we find
\begin{equation}
\gamma^{T(2)}_2=\frac{-3600  \gamma^{T(2)}_0+33600 \gamma^{T(2)}_1-18669 N+33362}{30000}\,.
\end{equation}
The anomalous dimensions for the other double-trace operators and in the antisymmetric sector can be found in a very similar way, in terms of two unknowns parameters. Instead of repeating the results here, we refer to section \ref{loop O(N)} for a closed form expression of these quantities.

\subsubsection{Contact term of degree 2 in $s$ and $t$}
Let us now consider the addition of a contact term which is a polynomial of degree 2 of $s$ and $t$. There are three independent unknowns which we fix by using the equations (\ref{eqnglobal}) corresponding to singlet sectors for $n=0,1,2$. We fix them in terms of unknowns $\g_0^{(1)},~\g_1^{(1)},~\g_2^{(1)}$, where $\D_0=2\Dphi+\g_0^{(1)} g,~\D_1=2\Dphi+2+\g_1^{(1)}g,~\D_2=2\Dphi+4+\g_2^{(1)}g$. For illustration purpose, we solve here for the singlet sector of the consistency conditions at tree level. A similar method is applicable for the symmetric traceless and anti-symmetric sector. From the identity contribution, we immediately find the OPE at leading order, with the same result as in eq.\eqref{ope0ON_$1d$egree}. In the singlet sector, the consistency conditions for $n=0,1,2$ at order $O(g)$ allow to solve for $c_{11}, c_{01}, c_{02}$ in terms of the unknowns $\g_0^{S(1)},~\g_1^{S(1)},~\g_2^{S(1)}, c_{00}$. Then replacing this values into the equation for $n=3$, we get for instance (for $\Dphi=1$)
\be
\g_3^{S(1)}=\frac{1}{980} \left(143 \gamma _0^{S\text{(1)}}-1638 \gamma _1^{S\text{(1)}}+2475 \gamma _2^{S\text{(1)}}\right),~~\Dphi=1\,.
\ee
For general $\Dphi, N$, we find after substituting $\g_0^{S(1)}=0, \g_1^{S(1)}=0,\g_2^{S(1)}=1$,
\be
\begin{split}
\g_3^{S(1)}=\frac{5 \left(\Delta _{\phi }+2\right) \left(4 \Delta _{\phi }+7\right) \left(\Delta _{\phi } \left(2 \Delta _{\phi } \left(2 (N+3) \Delta _{\phi }+13 N+41\right)+51 N+173\right)+3 (9 N+37)\right)}{2 \left(2 \Delta _{\phi }+5\right) \left(2 \Delta _{\phi }+7\right) \left(\Delta _{\phi } \left(2 \Delta _{\phi } \left(2 (N+3) \Delta _{\phi }+9 N+29\right)+25 N+87\right)+9 N+39\right)}\,.
\end{split}
\ee 

\paragraph{}
For an application of the PM bootstrap in $O(N)$ theories, let us consider the case $N=5$. This is relevant for the 1/2-BPS Wilson-Maldacena line in $\mathcal{N}=4$ super Yang-Mills, considered for instance in  \cite{Giombi:2017cqn} and, from the bootstrap perspective, in \cite{Liendo:2018ukf}.
To compare the results we set $\g_0^{S(1)}=-5,~ \g_1^{S(1)}=-10,~\g_2^{S(1)}=-19$. Then at tree level we find, for $\Dphi=1$,
\begin{equation}
\begin{split}
& \gamma^{S(1)}_n=-2n^2-3n-5\,,\\
& \gamma^{T(1)}_n=-2n^2-3n\,,\\
& \gamma^{A(1)}_n=-2n^2-5n-4.
\end{split}
\end{equation}
Similarly one can find $O(g)$ OPE coefficient and this is exactly same as given in \cite{Giombi:2017cqn}.



\subsection{Effective field theory--exchange interaction}
In this section we consider an interaction of the following form
\begin{equation}
\mathcal{L}=\lambda_4 \phi^4 +\lambda_{\mathcal{O}} \phi^2 \mathcal{O},
\end{equation} 
and we solve the corresponding PM bootstrap equations:
\be\label{pmeq} 
\sum_{\D}C_{\D}N_{\D,0}f_{\D}(\D_\phi+n)=0~;~n= 0,1,2,3\dots;
\ee
\be
\sum_{\D}\left(C_{\D}N_{\D,0}f_{\D}'(\D_\phi+n)\right)+ q_{dis}'(\D_\phi+n)=0~;~n= 0,1,2,3\dots ;
\ee
where
\be\label{functional} 
\begin{split}
f_{\D}(s)=& \Bigg[\left(\sum_{r=0}^{\infty}\sum_{\ell'=0}^{r}(-1)^r(\Dphi-s)^2_{\frac{r}{2}}  ~~ d _{r,\ell'}(s-\frac{r}{2})\right)\\
&\left\lbrace q^{(s)}_{\D,\ell'|0}(s-\frac{r}{2})\d_{\ell',0}+\lambda_4~\d_{\ell',0}+\left(1+(-1)^{\ell'}\right)q^{(t)}_{\D,\ell'|0}(s-\frac{r}{2})\right\rbrace\k_{\ell'}(s-\frac{r}{2})\Bigg]\,,
\end{split}
\ee

Unlike the previous cases, where the spectrum contains only double trace operators, now we have the exchange of an operator of dimension $\Delta$. Therefore in the leading order this operator will contribute in both the channels. In our normalization, we take the OPE coefficient of $\mathcal{O}$ in the $\phi \times \phi$ OPE to be $\pi^2$, and choose $\Delta=3$. Then, we get the following leading anomalous dimension for the double field operators:
\begin{equation}
\begin{split}
\gamma_n^{(1)}&=\frac{1}{4 g (3 - 2 n)^2 n (1 + n) (-1 + 2 n)}60 g (1 + 2 n) (-9 + 
   2 n (11 + 2 n (-6 + n (7 + 2 n \\
   & \times (1 - 6 n + 4 n^2))))) 40 g n^2 (n+1) (4 (n-2) n+3)^2 \left(4 n^2-2 n+1\right) \left(\psi ^{(1)}(n)-\psi ^{(1)}\left(n-\frac{3}{2}\right)\right)\\
   & -\pi  \lambda_4 (3-2 n)^2 (n+1)\,.
\end{split}
\end{equation}
In the large-$n$ limit, this falls of as $1/n^2$, which is expected since $\gamma^{(1)}_{n,0}$ should depend on the dimensionless combination  $(\lambda_{\mathcal{O}}/n)^2$. We can also demand a softer fall of for large $n$, and that will fix the undetermined coefficient $\lambda_4$. With this requirement, the final expression is
\begin{equation}
\begin{split}
\gamma_n^{(1)} &=\frac{5 (n+1) (2 n (6 n (2 n (2 n (2 n+9)+29)+43)+97)+65)+90}{(1-2 n)^2 (n+1) (n+2)}\\
& -\frac{10 (2 n+1) \left(2 n^2+n-1\right)^2 \left(4 n^2+6 n+3\right)}{(1-2 n)^2 (n+1)} \left(\psi ^{(1)}\left(n-\frac{1}{2}\right)-\psi ^{(1)}\Bigg(n+1\Bigg)\right).
\end{split}
\end{equation}

A similar exercise for $\Delta=5$ gives us,
\begin{equation}
\begin{split}
\gamma_n^{(1)} &=\frac{7}{2}\bigg(-\frac{1}{n+3}+\frac{2}{2 n-3}+\frac{38880}{2 n-1}+\frac{15120}{(1-2 n)^2}+\frac{18}{n+1}+11316\\
&+(n+1) ((n+1) (42 n (4 n (n (2 n+9)+23)+149)+7279)+6442)\bigg)\\
& -7(n+1) (2 n+1) (n (2 n+3) (7 n (2 n+3) (n (2 n+3)+8)+130)+90)\\
&  \left(\psi ^{(1)}\left(n-\frac{1}{2}\right)-\psi ^{(1)}(n+1)\right).
\end{split}
\end{equation}
\begin{figure}[h!]
  \centering
  \begin{subfigure}[b]{0.48\linewidth}
    \includegraphics[width=\linewidth]{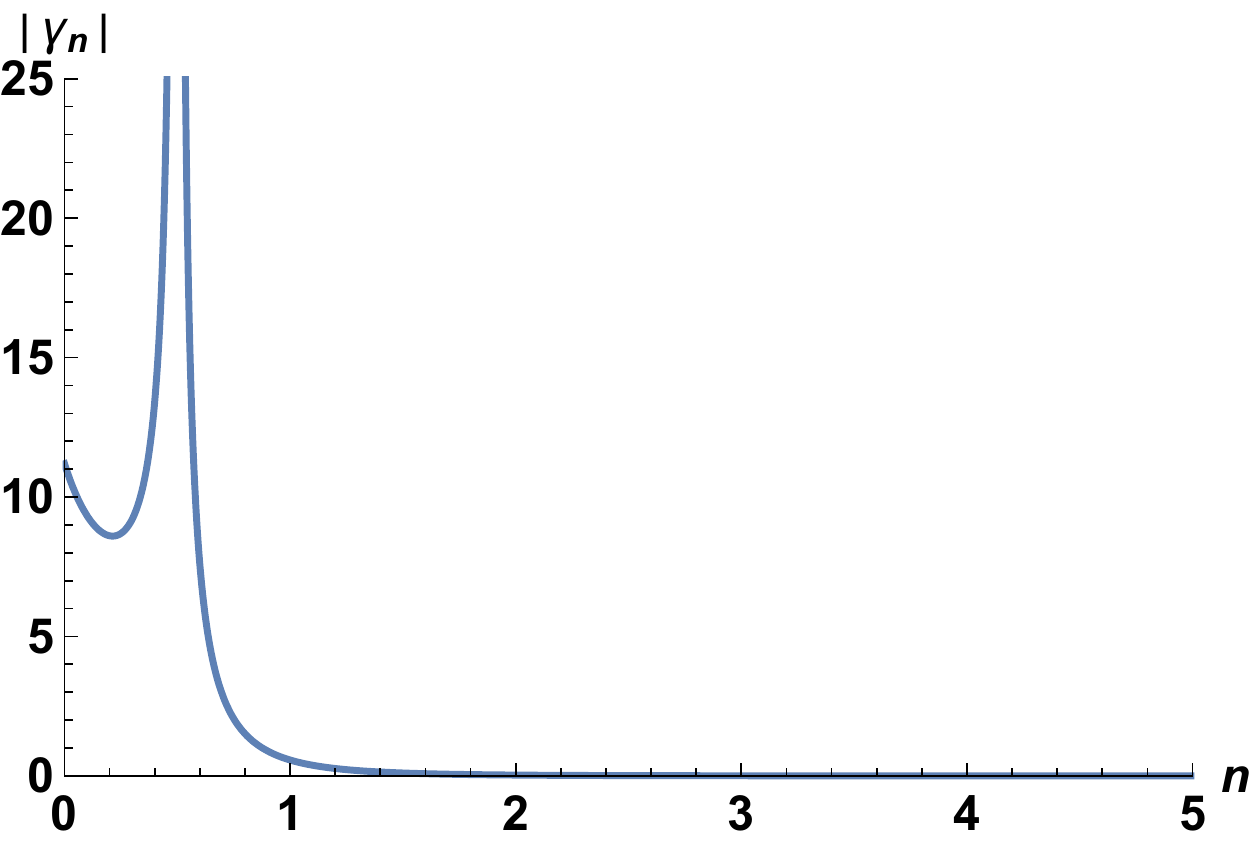}
    \caption{$\Delta=3$}
  \end{subfigure}
   \begin{subfigure}[b]{0.48\linewidth}
    \includegraphics[width=\linewidth]{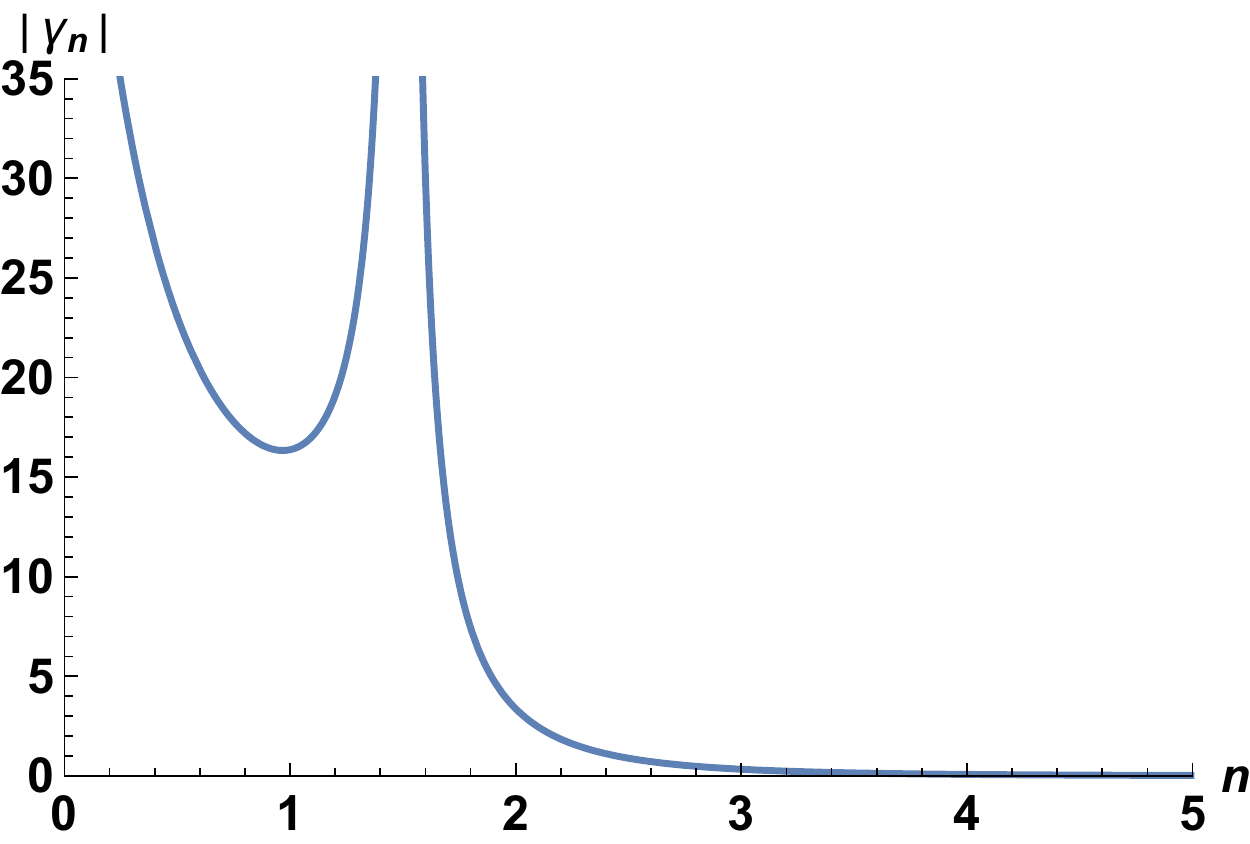}
    \caption{$\Delta=5$}
  \end{subfigure}
  \caption{$\left|\gamma_n\right|~vs~n$}
  \label{gamma}
\end{figure}
It can be seen from figure \ref{gamma} that the anomalous dimension blows up when $n$ is such that one of the double trace operators has dimension of order $\Delta$, this is analogous to a resonance for scattering amplitudes. We also checked that if we solve the equations in large $\Delta$ limit, each order in $1/\Delta$ grow with higher powers of $n$. This is consistent with the fact that in large $\Delta$ limit we get an effective theory where at each order we generate new derivative contact diagrams. We shall discuss this in more detail in section \ref{sec:treeexchanges}.

\subsection{Tower of exchange operators}

In this section, we consider a scenario where, in addition to the ``double-trace'' operators, an infinite number of ``single-trace'' operators is exchanged in the OPE. The goal is to understand whether the requirement of convergence of the infinite sums over these exchanged operators gives sum constraints on their OPE coefficients.

In order to do so, we solve the PM bootstrap equations, and compute the contribution of this tower of exchanged operators to the anomalous dimensions of the double-trace operators. If we let $C_{\Delta}$ be the OPE coefficients of the exchange with dimension $\Delta$, from \eqref{pmeq} we find
\begin{small}
\be
\begin{split}
& \gamma^{(1)}_0=-\sum_{\Delta}\frac{1}{4 (\Delta -2) (\Delta +1)}\bigg(\frac{\pi  \lambda_4 \left(\Delta ^2-\Delta -2\right)}{g}+\frac{C_{\D} 4^{2-\Delta } \Gamma (2 \Delta )}{\Gamma^2 \left(1-\frac{\Delta }{2}\right) \Gamma^4  \left(\frac{\Delta }{2}\right)\Gamma^2 \left(\frac{\Delta +1}{2}\right)}\\
& \times  \left(\pi  \left(\Delta ^2-\Delta -2\right) \left(\psi ^{(1)}\left(\frac{\Delta }{2}\right)-\psi ^{(1)}\left(\frac{\Delta +1}{2}\right)\right)+\pi \right)\bigg)\,,\\
&\gamma_{1}^{(1)}=\frac{1}{24(\Delta-4)(\Delta+3)}\Bigg\{-\frac{\pi \lambda_4\left(\Delta^{2}-\Delta-12\right)}{g}+\frac{C_{\D} 4^{2-\Delta} \Gamma(2 \Delta)}{\Gamma^{2}\left(1-\frac{\Delta}{2}\right) \Gamma^{4}\left(\frac{\Delta}{2}\right) \Gamma^{2}\left(\frac{\Delta+1}{2}\right)}\\
&\bigg[\pi\left(5 \Delta^{4}-10 \Delta^{3}-40 \Delta^{2}+45 \Delta-181\right)+\frac{1}{2} \pi\left(5 \Delta^{6}-15 \Delta^{5}-25 \Delta^{4}+75 \Delta^{3}-268 \Delta^{2}+228 \Delta-144\right)\\
&\left(\psi^{(1)}\left(\frac{\Delta+1}{2}\right)-\psi^{(1)}\left(\frac{\Delta}{2}\right)\right)\bigg] \Bigg\}\,.
\end{split}
\ee
\end{small}

In the previous expressions, $\lambda_4$, comes from the contact term that we added to the basis of Polyakov blocks in order to have a complete basis with convergent sums. In order to understand the behaviour of $C_{\Delta}$ that is required for convergence of the sum of eq. \eqref{pmeq}, we use for $C_{\Delta}$ the values of the mean field theory OPE coefficients. If we expand for large $\Delta$, we find
\begin{equation}\label{diveq}
\gamma^{(1)}_0\sim \left(-\frac{48}{\pi ^2 \Delta ^4}-\frac{32}{\pi ^2 \Delta ^3}-\frac{12 \Delta }{\pi ^2}+\frac{6}{\pi ^2}\right) \sin ^2\left(\frac{\pi  \Delta }{2}\right)\,,
\end{equation}
which grows linearly in $\Delta$. Hence, the sum over $\Delta$ is divergent and we conclude that we need OPE coefficients of the tower of exchanged operators to decay faster than the MFT OPE coefficient. However, so far we have neglected the presence of the free parameter $\lambda_4$, which multiplies the contact term added to the basis of Polyakov blocks\footnote{This is the same ambiguity that we found in PM bootstrap.}. As already done previously, we work with subtracted equations n such a way that the dependence on $\lambda_4$ cancels. Doing so, we find
\begin{align*}
& \frac{1}{6}\gamma^{(1)}_0-\gamma^{(1)}_1\\
& =\sum_\Delta \frac{5 C_{\Delta} \Gamma (2 \Delta )}{12 \pi ^2 (\Delta -4) (\Delta -2) (\Delta +1) (\Delta +3) \Gamma (\Delta )^2}\left(-2(70+(\Delta-1)\Delta(-18+(\Delta-1)\Delta(-11\right.\\
&+(\Delta-1)\Delta)))\left. +(\Delta-4)(\Delta-2)(\Delta+1)(\Delta+3)(2+(\Delta-1)\Delta(4+(\Delta-1)\Delta))\right.\\
& \times  \left.  \left(\psi^{(1)}(\frac{\Delta}{2})-\psi^{(1)}(\frac{1+\Delta}{2})\right)\right) \sin ^2\left(\frac{\pi  \Delta }{2}\right)\,.
\end{align*}
If we expand this in the large $\Delta$ limit, using again the MFT OPE coefficients for $C_{\Delta}$, we find
\begin{equation}
\frac{1}{6}\gamma^{(1)}_0-\gamma^{(1)}_1\sim \left(\frac{280}{\pi ^2 \Delta ^4}+\frac{560}{3 \pi ^2 \Delta ^3}\right) \sin ^2\left(\frac{\pi  \Delta }{2}\right)\,,
\end{equation}
which falls off faster than \eqref{diveq}, thus leading to a convergent expression. This turns out to be true for the anomalous dimensions of all double-trace operators. Hence, we conclude that if there is a tower of exchanged operators, then their OPE coefficients $C_{\Delta}$ must grow slower than $C^{MFT}_{\Delta}\Delta^2$.

\section{Transcendentality ansatz - tree level}\label{transc - tree}

In this section, we implement an alternative technique, which allows to find analytic expressions for one-dimensional correlators $\cA(z)$ up to $\cO(g^2)$, under certain assumptions. While working only for integer values of the (external) operators dimension $\D_{\f}$, this provides an independent check of the validity of some results obtained with PM bootstrap. Furthermore, this approach will enable us to find analytical results for the anomalous dimensions and OPE coefficients up to $\cO(g^2)$ for some values of $\D_{\f}$. Our work also extends and generalises the approach of \cite{Liendo:2018ukf, Gimenez-Grau:2019hez} to one loop and more general types of interactions\footnote{In both \cite{Liendo:2018ukf} and \cite{Gimenez-Grau:2019hez} supersymmetric theories were considered, and this puts restrictions on the type of vertices and the relations between themselves. In particular, in \cite{Liendo:2018ukf} a one-loop correlator was computed, but the high amount of supersymmetry of their model notably simplifies the transcendentality structure, allowing at most transcendentality two (as opposed to four, for generic one-loop correlators with integer external dimensions).}.

The strategy, which will be described in more detail in the next subsections, is to make an ansatz for the correlator $\cA(z)$ in terms of products of rational and transcendental functions, up to some maximal transcendentality which is fixed order by order in perturbation theory. The idea that in some theories the perturbative expansion is governed by transcendentality principles dates back to \cite{Kotikov:2001sc}, where drawing from some massive Feynman diagrams computations such as \cite{Fleischer:1998nb, Fleischer:1998nc}, such principles were first observed in the case of $\cN=4$ super Yang-Mills (SYM) in four dimensions. The idea was later expanded in \cite{Bern:2005iz, Beisert:2006ez}, and it enabled computations to be carried out up to seven loops for the anomalous dimensions of twist-two operators in the planar version of $\cN=4$ SYM \cite{Marboe:2016igj}. One may argue that $\cN=4$ SYM is a rather special theory, being the maximally supersymmetric theory in four dimensions and supposed to be integrable. However, it was then found in \cite{Blumlein:1998if, Vogt:2004mw, Moch:2004pa} that the QCD splitting function can be expressed, in perturbation theory, in terms of harmonic sums of given weight, that appear in the expansion of harmonic polylogarithms. Interesting observations about transcendentality in a CFT context were made in \cite{aldayeps}, where the Lorentzian inversion formula of \cite{caronhuot} was applied, which in $d>2$ allows to reconstruct the CFT data from the double discontinuity of correlators. One of the upshots of \cite{aldayeps} is that there is a dictionary between the transcendentality of the double discontinuity and that of the corresponding CFT data. Finally, let us also mention that interesting structures related to the transcendentality of AdS$_3$ correlators were recently found in \cite{Carmi:2019ocp} at one loop, from a computation of individual Witten diagrams\footnote{Note that the technique applied in this case is very different from that of the present paper: for instance, individual Witten diagrams are not crossing-symmetric.}.

In the present paper, we shall observe that similar ideas apply to weakly-coupled one-dimensional CFTs dual to scalar Effective Field Theories (EFTs) on a fixed AdS$_2$ background. In particular, we shall find that for integer dimension $\df$ of the external operators, and order by order in perturbation theory, the class of functions that can appear in a correlator is rather restricted: it amounts to products of rational functions and harmonic polylogarithms, with a maximal transcendentality that is fixed by the perturbative order. This is found to be true at tree level and at one loop both in theories with only one, self-interacting field, and in theories with a multiplet of fields and an $O(N)$ global symmetry. In all cases that we consider, we were also able to express the CFT data corresponding to these correlators in terms of harmonic sums, and the upshot is that a very similar transcendentality principle is satisfied by the CFT data.
\subsection*{The strategy}
Let us now illustrate the principles that we use to bootstrap correlators using this principle of maximal transcendentality. First, based on some external input or on guesswork, one has to establish the maximal transcendentality of the correlator under inspection. Then, an ansatz is given in terms of products between rational functions and a basis of suitable transcendental functions. Once an appropriate ansatz is established, the following principles are employed to fix the correlator\footnote{Up to some ambiguities, whose meaning will be discuss later and is completely analogous to the contact terms that one needs to add to the PM basis.}:
\begin{itemize}
\item {\bf Crossing symmetry}. Very much like in higher dimensions, we demand that the correlator $\cA(z)$ of four scalars with identical scaling dimension $\D_{\f}$ satisfies
\be 
(1-z)^{2\,\D_{\f}} \, \cA(z) =
z^{2\,\D_{\f}} \, \cA(1-z),
\ee
when $z$ is in the so-called ``crossing region'', where both sides are analytic \cite{mp}. While being one of the pillars of the conformal bootstrap, this alone is not sufficient to fix the correlator.
\item {\bf Symmetry of the conformal blocks}. We shall exploit that the one-dimensional conformal blocks
\be 
G_{\D}(z) \, = \, z^\D \, {}_2F_1(\D, \, \D; \, 2\D; \, z),
\ee
satisfy
\be \label{CB_exchange12}
 G_{\D}\left( \frac{z}{z-1} \right)\, = \, \frac{(1-z)^{\D}}{(z-1)^{\D}} \, G_{\D}(z).
\ee
Note that since the argument of the first conformal block is on the branch cut for $z\in (0,1)$, one has to specify an analytical continuation in $z$. In particular, we have
\be 
\begin{cases}
z=x+i\,\e \,\ra \, \frac{(1-z)^{\D}}{(z-1)^{\D}}=e^{-i\pi\,\D},\\
z=x-i\,\e \,\ra \, \frac{(1-z)^{\D}}{(z-1)^{\D}}=e^{+i\pi\,\D},
\end{cases}
x\in (0,1), \,\, \e \in \mathbb{R}^+.
\ee
We will only be able to find analytic expressions for the correlators when $\df$ is integer, and the relevant operators in the OPE are double trace operators of dimension $\D=2\df+2n+\g_n$. Hence, for integer $\df$, we actually have $e^{\pm i\pi\,\D}=e^{\pm i\pi\,\g_n}$. Then, we have to expand this in perturbation theory, and if we choose for instance the continuation $z=x-i\,\e$, we get\footnote{We shall express the four-point functions in terms of harmonic polylogarithms, and the analytical continuations around the branch cuts of these functions must agree with the choice that one makes for this expansion.}
\be 
e^{+i\pi\,\g_n}=1 + g \, \left( i \, \pi  \, \g^{(1)}_n \right) + g^2 \, \left( i \, \pi  \, \g^{(2)}_n - \frac{\pi^2}{2} \, \left( \g^{(1)}_n \right)^2 \right)  + \cO(g^3).
\ee
Therefore, we can constrain the correlator using
\be 
\cA \left( \frac{z}{z-1} \right)=\cA(z)+\sum_n C_n\,\left[ e^{i\pi\,\g_n}-1 \right]\,G_{2\df+2n+\g_n}(z),
\ee
expanded order by order in perturbation theory.
\item {\bf ``AdS unitarity method''}. That is, we recall that, at least in a small $z$ expansion, at fixed order $\cO(g^{L+1})$ ($L$ loops in AdS) in perturbation theory, the terms proportional to $\log^{L+1} (z)$ are fixed by the solution at tree level ($L=0$). This can be easily seen with a perturbative expansion of the sum over conformal blocks around MFT. In particular, at one loop ($L=1$) the part of the correlator that is proportional to $\log^2 (z)$ is
\be 
\cA(z)|_{\log^2 (z)}=\frac{1}{2}\,\sum_{n=0}^{\infty}C^{(0)}_n\,\left( \g^{(1)}_n \right)^2 \, G_{2\D_{\f}+2n}(z),
\ee
which is entirely fixed in terms of MFT OPE coefficients and tree-level anomalous dimensions. This was first employed in \cite{Aharony:2016dwx} to find some one-loop correlators in $d=2$ and $d=4$, and later applied to $\cN=4$ SYM in \cite{Alday:2018pdi, Alday:2018kkw}, allowing to compute one-loop superstring amplitudes via the AdS/CFT correspondence.
\item {\bf OPE expansion}. It must be possible to write the correlator $\cA(z)$ as a sum over conformal blocks. In particular, since the sum over conformal blocks is done perturbatively around the GFF double trace dimensions $2\,\D_{\f} + 2\,n$, we shall use that in a small $z$ expansion of $\cA(z)$ the lowest power of $z$ that can be present is $z^{2\D_{\f}}$.
\end{itemize}
In the remainder of this section, we shall apply this method to tree-level correlators, both in the case of one single trace operator and of models with $O(N)$ global symmetry. Finally, we will also consider the exchange of one single trace operator at tree level.

\subsection{Contact terms, single field}\label{treesinglefield}

At tree-level, a correlator $\cA^{(1)}(z)$ can be written as 
\begin{align}
\begin{split}
\cA^{(1)}(z)\, &= \,
\sum_{n}\left(C^{(1)}_n+\12 C^{(0)}_n\,\g^{(1)}_n\,\frac{\pr}{\pr n}\right)\,G_{2\D_{\f}+2n}(z)\\
&= \,\sum_{n} z^{2\D_{\f}+2n} \left(
\left( C^{(1)}_n+\12 C^{(0)}_n\,\g^{(1)}_n\,\frac{\pr}{\pr n} \right) + C^{(0)}_n\,\g^{(1)}_n \log(z)\right)\,F_{2\D_{\f}+2n}(z),
\end{split}
\end{align} 
where we have introduced the notation $F_{\b}(z)={}_2F_1(\b,\b;2\b;z)$.

We want to construct contact term solutions, and in $d>2$ these correspond to solutions whose singularities at $z=0$ and $z=1$ are {\it not} enhanced with respect to that of a single conformal block \cite{PolchinskiPenedones}. In fact, these are often referred to as truncated solutions, as the corresponding CFT data have compact support in the spin, and therefore the sum over spin actually truncates. In $d=2$ and $d=4$, the conformal blocks are found to be essentially products of hypergeometric functions, and for integer external dimensions, twist and spin they reduce to functions of transcendentality two. In $d=1$, the conformal blocks contain only one hypergeometric function, and therefore we propose an ansatz of transcendentality one, of the form
\be 
\cA^{(1)}(z)=\frac{1}{(1-z)^{2\df}}\,\frac{1}{z^k\,(1-z)^k}\, 
\Big( p_1(z)+p_2(z)\,\log(z)+p_3(z)\log(1-z) \Big),
\ee
where $p_m(z)$ ($m=1,2,3$) are polynomials in $z$, and $k\in\mathbb{Z}$. With this, one can implement the strategy outlined above, and fix the polynomials varying the integer $k$ and their degree. 

For any integer $\df>0$, we find an infinite number of solutions. This is expected, since an AdS effective field theory can have contact terms with arbitrarily high number of derivatives. To classify such solutions, we note that the values of the exponent $k$ allowed by our constraints are $k=2q-1$, with $q \in \mathbb{N}$. The corresponding anomalous dimensions $\g^{(1)}_n$ have the following behaviour for $n \ra \infty$:
\be \label{ReggeGammaTreeDer}
\g^{(1)}_n \, \sim \, n^{4q-2} \qq (n \ra \infty). 
\ee
As we shall discuss at length in Section \ref{sec:Reggelimit}, this is closely related to the behaviour of the correlator in the Regge limit, as defined in \cite{mp}, and we find that for solutions with $k=2q-1$ the Regge limit of the correlator is
\be 
\left(\frac{1}{2}+i\,t\right)^{-2\df}\,\cA\left(\frac{1}{2}+i\,t\right) \, \sim
\, t^{2q-1} \qquad (t \ra \infty). 
\ee

This, however, does not fix all the free parameters. In particular, at fixed $q=Q$, there are still exactly $Q+1$ free parameters. This comes from the possibility to add any linear combinations of solutions with $q<Q$ without affecting the large $n$ behaviour of the anomalous dimensions. It also corresponds to the number of contact terms that one has to add to the PM basis for contact terms with derivatives at tree level. To fix this ambiguity, as a convention we set to zero the anomalous dimensions of the first $q$ double trace operators
\be
\g_n^{(1)}|_q=0 \quad (0\le n \le q-1).
\ee
This amounts to fixing $q$ free parameters, while the last one can be seen as an overall normalization, which we fix in such a way that
\be
\g_n^{(1)}|_q=1 \quad \text{for} \quad n=q.
\ee
With these conventions, we can express the correlators for any integer $\df\ge 1$ and $q\ge 0$ as
\begin{align}
\begin{split}
\cA^{(1)}(z)=\frac{1}{(1-z)^{2\df}}\Big\{&\frac{z^{2\df+2q}}{(1-z)^{2q-1}}P_1^{2\df+2q-2}\,\log(z)+\frac{(1-z)^{2\df+2q}}{z^{2q-1}}P_2^{2\df+2q-2}\,\log(1-z)\\&+\frac{1}{\left( z(1-z) \right)^{2q-2}}P_3^{4\df+6q-6}\Big\},
\end{split}
\end{align}
where $P_i^n$ are polynomials of degree $n$ in the variable $z$.

We shall not write such polynomials explicitly here, but we can observe that they are closely related to the $D$-functions introduced in \cite{DHoker:1999kzh}\footnote{The $D$-functions include a kinematical prefactor that serves to guarantee conformal invariance. One can then define $\bar{D}$-functions, which can be computed following \cite{Arutyunov:2002fh}, which are only functions of conformal invariant cross ratios.}. Indeed, the analogous problem was studied in \cite{PolchinskiPenedones}, for $d=2$ and $d=4$. The tree-level solutions that result from their analysis can be expressed in terms of $\bar{D}$ functions, corresponding to tree-level Witten diagrams. The authors found that only quartic interactions with even spin $L$ are allowed, and for fixed spin $L=2a$ there are $a+1$ solutions, given by interactions with $2k$ derivatives for $k=2a,\,2a+1,z,...,3a$. We can then study the diagonal limit ($z=\bar{z}$) of those solutions, and we find that in the diagonal limit the $L/2+1$ solutions with spin $L$ all collapse to a unique solution in $d=1$, with $L/2$ corresponding to our label $q$. This can be explained in terms of two-dimensional scattering in AdS$_2$. Indeed, in higher dimensions a basis of solution for scalar scattering amplitudes is given in terms of
\be 
\s_2^p\,\s_3^q,
\ee
where
\be 
\s_2=s^2+t^2+u^2, \q \s_3=s^3+t^3+u^3,
\ee
and $s$, $t$, $u$ are Mellin space variables, analogous to the Mandelstam variables for four-particles scattering. The spin $L$ of the solution is then given by $2(p+q)$, so that for a given even spin $L=2a$ there are indeed $a+1$ solutions. However, in $D=2$ one finds that $u=0$, hence the variables $s$ and $t$ are not independent. In particular, we have
\be 
\begin{cases}
s+t+u=4m^2, \\
u=0,
\end{cases}
\Rightarrow t=4m^2-s,
\ee
and therefore
\be 
\s_3=s^3+t^3+u^3=s^3+(4m^2-s)^3=4 m^2 \, \left(16 m^4-12 m^2 s+3 s^2\right),
\ee
which is actually of degree two in $s$, like $\s_2$. Therefore, in $D=2$, a basis of solutions can be simply written in terms of $s^{2q}$, and the condition $l=2(p+q)$ is now $L=2q$, which gives $q=L/2$ as claimed above. Furthermore, having only one solution per each value of the spin, we can limit ourselves to consider interactions with exactly $2L=4q$ derivatives, which act symmetrically on the fields, so that the Lagrangian contact terms are written as
\be 
\left((\pr_{\m_1}...\pr_{\m_q}\f)\,(\pr^{\m_1}...\pr^{\m_q}\f)\right)^2,
\ee
and the corresponding $\bar{D}$ functions are
\be  \label{eq:Dbarq}
\lim_{\bar{z}\ra z}(z \bar{z})^{\D_{\f}}\,\left( 1+(z\,\bar{z})^q+((1-z)\,(1-\bar{z}))^q \right)\,\bar{D}_{\D_{\f}+q\,\D_{\f}+q\,\D_{\f}+q\,\D_{\f}+q}(z,\bar{z}).
\ee
We have checked that this choice reproduces, in the diagonal limit, all the solutions that we have found. In particular, eq. \eqref{eq:Dbarq} for $q=Q$ is a linear combination of our solutions with $0\le q \le Q$. Also, on a technical note, we can observe that the $\bar{D}$ functions have transcendentality two, which is however reduced to one upon taking the diagonal limit $\bar{z}\ra z$: this acts as a derivative, and lowers by one unit the transcendentality of the functions.

We can now turn to examine the CFT data corresponding to these solutions. Having found results for enough values of $\df$, we were able to guess closed-form expressions for the anomalous dimensions corresponding to $q=0,1,2,3$ as analytical functions of $n$ and $\df$, that agree with the results obtained with PM bootstrap. We can express the anomalous dimensions in terms of the functions
\be 
\h(n,q,\df)=\frac{(n-q+1)_{\df-1} \left(\df+n+q+\frac{1}{2}\right)_{\df-1}}{\left(n+\frac{1}{2}\right)_\df (\df+n)_\df}, 
\ee
and
\be 
\cN(q,\df)=2^{2 q}(\df)_{3 q} \left(2 \df+q-\frac{1}{2}\right)_{2 q} ,
\ee
as
\be \label{gammastree}
\g^{(1)}_n=\frac{\h(n,q,\df)}{\h(q,q,\df)\,\cN(q,\df)}\,P_q(n,\,\df),
\ee
where $P_q(n,\,\df)$ are polynomials of degree $4q$ in $n$, whose coefficients are polynomials in $\df$. Note that the denominator in the previous formula does not depend on $n$, and is therefore simply due to our choice of normalisation. The polynomials $P_q(n,\,\df)$ are collected in Appendix \ref{sec:DerTree}, and for instance
\be 
P_{q=0}(n,\,\df)=1.
\ee
For cases in which the same anomalous dimensions have been computed with PM bootstrap (such as $q=0$ in \eqref{anm1}, $q=1$ for $\df=1$ in \eqref{gammatreeq=$1d$=1} and $q=2$ for $\df=1$ in \eqref{gammatreeq=2d=1}), we have found exact agreement. For the OPE coefficients, we found that the  derivative relation of \cite{PolchinskiPenedones} holds, that is\footnote{There is a caveat: for integer $\df$ one must first take the derivative, then choose the desired value of $n$ and only at the end fix $\df$. Otherwise, the first $q+1-\df$ values of $a^{(1)}_n$ would not reproduce the correct result.}
\be \label{derruletree}
C^{(1)}_n=\12 \frac{\pr}{\pr n}\left( C^{(0)}_n\,\g^{(1)}_n \right).
\ee
We also observe that the anomalous dimensions given in \eqref{gammastree} are rational functions of $n$ for integer $\df$. This will be the first entry of our dictionary: a correlator with transcendentality one translates into anomalous dimensions that are rational functions, {\it {\it i.e.},} have transcendentality zero. Via the derivative relation \eqref{derruletree}, the ratio $C^{(1)}_n/C^{(0)}_n$ is then found to have transcendentality one, since it contains harmonic numbers.

As a concluding remark, let us justify why higher powers of $(1-z)$ in the denominator of $\cA^{(1)}(z)$ correspond to a more divergent behaviour of $\g^{(1)}_n$ as $n \ra \infty$, since this is a common feature of all the solutions we have found, including results at one loop and for the $O(N)$ model. This can be seen looking at the action of the Casimir operator $\cC$ on the correlator\footnote{When acting on a four-point function in one dimension, the conformal Casimir can be written as $\cC=(1-z) z^2\pr_z^2-z^2 \pr_z$.}, in the limit $z\ra 1$: if in this limit the correlator has a singularity $\sim (1-z)^{-k}$, the action of the Casimir gives
\be 
\cC\left(\frac{1}{(1-z)^k}\right)\sim\frac{1}{(1-z)^{k+1}} \quad (z\ra 1),
\ee
so that the singularity is enhanced. On the other hand, when the Casimir hits a conformal block it multiplies the CFT data with the corresponding eigenvalue, that for double trace operators is $(2\D_{\f}+2n)(2\D_{\f}+2n-1)$, and therefore grows $\sim n^2$ as $n\ra \infty$. As a result, we can associate to the Casimir an action on the anomalous dimensions in this limit, that is 
\be 
\cC(\g_n)\sim n^2\,\g_n \quad (n\ra \infty),
\ee
and comparing with the action in the limit $z\ra 1$ we can deduce that increasing by one unit the power of $1-z$ in the denominator of $\cA^{(1)}(z)$ results in increasing by two units the power of $n$ in the behaviour of $\g_n$ as $n\ra \infty$. Hence, it is natural to find higher powers of $1-z$ in the denominator of solutions with higher $q$.

\subsection{Contact terms, $O(N)$ global symmetry}\label{sec:O(N)tree}

Let us now consider a model with $N$ scalar fields $\f_i$ and global $O(N)$ symmetry. Now intermediate states in the OPE of two fields decompose into irreducible representations of $O(N)$. Therefore, the four-point function of identical scalars with dimension $\D_{\f}$ reads
\be 
\langle\f_i(x_1)\,\f_j(x_2)\,\f_k(x_3)\,\f_l(x_4)\rangle=\frac{1}{x_{12}^{2\D_{\f}}\,x_{34}^{2\D_{\f}}}\,\cA_{ijkl}(z),
\ee
where 
\be \label{O(N)corr}
\cA_{ijkl}(z)=\left(\frac{\d_{ik}\,\d_{jl}+\d_{il}\,\d_{jk}}{2}-\frac{1}{N}\,\d_{ij}\,\d_{kl}\right) \cA_T(z)+\frac{\d_{ik}\,\d_{jl}-\d_{il}\,\d_{jk}}{2} \cA_A(z)+\d_{ij}\,\d_{kl}\,\cA_S(z),
\ee
and we denote with $T$ the two-indices symmetric traceless representation of $O(N)$, with $A$ the two-indices antisymmetric representation and with $S$ the singlet.

The MFT solution for this model is
\begin{align}
\cA_T^{(0)}(z)&=z^{2\D_{\f}}+\left( \frac{z}{1-z} \right)^{2\D_{\f}},\\
\cA_A^{(0)}(z)&=z^{2\D_{\f}}-\left( \frac{z}{1-z} \right)^{2\D_{\f}},\\
\cA_S^{(0)}(z)&=1+\frac{1}{N}\,G_T(z),
\end{align}
corresponding to double trace operators of dimension $\D^{T,S}_n=2\D_{\f}+2n$ for the $T$ and $S$ representations and $\D^{A}_n=2\D_{\f}+2n+1$ for the $A$ representation. The MFT OPE coefficients can be derived from
\be 
C^{(0)}_n=\frac{2 (-1)^n \Gamma^2 (2 \D_{\f}+n) \Gamma (4 \D_{\f}+n-1)}{\Gamma^2(2 \D_{\f}) \Gamma (n+1) \Gamma (4 \D_{\f}+2 n-1)},
\ee
via
\be 
C^{(0)}_{n,T}=C^{(0)}_{2n}, \quad C^{(0)}_{n,A}=C^{(0)}_{2n+1},
\quad C^{(0)}_{n,S}=\frac{1}{N}\,C^{(0)}_{2n},
\ee
where we have removed from the singlet the contribution of the identity operator.

As in the case of a single field, to find perturbative solutions, we formulate an ansatz and require the conditions outlined at the beginning of the present section. However, there are a few differences due to the presence of the $O(N)$ symmetry, that we are now going to discuss. Most importantly, let us notice that since crossing symmetry corresponds to the exchange of operators 1 and 3, and now the fields carry a ``flavour'' index, in addition to the exchange of the positions $x_1$ and $x_3$, but we also have to swap the indices $i$ and $k$ in eq. \eqref{O(N)corr}. The full crossing equation then reads
\be 
(1-z)^{2\D_{\f}}\,\cA_{ijkl}(z)=z^{2\D_{\f}}\,\cA_{kjil}(1-z),
\ee
and it can be decomposed requiring the equality of independent tensor structure. The result is better read in terms of 
\be \label{fdef}
f_i(z)=(1-z)^{2\D_{\f}}\cA_i(z),
\ee 
for $i=T,\,A,\,S$. We get only two independent equations, that are
\begin{align} \label{crossingO(N)}
\begin{split}
f_T(z)+f_A(z)&=f_T(1-z)+f_A(1-z),\\
f_T(z)-f_A(z)&=2f_S(1-z)-\frac{2}{N}\,f_T(1-z),
\end{split}
\end{align}
or equivalently
\begin{align}
\begin{split}
f_T(z)&=f_S(1-z)+\frac{N-2}{2N}\,f_T(1-z)+\frac{1}{2}\,f_A(1-z),\\
f_A(z)&=-f_S(1-z)+\frac{N+2}{2N}\,f_T(1-z)+\frac{1}{2}\,f_A(1-z),\\
f_S(z)&=\frac{1}{N}\,f_S(1-z)+\frac{N^2+N-2}{2N^2}\,f_T(1-z)+\frac{1-N}{2N}\,f_T(1-z),
\end{split}
\end{align}
where the last equation is not independent.

Another slight difference with the case of one single field comes from the study of the transformation $z \ra \frac{z}{z-1}$. Again, the conformal blocks will get a factor of $e^{\pm i\pi\,\D}$ under this transformation, but now we must make an important distinction between the antisymmetric and the other two representation. Indeed, for integer $\df$ we still have
\be \label{-1^TS}
e^{\pm i\pi\,\D_n^{T,S}}=e^{\pm i\pi\,\g_n^{T,S}},
\ee
but now since the double trace operators in the antisymmetric representation have an odd number of derivatives, and so
\be \label{-1^A}
e^{\pm i\pi\,\D_n^{A}}=-e^{\pm i\pi\,\g_n^{A}}.
\ee
Apart from these observations, the methods employed in the previous Section still apply in a very similar way. To find tree-level solutions to the $O(N)$ model, we make an ansatz for the functions $f_i(z)$ defined in eq. \eqref{fdef}, very much like the case with a single field:
\begin{align}
\begin{split}
f_T(z)&=\frac{1}{z^{k}\,(1-z)^{k}}\Big( P^T_{1}(z)+P^T_{2}(z)\,\log (z)+P^T_{3}(z)\,\log (1-z) \Big),\\
f_A(z)&=\frac{1}{z^{k}\,(1-z)^{k}}\Big( P^A_{1}(z)+P^A_{2}(z)\,\log (z)+P^A_{3}(z)\,\log (1-z) \Big),\\
f_S(z)&=\frac{1}{z^{k}\,(1-z)^{k}}\Big( P^S_{1}(z)+P^S_{2}(z)\,\log (z)+P^S_{3}(z)\,\log (1-z) \Big).
\end{split}
\end{align}
Again, our constraints fix the polynomials $P^x_{i}$ for $x=T,\,A,\,S$ and $i=0,\,1,\,2$ up to a finite number of ambiguities, which play a very similar role as in the case of a single field, discussed in the previous Section.

Having found a certain number of solutions, we observe the possibility of disentangling two distinct families, which correspond to
\begin{itemize}
\item Solutions where all the functions $f_T$, $f_A$ and $f_S$ are different from zero. These can be labeled with an integer $p=0,1,2,...$, and they have anomalous dimensions with a behaviour
\be 
\g^1_T\sim\g^1_A\sim\g^1_S\sim n^{2p} \quad (n\ra \infty).
\ee
Correspondingly, the value of the exponent in the denominators is $k=p$. 

\item Solutions with $f_A(z)=0$\footnote{While $f_A=0$ is allowed, there are no non-trivial solutions with either $f_T=0$ or $f_S=0$. Take for instance $f_T=0$: crossing symmetry then requires both $f_S$ and $f_A$ to be self-crossing symmetric, and proportional to each other. However, given the different transformation under $z \ra \frac{z}{z-1}$ (see \eqref{-1^TS} and \eqref{-1^A}), this is not possible.}. Then, the symmetry requirements of eqs. \eqref{crossingO(N)} demand that $f_T(z)$ and $f_S(z)$ be self-crossing symmetric ($f_T(z)=f_T(1-z)$, $f_S(z)=f_S(1-z)$) and proportional to each other ($f_S(z)=\frac{N+2}{2N}f_T(z)$). Hence, there is actually only one independent function, which satisfies the same constraints as those of a model without $O(N)$ symmetry. Therefore, these solutions can be labeled as in Section \ref{treesinglefield}, with an integer $q=0,1,2,...$, and the only independent anomalous dimension is 
\be 
\g^1_T\sim n^{4q-2} \quad (n\ra \infty).
\ee
\end{itemize}
This situation can be easily understood looking at AdS$_2$ contact terms. For instance, consider an interaction with two derivatives. Without $O(N)$ symmetry, we can integrate by parts and we get
\be 
(\pr_{\m}\f)^2\,\f^2=-\f^3\,\Box\,\f-2(\pr_{\m}\f)^2\,\f^2 \q
\Rightarrow \q
(\pr_{\m}\f)^2\,\f^2=-\frac{1}{3}\f^3\,\Box\,\f,
\ee
and using the equations of motion (or, equivalently, with a field redefinition) we can see that this interaction is actually trivial. However, when we add a flavour index to the fields, this is no longer true. For an arbitrary tensor structure $\cT^{ijkl}$, we have an interaction
\be 
\cT^{ijkl}(\pr_{\m}\f_i\,\pr^{\m}\f_j\,\f_k\,\f_l)
=-\cT^{ijkl}\left(
\Box\f_i\,\f_j\,\f_k\,\f_l
+\pr_{\m}\f_i\,\f_j\,\pr^{\m}\f_k\,\f_l
+\f_i\,\pr^{\m}\f_j\,\f_k\,\pr_{\m}\f_l,
\right)
\ee
and the second and third term on the l.h.s. are in general not dependent on the r.h.s. interaction.  Hence, we can conclude that in the case with $O(N)$ symmetry there are more types of interactions to be taken into account, and the family of solutions that we labeled with $p$ corresponds precisely to those interactions that would {\it not} be independent in the $N=1$ case. Again, all solutions can be written as a sum over the diagonal limit of appropriate $\bar{D}$ functions.

Let us now discuss general expressions for the solutions in the family that we labeled with $p$, since the other solutions have already been discussed. For given $p=P$, we find ambiguities corresponding to solutions with $p<P$ or $q\le P/2$, much like in the case without $O(N)$ symmetry. Again, these ambiguities precisely match the contact terms that one must add to the PM bootstrap basis. In order to fix them we make an arbitrary choice, and we use the solutions with $p<P$ to set
\be 
\g^{(1),A}_n=0 \quad (0\le n \le P-1),
\ee
and then the solutions with $q\le P/2$ to set
\be 
\g^{(1),T}_n=0 \quad (0\le n \le \lceil P/2\rceil),
\ee
and finally we normalise the diagram with
\be 
\g^{(1),T}_{n=1+\lceil P/2\rceil}=1.
\ee
With these conventions, we can express our results for the functions $f_x(z)$ introduced in \eqref{fdef} for any integer $\df\ge 1$ and $p\ge 0$, and to do so we distinguish two cases:
\begin{itemize}
\item $p=2k$. In this case, we found
\begin{align}
\begin{split}
f_T(z)=&\frac{z^{2\df+p+2}}{(1-z)^{p}}\,T_1^{2\df+p-2}\,\log(z)
+\frac{(1-z)^{2\df}}{z^{p-1}}\,T_2^{2\df+2p-1}\,\log(1-z)\\&
+\frac{T_3^{4\df+3p-4}}{z^{p-2} (1-z)^{p-1}},\\
f_A(z)=&z^{2\df+2p+1}\,(z-2)\,A_1^{2\df-2}\,\log(z)
+\frac{(1-z)^{2\df}}{z^p}\,A_2^{2\df+2p}\,\log(1-z)\\&
+\frac{(z-2)\,A_3^{4\df+3p-4}}{ \left( z (1-z)\right)^{p-1}},\\
f_S(z)=&\frac{1}{N}f_T(z)+\frac{z^{2\df}}{(1-z)^p}S_1^{2\df+2p}\,\log(z)+\frac{(1-z)^{2\df+p+2}}{z^{p-1}}S_2^{2\df+p-3}\,\log(1-z)\\&+\frac{S_3^{4\df+3p-4}}{z^{p-2}(1-z)^{p-1}}.
\end{split}
\end{align}
\item $p=2k+1$. In this case, we found
\begin{align}
\begin{split}
f_T(z)=&\frac{z^{2\df+p+3}}{(1-z)^{p}}\,T_1^{2\df+p-3}\,\log(z)
+\frac{(1-z)^{2\df}}{z^{p}}\,T_2^{2\df+2p}\,\log(1-z)\\&
+\frac{T_3^{4\df+3p-3}}{z^{p-1} (1-z)^{p-1}},\\
f_A(z)=&\frac{z^{2\df+2p+1}}{(1-z)^p}\,A_1^{2\df-2}\,\log(z)
+\frac{(1-z)^{2\df}}{z^{p-1}}\,A_2^{2\df+2p-1}\,\log(1-z)\\&
+\frac{(z-2)\,A_3^{4\df+3p-5}}{z^{p-2} (1-z)^{p-1}},
\end{split}
\end{align}
\begin{align}
\begin{split}
f_S(z)=&\frac{1}{N}f_T(z)+\frac{z^{2\df}}{(1-z)^p}S_1^{2\df+2p}\,\log(z)+\frac{(1-z)^{2\df+p+3}}{z^{p}}S_2^{2\df+p-3}\,\log(1-z)\\&+\frac{S_3^{4\df+3p-3}}{\left( z (1-z)\right)^{p-1}}.
\end{split}
\end{align}
\end{itemize}
As for the CFT data, we find that the derivative relation for the OPE coefficients is always satisfied, and for instance we provide the anomalous dimensions in closed form for the $p=0$ case:
\begin{align}
\begin{split} 
\g^{(1),T}_n|_{p=0}=&\frac{1}{\cN(\df)}\,\frac{(n)_\df\, \left(\df+n+\frac{1}{2}\right)_\df}{(\df+n)_\df\,\left(n+\frac{1}{2}\right)_\df},\\
\g^{(1),A}_n|_{p=0}=&-\frac{1}{\cN(\df)}\,\frac{(n+1)_{\df-1} \left(\df+n+\frac{3}{2}\right)_{\df-1}}{(\df+n+2-1)_{\df-1} \left(n+\frac{3}{2}\right)_{\df-1}},\\
\g^{(1),S}_n|_{p=0}=&\left(\g^{(1),T}_n|_{p=0}\right)+\frac{N}{4}\frac{1}{\cN(\df)}\, \frac{(n+1)_{\df-1} \left(\df+n+\frac{1}{2}\right)_{\df-1}}{(\df+n)_\df \left(n+\frac{1}{2}\right)_\df}\\
&\times \left(\df (1-4 \df)+ 2 (1-4 \df) n-4 n^2 \right),
\end{split}
\end{align}
where
\be 
\cN(\df)=\frac{(1)_\df \left(\df+\frac{3}{2}\right)_\df}{\left(\frac{3}{2}\right)_\df (\df+1)_\df}.
\ee
The OPE coefficients are found to satisfy the derivative relation in all cases. As in the case of a single field, for integer $\df$ the anomalous dimensions are rational functions of $n$, for every value of $p$.

Finally, let us observe that the results of \cite{Giombi:2017cqn} for tree-level correlators on the 1/2-PBS Wilson-Maldacena loop in $\cN=4$ SYM, already reproduced in Section \ref{theoryo(n)global} using the PM bootstrap, correspond in the language of this section to a linear combination of the solutions with $p=0$ and $p=1$.

\subsection{Exchanges}\label{sec:treeexchanges}

In this section we turn to the study of tree-level exchange diagrams: we consider a new single-trace primary operator $\cO$ of dimension $\D_E$, which appears in the OPE of $\f\times\f$. This problem was already considered, from the point of view of bootstrap, in \cite{Alday:2017gde}, where the double-trace CFT data due to a single-trace exchange at tree level in $d=4$ were considered. Exchange Witten diagrams in any dimension were also studied in \cite{Zhou:2018sfz}, where recursion relations were given for the coefficients in the conformal blocks expansion of such diagrams in every channel.

In the presence of an exchanged operator with dimension $\D_E$, the conformal blocks decomposition of a tree-level correlator reads
\begin{align} \label{CBexpexch}
\begin{split}
\cA^{(1)}_E(z)=&C_{\D_E}\,G_{\D_E}(z)+\cA^{(1)}_{DT}(z)\\
=&C_{\D_E}\,G_{\D_E}(z)+\sum_{n}\left(C^{(1)}_n+\12\,C^{(0)}_n\,\g^{(1)}_n\,\frac{\pr}{\pr n}\right)\,G_{2\D_{\f}+2n}(z)\\
=&C_{\D_E}\,G_{\D_E}(z)+\sum_{n}z^{2\D_{\f}+2n}\,\left(C^{(1)}_n+\12\,C^{(0)}_n\,\g^{(1)}_n\,\left(2\,\log(z)+\frac{\pr}{\pr n}\right)\right)\,F_{2\D_{\f}+2n}(z),
\end{split}
\end{align} 
where $G_{\D_E}(z)$ is a conformal block of dimension $\D_E$. Again, the constraints that we are going to apply are completely analogous to the case of the contact diagrams for a single field, but there are a few important differences, which are listed below.
\begin{itemize}
\item We will still consider integer values for the conformal dimensions, both for the external and for the exchanged operators. However, we notice that when the dimension of the exchanged operator satisfies $\D_E\ge 2\D_{\f}$ and is even, there is one double trace operator with the same dimension at the MFT level. Therefore, there is mixing between the two operators, and we cannot  solve for the correlator. Therefore, we shall consider only odd values of $\D_E$ when $\D_E\ge 2\D_{\f}$.
\item We recall that under the transformation $z\ra \frac{z}{z-1}$ the conformal blocks transform as\footnote{Recall that the sign in $e^{\pm i\pi\,\D}$ is fixed by the choice of analytical continuation.}
\be 
G_{\D}\left( \frac{z}{z-1} \right)=e^{\pm i\pi\,\D}\,G_{\D}(z).
\ee
Hence, we must take into account the different transformation properties of $G_{\D_E}(z)$ and $G_{DT}(z)$ under this symmetry.
\item $C_{\D_E}$ is the (square of the) OPE coefficient of the exchanged primary $\cO$ with two fields $\f$, and in the AdS theory it is proportional to the coupling $\l_{\cO}$ in the three-point interaction $\l_{\cO}\,\f^2\,\cO$. It is therefore the only free parameter, and all the double trace CFT data should be fixed in terms of $\l_{\cO}$ and a finite number of contact-terms ambiguities. We will make the convenient choice $C_{\D_E}=\pi^2$, that simplifies the structure of the CFT data. Note, however, that this choice is completely arbitrary.
\item The transcendentality of the correlator is, in principle, not fixed by any constraint. To have an idea of what this might be, we make two observations. The first is that when 
\be \label{exchsimple}
\frac{2\,\D_{\f}-\D_E}{2}\in \mathbb{N},
\ee
it is known that the exchange correlator is given by a finite sum of $\bar{D}$-functions, with known coefficients \cite{DHoker:1999mqo}. Since, as discussed, in $d=1$ the $\bar{D}$-functions have transcendentality one (corresponding to tree-level interactions, possibly with derivatives), we expect the exchange correlator to have transcendentality one as well in this case. To develop some intuition for the other cases (still with integer dimensions), we solved the recursion relation of \cite{Zhou:2018sfz} in the simplest case that does not respect \eqref{exchsimple}, that is $\D_{\f}=\D_E=1$. It turns out that in this case the transcendentality of the correlator is three\footnote{Despite the high degree of transcendentality, we are still dealing with a tree-level correlator. Hence, the solution cannot contain powers of $\log(z)$ higher than one in the small $z$ expansion, as dictated by \eqref{CBexpexch}.}, so we proceed with an ansatz of the same transcendentality for general exchanges. This also justifies our choice of $C_{\D_E}=\pi^2$: $G_{\D_E}(z)$ has transcendentality one for integer $\D_E$, and multiplication by $\pi^2$ gives transcendentality three, {\it {\it i.e.}} the maximal one for this type of correlator, according to our ansatz.
\end{itemize}
Hence, we take the ansatz for $\cA^{(1)}_{DT}$ to be
\be \label{ansatzexchange}
\cA^{(1)}_{DT}=\frac{1}{(1-z)^{2\df}}\frac{1}{z^k\,(1-z)^k}\left( \sum_{i=1}^{12} p_i(z)\,T_i(z) \right),
\ee
where $p_i(z)$ are polynomials in $z$ and $T^i(z)$ are transcendental functions from the following basis:
\begin{itemize}
\item Transc. 0: 1.
\item Transc. 1: $\log z$, $\log(1-z)$.
\item Transc. 2: $\Li_2(z)$, $\log(z)\,\log(1-z)$, $\log^2(1-z)$.
\item Transc. 3: $\Li_3(z)$, $\Li_3(1-z)$, $\Li_2(z)\,\log(z)$, $\Li_2(z)\,\log(1-z)$, $\log(z)\,\log^2(1-z)$, $\log^3(1-z)$.
\end{itemize}
This guarantees that there are no terms in $\log^2(z)$ in the small $z$ expansion.

Once again, we first of all characterize our solutions with the power $k$ that appears in the denominator of \eqref{ansatzexchange}. It turns out that to find solutions we need at least $k=\D_E-1$. Higher values of $k$ would still lead to non-trivial solutions, which however all differ by the addition of derivative contact terms\footnote{From the point of view of crossing symmetry, one could always add self-crossing symmetric terms to $\cA^{(1)}_{DT}$ and get a new solution.}. We can therefore restrict to $k=\D_E-1$. Once this is fixed, there are still parameters corresponding to tree-level derivative contact terms with $q<\D_E/2$. We can fix these terms simply by requiring that the correlator is Regge-bounded, which sets to zero all the coefficients of the derivative contact terms. At this point, one is left with the only ambiguity of adding a $\f^4$ contact term, which cannot be fixed in general. 

In order to discuss the corresponding corrections to the double-trace CFT data, let us first define the OPE coefficients in terms of a deviation from the derivative rule, as
\be 
C^{(1)}_n=\12 \frac{\pr}{\pr n}\left( C^{(0)}_n\g^{(1)}_n \right)+C^{(0)}_n\,\d C^{(1)}_n.
\ee
We distinguish two cases, according to whether $\D_E$ is smaller or larger than $2\D_{\f}$, that is the dimension of the lowest double trace operator built out of two primaries $\f$.
\begin{itemize}
\item $\boxed{\D_E<2\D_{\f}}$  In this case we can find the result for both even and odd $\D_E$. When the condition \eqref{exchsimple} is satisfied, which is possible only for even $\D_E<2\D_{\f}$, the solution is a sum of $D$-functions, and the anomalous dimensions are rational functions of $n$. Their behaviour as $n\ra\infty$ is generically $\sim n^{-2}$, but (except for a finite number of cases) it is possible to fix the coefficient of the $\f^4$ contact term in such a way that $\g^{(1)}_n\sim n^{-6}$. On the other hand, when $\D_E$ is odd the correlator has transcendentality three, and correspondingly the anomalous dimensions are more complicated. In general, they are of the form
\be 
\g^{(1)}_n=Q(n)+R(n)\,H_{2n+2\D_{\f}-1}+P(n)\,\left(2\,\z(2)+\p^{(1)}(n+\D_{\f})- \psi^{(1)}(n+\12+\D_{\f}) \right),
\ee
where $H_{n}$ is the $n$-th harmonic number, and $\psi^{(1)}(z)$ is the first derivative of the polygamma function $\psi(z)=\frac{d}{dz}\G(z)$. $Q(n)$ and $R(n)$ are rational functions, whereas $P(n)$ is a polynomial. In this case, the OPE coefficients do {\it not} satisfy the derivative rule, and we find
\be 
\d C^{(1)}_n=\frac{\Gamma (2 n+1)}{\Gamma (4 \D_{\f}+2 n-1)}\,P_{4\D_{\f}-2\D_E-2}(n),
\ee
where $P_{4\D_{\f}-2\D_e-2}(n)$ is a polynomial of degree $4\D_{\f}-2\D_E-2$ in $n$. Let us observe that in the large $n$ limit these corrections to the derivative rule satisfy
\be \label{derruleexch}
\d C^{(1)}_n \sim \frac{1}{n^{2\D_E}} \quad (n\ra \infty),
\ee
for any $\df$. This was already observed in higher dimensions, at tree level, in \cite{Alday:2017gde}, and will be justified (to all orderds) by a careful analysis of the Regge limit for one-dimensional CFTs in Section \ref{sec:Reggelimit}.
\item $\boxed{\D_E>2\D_{\f}}$ In this case it turns out that we can always fix the coefficient of the non-derivative contact term in such a way that
\be \label{gammaexchlargen}
\lim_{n\ra \infty}\g^{(1)}_n \sim n^{-6}.
\ee
Using this criterion to fix the only free parameter left, the general expression for $\g^{(1)}_n$ is of the form
\be 
\g^{(1)}_n=Q(n)+P(n)\,\left(2\,\z(2)+\p^{(1)}(n+\D_{\f})- \psi^{(1)}(n+\12+\D_{\f}) \right),
\ee
where $Q(n)$ is a rational function, while $P(n)$ is a polynomial. On the other hand, in this case the OPE coefficients do satisfy the derivative rule, and we simply have $\d C^{(1)}_n=0$.
\end{itemize}
Let us comment on the transcendentality of these results. What we have observed is that an ansatz of transcendentality three \eqref{ansatzexchange} leads to anomalous dimensions which contain at most $\psi^{(1)}(n)$, and have therefore transcendentality two. For the CFT data, if we consider the ratio $C^{(1)}_n/C^{(0)}_n$, we can argue that the transcendentality is three, the same as the correlator.

Finally, we can justify the large $n$ behaviour of $\g^{(1)}_n$ for the exchange diagrams looking at the Mellin transformed correlators in higher dimensions (or equivalently, given the similarity in the structure, at flat space scattering). For a scalar exchange, we have schematically
\be 
M_E(s,t,u)\sim \frac{1}{s-M^2}+\frac{1}{t-M^2}+\frac{1}{u-M^2},
\ee
where $M$ is the mass of the exchanged field and $s$, $t$ are independent (Mandelstam) variables. In the high energy limit $s\ra\infty$, we have
\be  \label{exchmellin}
\lim_{s\ra \infty}M_E(s,t)\sim a+\frac{b}{s},
\ee
for some contants $a$ and $b$, whereas for the non-derivative contact term $\f^4$
\be 
\lim_{s\ra \infty}M_{\f^4}(s,t,u)\sim c,
\ee
for some constant $c$. As we also discussed in the study of the tree level solutions, every derivative increases the power of $n$ in $\lim_{n\ra \infty}\g^{(1)}_n$ by one unit, and therefore every inverse derivative (like the propagator) decreases it by one unit. Therefore, we expect the contribution of the constant term proportional to $1$ in \eqref{exchmellin} to give a contribution that is $\sim n^{-2}$, and the term proportional to $\frac{1}{s}$ to give a contribution that is further suppressed by $n^{-2}$, and therefore that is $\sim n^{-4}$. However, in two dimensions (looking at scattering in AdS$_2$ for the Mandelstam variables) $s$ and $t$ are not independent, and in particular $s+t=4m^2$, with $m$ the mass of the external fields. Therefore, for a CFT$_1$ we actually have to study 
\be 
\lim_{s\ra \infty}M_E(s,4m^2-s,0)\sim a+\frac{b}{s^2},
\ee
and therefore, after the addition of an appropriate constant term, there is a further suppression by two powers of $n$ in the anomalous dimensions, so that \eqref{gammaexchlargen} is actually justified. This fact was already observed in \cite{Mazac:2018qmi} in the study of the one-dimensional Lorentzian inversion formula.

\subsubsection{Large $\D_E$: EFT expansion}

As it is well-known, in QFT one can see a tree-level exchange diagram as arising from the sum of an infinite number of contact interactions, with an increasing number of derivatives. This comes, intuitively, from the expansion of the Feynman propagator when the mass of the exchanged particle is very large, and leads to the usual notion of Effective Field Theories (EFTs), in which one integrates out the heavy modes and focuses on the low-energy physics. As it was already argued in \cite{Alday:2014tsa} using Mellin space techniques, the situation in AdS is very similar. However, in one dimension the Mellin transform is not uniquely defined since the two cross ratios are not independent, and one would like to recover the EFT expansion with other techniques. One option would be to find closed-form expressions as functions of $\D_E$ for the correlators that we have just discussed, but we were not able to do so. On the other hand, we can consider the recursion relation of \cite{Zhou:2018sfz}, and solve it order by order in a $1/\D_E$ expansion. Looking at (B.21) in \cite{Zhou:2018sfz}, we can see that the large $\D_E$ expansion of the recursion relation itself is trivial, with all the non-trivial dependence on $\D_E$ encoded in the $n=0$ value, given in (C.31). Such expression is however very hard to expand in $1/\D_E$, given that it contains the sum of two terms, both singular for even $\D_E$, while the sum of the two is regular. However, we observe that eq. (C.31) of \cite{Zhou:2018sfz} can be re-written as
\begin{align} \label{5F4}
\begin{split}
&C^{(t)}_0=-\frac{\sqrt{\pi } \Gamma (\D_E) \Gamma^2 \left(\D_{\f}+\frac{\D_E}{2}-\frac{1}{2}\right)  }{\D_E \Gamma \left(\D_E+\frac{1}{2}\right) \Gamma^2 \left(\D_{\f}+\frac{\D_E}{2}\right)} \\ &  _5F_4\left(\frac{1}{2},\frac{\D_E}{2},\D_E,-\D_{\f}+\frac{\D_E}{2}+1,-\D_{\f}+\frac{\D_E}{2}+1;\frac{\D_E}{2}+1,\D_E+\frac{1}{2},\D_{\f}+\frac{\D_E}{2},\D_{\f}+\frac{\D_E}{2};1\right).
\end{split}
\end{align} 
This expression is obtained looking at the expression for the Witten exchange diagram in the $t$-channel given in \cite{gs}. The hypergeometric function can be evaluated at integer values of $\D_E$ for every fixed integer $\D_{\f}$, and analytically continuing on the positive even integers the result can be expressed in terms of
\be 
S_{-2}(x)=\frac{1}{4}
\left( \p^{(1)}\left( \frac{x+1}{2} \right)-
\p^{(1)}\left( \frac{x+2}{2} \right) \right)-\12 \, \z(2).
\ee
For instance, for $\df=1$, we find
\begin{align}
\begin{split} 
&_5F_4\left(\frac{1}{2},\frac{\Delta_E }{2},\frac{\Delta_E }{2},\frac{\Delta_E }{2},\Delta_E ;\frac{\Delta_E }{2}+1,\frac{\Delta_E }{2}+1,\frac{\Delta_E }{2}+1,\Delta_E +\frac{1}{2};1\right)\\
=&\frac{\pi  2^{-\Delta _E-3} \Delta _E^3 \Gamma \left(\frac{\Delta _E}{2}\right) \Gamma \left(\Delta _E+\frac{1}{2}\right) \left(\psi ^{(1)}\left(\frac{\Delta _E}{2}\right)-\psi ^{(1)}\left(\frac{1}{2} \left(\Delta _E+1\right)\right)\right)}{\Gamma^3 \left(\frac{1}{2} \left(\Delta _E+1\right)\right)},
\end{split}
\end{align} 
and this allows to find an expansion in $1/\D_E$, that for every $\D_{\f}$ starts with $1/\D_E^2$. Alternatively, we can make an ansatz for $C^{(t)}_0$ as
\be \label{C0exp}
C^{(t)}_0=\sum_{n=0}^{\infty}\frac{a_n}{\D_E^{2+n}},
\ee
and fix the coefficients $a_n$ requiring that all the $C^{(t)}_n$ have an expansion in $1/\D_E$ starting with $1/\D_E^2$, as in \eqref{C0exp}. Indeed, given the structure of the recurrence relation, for general $C^{(t)}_0$, $C^{(t)}_n$ also contains positive powers of $\D_E$, with coefficients that depend on $C^{(t)}_0$. Requiring the coefficients of undesired powers to cancel, we get conditions that constrain $C^{(t)}_0$. This allows to solve both for the coefficients $a_n$ and for $C^{(t)}_n$, order by order in $1/\D_E$, to arbitrarily high order. The resulting expansion for $C^{(t)}_0$ precisely matches the result coming from \eqref{5F4}. Now, $C^{(t)}_n$ represents a contribution to the conformal blocks expansion of a $t$-channel exchange Witten diagram, which contains operators with dimension $2\df+n$. For the $u$-channel, we have $C^{(t)}_n=(-1)^n\,C^{(t)}_n$, so that when we sum them only even values of $n$ contribute and we get the usual sum over double trace operators. Once we sum with $C^{(s)}_n$, which was given in closed form in \cite{Zhou:2018sfz}, we get
\be 
C^{(s)}_n+2C^{(t)}_{2n}=C^{(0)}_n\,\g^{(1)}_n,
\ee
therefore expanding $C^{(s)}_n$ and $C^{(t)}_n$ in powers of $1/\D_E$ we can find the large $\D_E$ expansion of the anomalous dimensions. This is expected to reproduce the usual EFT expansion of exchange amplitudes for large value of the exchanged mass, and therefore each term in the expansion should correspond to a linear combination of contact term anomalous dimensions. Qualitatively, this can be seen from the expansion of the propagator:
\be \label{propexp}
\frac{1}{\Box-\Delta_E \left(\Delta_E-1\right)}=-\frac{1}{\Delta_E^2}-\frac{1}{\Delta_E^3}+\frac{-\Box-1}{\Delta_E^4}+\cO\left(\frac{1}{\Delta_E^5}\right).
\ee
We found that all the contact terms with arbitrarily high number of derivatives indeed contribute to the expansion of $\g^{(1)}_n$ in $1/\D_E$, but contact terms with higher number of derivatives enter the expansion at higher orders, as expected from \eqref{propexp}. In particular, the expansion reads
\be 
\g^{(1)}_n|_E=\frac{1}{\D_E^2}\,
\sum_{k=0}^{\infty} \frac{\a_{0,k}}{\D_E^k} \,\g^{(1)}_n|_{q=0}
+\frac{1}{\D_E^6}\,
\sum_{k=0}^{\infty} \frac{\a_{1,k}}{\D_E^k} \,\g^{(1)}_n|_{q=1}
+\frac{1}{\D_E^{10}}\,
\sum_{k=0}^{\infty} \frac{\a_{2,k}}{\D_E^k} \,\g^{(1)}_n|_{q=2}+...\,\,\, .
\ee
The power of $\D_E$ at which a given contact term enters the expansion can be read from \eqref{propexp}, recalling that some derivative interactions are equivalent up to integration by parts and field redefinitions.

If we now consider a limit where both $n$ and $\D_E$ are large (and of the same order), each group of terms is dominated by the first, which contains the lowest power of $\D_E$ in the denominator, and recalling that $\g^{(1)}_n|_{q}\sim n^{-2+4q}$ we find an expansion of the type
\be 
\g^{(1)}_n|_E\sim \frac{1}{\D_E^2}\,f\left(\frac{n}{\D_E}\right).
\ee
We were able to find the coefficients in the expansion and re-sum it exactly in all cases in which we computed $\g^{(1)}_n|_E$. The result is that for large $n$ and $\D_E$, with $n/\D_E$ fixed, we have (up to the addition of a non-derivative contact term)
\be 
\g^{(1)}_n|_E=\frac{c(\D_{\f})}{(\D_E\,n)^2}\frac{1}{1-\left( \frac{2n}{\D_E} \right)^4},
\ee 
where $c(\D_{\f})$ is a constant that only depends on the external dimension. This result diverges when the twist $\t=2n$ of the double trace operators is of order $\D_E$, as expected.

\section{Intermezzo: Regge limit in $1d$ CFTs}\label{sec:Reggelimit}

As we have discussed in the previous section, the Regge limit plays an important role in the study and classifications of solutions to the bootstrap equation in one dimension. For the moment, we have only heuristically implied a connection, at tree level, between the Regge limit of correlators and the large $n$ behaviour of anomalous dimensions. In this section we will be more precise about this link, systematically studying the Regge limit in one dimension. One can also wonder whether looking at the crossing equation in the Regge limit can put some constraint on the CFT data. Unfortunately, it will turn out that this is not enough to completely fix the correlators, but we will still be able to make some universal statements which hold at all orders in perturbation theory, provided some assumptions are satisfied.\\

The Regge limit of CFTs was already considered for dimensions $d>1$, both from the CFT and from the AdS perspective. The case of CFTs dual to pure Einstein gravity was first considered in \cite{Cornalba:2006xm, Cornalba:2007zb}, and an extension to more general gravity duals was provided in \cite{Li:2017lmh, Kulaxizi:2017ixa}. While in $d=1$ we shall find that the position space Regge limit corresponds to a large twist ($n$) limit of the CFT data, in higher dimensions this is related to a limit in which both $n$ and the spin $j$ are large, with their ratio $n/j$ being kept fixed.

\subsection{The Regge limit of conformal blocks}

Let us first recall that, following \cite{mp}, we have defined the Regge limit of a correlator of identical scalars with scaling dimension $\df$ in a $1d$ CFT to be 
\be 
\lim_{t\ra \infty}\left( \12+i\,t \right)^{-2\df}\,\cG	\left(\12+i\,t \right).
\ee
To motivate the present discussion let us recall that, at least for tree-level correlators, a Regge behaviour of the type 
\be 
\lim_{t\ra \infty}\left( \12+i\,t \right)^{-2\df}\,\cG	\left(\12+i\,t \right)\sim t^a
\ee
implies that the anomalous dimensions satisfy
\be 
\lim_{n\ra\infty}\g^{(1)}_n\sim n^{2a}.
\ee
Therefore, we can see that at least in this case the large $t$ behaviour of the correlator and the large $n$ behaviour of the anomalous dimensions are closely related. In particular, one can also argue that $n^2\sim t$: while for the moment this is just heuristic, we shall soon derive this relation. 

\begin{figure}
\begin{subfigure}{.5\textwidth}
  \centering
  \includegraphics[width=.8\linewidth]{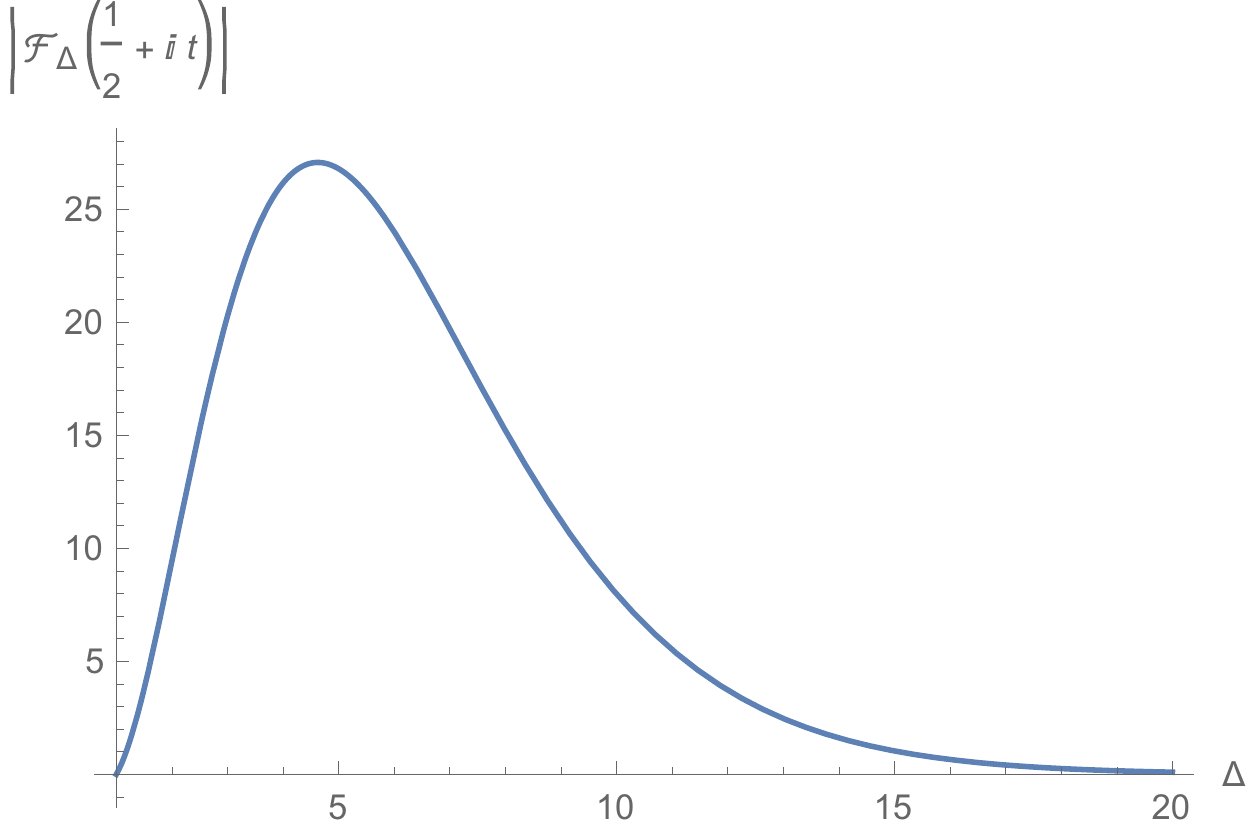}
  \caption{$\df=1$, $t=5$}
  \label{fig:t=5}
\end{subfigure}%
\begin{subfigure}{.5\textwidth}
  \centering
  \includegraphics[width=.8\linewidth]{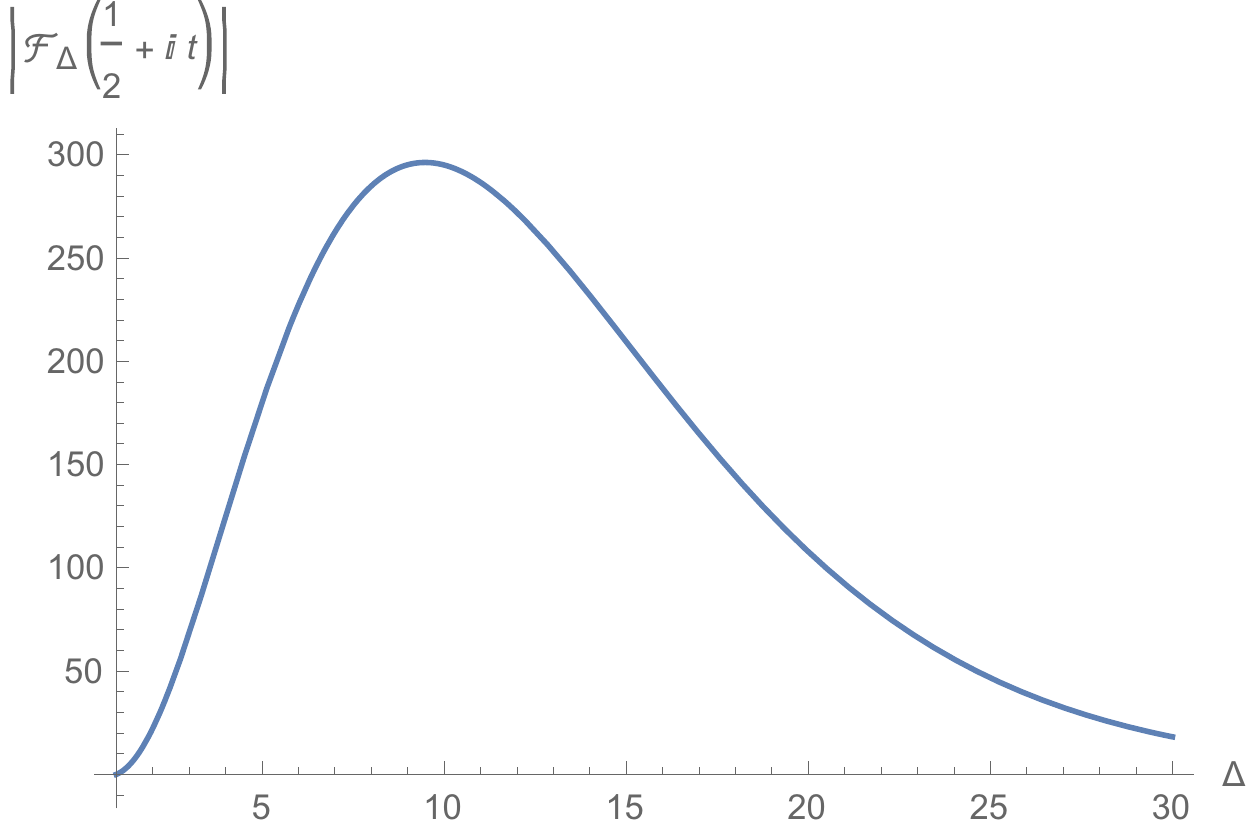}
  \caption{$\df=1$, $t=25$}
  \label{fig:t=25}
\end{subfigure}
\caption{Plots of $\vline\, \cF_{\D}(1/2+it)\,\vline$ vs $\D$ for $\df=1$ and two values of $t$. One can see that the plot is peaked for a value of $\D$ that increases roughly with $\sqrt{t}$.}
\label{fig:reggeCB}
\end{figure}

To make this connection more precise, let us study the behaviour of a single conformal block in the Regge limit. To this end, let us first redefine for convenience the CFT data, in terms of the MFT ones. We shall write the OPE as
\be \label{opeexp}
\cA(z)=\sum_{n}\hat{C}_{\D}\,C^{(0)}_{\D}\,G_{\D}(z),
\ee
where $G_{\D}(z)$ are the conformal blocks, $\hat{C}_{\D}$ are rescaled OPE coefficients and $C^{(0)}_{\D}$ are the MFT OPE coefficients written in terms of the physical dimension $\D$ of the intermediate operators:
\be 
C_{\Delta}^{(0)}=\frac{2\, \Gamma^{2}(\Delta)\, \Gamma\left(2 \Delta_{\varphi}+\Delta-1\right)}{\Gamma^{2}\left(2 \Delta_{\varphi}\right)\, \Gamma(2 \Delta-1) \,\Gamma\left(-2 \Delta_{\varphi}+\Delta+1\right)}.
\ee
Now, we can define rescaled blocks to be
\be 
\cF_{\D}:=C_{\Delta}^{(0)}\,G_{\D}(z),
\ee
and study their behaviour in the Regge limit. To begin with, we can study the behaviour of $\vline \, \cF_{\D}\left( 1/2+i\,t \right) \,\vline$ at fixed $\df$ and $t$ as a function of $\D$, for increasing $t$ (in the end, we want $t\ra \infty$). As it is clear from Figure \ref{fig:reggeCB}, for fixed $t$ and $\df$ this  is peaked at some value of $\D$, which increases as one increases $t$. In particular, it is possible to show that the peak is for $\D\sim \a \,\sqrt{t}$, for some real number $\a$. This means that in the Regge limit the contribution of conformal blocks with large scaling dimension is enhanced with respect to the others. Since double-trace operators have dimension $\D=2\df +2n +\g_n$, we can argue that operators with large $n$ dominate the Regge limit, with $n$ being of order $\sqrt{t}$, as claimed at the beginning of this Section.

To explicitly compute the large $t$ expansion of the conformal blocks, we use a well-known integral representation of hypergeometric functions, and we get
\be \label{intrep}
\cF_{\D}(z)=\int_0^1 ds\,
\frac{2 (2 \Delta -1) ((1-s) s)^{\Delta -1} z^{\Delta } \Gamma (2 \df+\Delta -1) (1-s z)^{-\Delta }}{\Gamma^2(2 \df) \Gamma (-2 \df+\Delta +1)}.
\ee
Studying the integrand as a functions of $s$, one can argue that when $z=1/2+i t$ for large $t$, the main contribution comes from the region with $s\sim 1/\sqrt{t}$. Hence, we can make a change of variable $s=\l/\sqrt{t}$, as well as $\D=\a \sqrt{t}$. Now, we can expand the integrand in $1/t$ and integrate order by order, with the result
\be 
\cF_{\D}(1/2+i t)=e^{i\pi\,\D}\,\cK(\a,\,t),
\ee
where 
\be \label{kernel}
\cK(\a,\,t)=t^{2\df}\,\left(\frac{8}{\sqrt{t}} \frac{\alpha^{4 \Delta_{\varphi}-1}}{\Gamma^{2}\left(2 \Delta_{\varphi}\right)} K_{0}\left(2 e^{i \pi / 4} \alpha\right)+\cdots\right).
\ee
The expansion is in {\it half-integer} powers of $1/t$\footnote{Despite this, correlators admit an expansion in {\it integer} powers of $1/t$ (times $t^{2\df})$. The absence of half-integer powers can be seen as a constraint on the CFT data.}, and we have defined the kernel $\cK(\a,\,t)$ collecting an overall factor of $e^{i\pi\,\D}$, which will be crucial in the following. Finally, $K_{0}\left(z\right)$ is a modified Bessel function, and our expansion is well-defined provided $t$ has a large and positive real part.

Now that we have an expansion for the individual conformal blocks in the Regge limit, and we know that in such a limit the OPE is dominated by operators with large dimension, we can also find an expansion for the OPE itself. To this end, we convert the sum over $n$ in \eqref{opeexp} into an integral over $\a$. If we think of the anomalous dimensions as functions of $\D$ (rather than $n$), we can write 
\footnote{We have $2n = \D-\g(\D)-2\df$, hence $\sum_n\sim \int dn=\int d\D \,\frac{\pr n}{\pr \D}=\int d\a\,\frac{\sqrt{t}\,(1-\g'(\D))}{2}:=\int [d\a]$.}
\be 
\sum_n=\int [d\a], \qquad
[d\a]=d\a\,\frac{\sqrt{t}\,(1-\g'(\D))}{2}.
\ee
This allows to express \eqref{opeexp} in the Regge limit as
\be \label{reggeope}
\cA \left(1/2+i t \right)=e^{2i\pi\,\df}\,\int_0^{\infty} [d\a]\,\hat{C}(\a\sqrt{t})\,e^{i\pi\,\g(\a\sqrt{t})}\,\cK(\a,t),
\ee
and if we think the CFT data $\hat{C}(\D)$ and $\g(\D)$ in a $1/\D$ expansion this gives a $1/t$ expansion for the correlator $\cA(z)$.

Before proceeding, as a consistency check of our expansion, let us notice that if we insert the MFT CFT data $\hat{C}(\D)=1$ and $\g(\D)=0$, we get
\be  \label{mftregge}
e^{2\,i\pi\,\df} \int_{0}^{\infty}[d \alpha]\, \cK(\alpha, t)=\left(\frac{1}{2}+i t\right)^{2 \df}.
\ee
When compared with the MFT correlator
\be \label{mft}
\cA(z)=1+z^{2\df}+\left( \frac{z}{1-z} \right)^{2\df},
\ee
this result correctly reproduces the divergent term $z^{2\df}$, but not the first and the last term of \eqref{mft}, which are regular in the Regge limit. This allows us to highlight an important fact about our Regge limit expansions: they are only sensitive to the divergent part of correlators, {\it {\it i.e.},} the part that scales with $t^{2\df -n}$, with $n$ an integer number. All terms that are suppressed by non-analytic factors of $t^{-2\df}$ with respect to the latter, like the contribution of the identity in \eqref{mft}, cannot be reproduced by the expansion \eqref{reggeope}, which is therefore only asymptotic.

\subsection{OPE limits and crossing symmetry}

A key fact for the development of the analytic bootstrap was the observation that the light-cone OPE is dominated by operators with large spin \cite{Fitzpatrick:2012yx, Komargodski:2012ek, usl1, usl2}. As we briefly discussed, this has allowed to systematically constrain the CFT data expanded in inverse powers of the spin, using crossing symmetry \cite{Alday:2015eya, Alday:2015ewa, alday1}. In one dimension, $z$ and $\bar{z}$ actually coincide, and there is no notion of light-cone limit. However, one can wonder whether a limit exist, in the cut $z$ plane where $\cA(z)$ is analytic \cite{Pappadopulo:2012jk, mp}, in which the OPE is dominated by some specific kind of operators. As we have discussed in the previous section, at least one such a limit exists: the Regge limit, where the OPE is dominated by operators with large dimension.

The Regge limit can be seen as a special case of the so-called OPE limits, in which two operators become arbitrarily close to each other. In particular, from the definitions
\be 
z=\frac{x_{12}\,x_{34}}{x_{13}\,x_{24}},
\quad 
1-z=\frac{x_{14}\,x_{23}}{x_{13}\,x_{24}}
\ee
we can distinguish three OPE limits\footnote{Note that in one dimension operators live on a line, or its compactified version, {\it i.e.} a circle. Therefore, while starting from an ordering ``1234'' one can bring 2 close to 1 ($s$-channel) or to 3 ($t$-channel), physically it is not possible to bring 1 close to 3. This is also related to the exchange of 1 and 2 not being an actual symmetry (due to the factor of $e^{\pm i\,\pi\,\D}$ in \eqref{CB_exchange12}). Nonetheless, one can consider the analytical continuation to complex $z$ and find a regime where 1 and 3 are close: we shall call this configuration $u$-channel limit, but it does not imply the existence of a standard $u$-channel OPE (precisely due to the aforementioned $e^{\pm i\,\pi\,\D}$).}
\begin{itemize}
\item $s$-channel OPE, where operators 1 and 2 become close. This corresponds to $x_{12}\ra 0$, and therefore $z\ra 0$.
\item $t$-channel OPE, where operators 1 and 4 become close. This corresponds to $x_{14}\ra 0$, and therefore $z\ra 1$.
\item $u$-channel OPE, where operators 1 and 3 become close. This corresponds to $x_{13}\ra 0$, and therefore $z\ra \infty$.
\end{itemize}
Hence, we can see that the Regge limit is actually a $u$-channel OPE limit. But then, a question naturally arises: can we put constraints on the CFT data of $1d$ CFTs by inspecting these three OPE limits?

Let us first discuss the constraints arising from the Regge limit. We can write the crossing equation in such regime as
\be 
\left( \12-it \right)^{2\df}\,\cA\left( \12+it \right)=
\left( \12+it \right)^{2\df}\,\cA\left( \12-it \right),
\ee
which, for $t\in \mathbb{R}$, can be seen as the condition
\be 
\left( \12-it \right)^{2\df}\,\cA\left( \12+it \right) \in \mathbb{R}.
\ee
For large $t$, we can use our expansion \eqref{reggeope} and get our first Regge limit constraint, to be interpreted order by order in $1/t$:
\be 
\left( \12-it \right)^{2\df}\,e^{2\,i\pi\,\df}\,\int_0^{\infty} [d\a]\,\hat{C}(\a\sqrt{t})\,e^{i\pi\,\g(\a\sqrt{t})}\,\cK(\a,t)\in \mathbb{R}.
\ee
Now, let us turn to the $t$-channel OPE limit\footnote{This is related by crossing symmetry to the $s$-channel limit, and therefore the latter will give no new constraints.}. In order to study this limit still using our expansion \eqref{reggeope}, we note that when $z=1/2+it$ and $t\ra\infty$, we have
\be 
\frac{z}{z-1}=\frac{it+1/2}{it-1/2}\ra 1 \quad (t\ra \infty).
\ee
Therefore, we can reach the $t$-channel OPE limit looking at $\cA(z/(z-1))$ with $z=1/2+it$ and large $t$, which is very reminiscent of the Regge limit. Furthermore, as already discussed, the conformal blocks in one dimension have a very simple transformation property under $z\ra (z/(z-1))$, given by\footnote{Note that we are choosing $z$ to have a {\it positive} imaginary part ($t$), and the sign in $e^{-i\pi\,\D}$ is fixed by this choice, as discussed in Section \ref{transc - tree}.}
\be 
G_{\D}\left(\frac{z}{z-1}\right)=e^{-i\pi\,\D}\,G_{\D}(z).
\ee
Therefore, we can write the OPE as
\be 
\mathcal{A}\left(\frac{z}{z-1}\right)=\sum_{\D} \hat{C}_{\Delta}\,e^{-i\pi\,\D}\, \cF_{\Delta}(z),
\ee
and if we plug in $z=1/2+it$ we get
\be \label{tch}
\mathcal{G}\left(\frac{i t+1 / 2}{i t-1 / 2}\right)=\int_{0}^{\infty}[d \alpha]\, \hat{C}(\alpha \sqrt{t})\, \cK(\alpha, t).
\ee
Therefore, we can conclude that the $t$-channel OPE limit is also dominated by operators with large dimension, and leads to an expansion of the type $t^{2\df-n}$, for integer $n$. Now, we can use again crossing symmetry, which relates $z/(z-1)$ to $1/(1-z)$. When $z=1/2+it$ for large $t$, we have that $1/(1-z)=1/(1/2-it)\ra 0$, and therefore we get the $s$-channel OPE limit, as expected. As opposed to the two previous ones, this limit is dominated by the contribution of the identity operator, and we have\footnote{We are assuming that, as in unitary CFTs in $d>2$, there is a gap between the dimension of the identity operator ($\D_{\mathbb{1}}=0$) and all the other operators of the theory.} 
\be \label{sch}
\mathcal{A}\left(\frac{1}{1 / 2-i t}\right)=1+\mathcal{O}\left(t^{-\D_{min}}\right),
\ee
where $\D_{min}$ is the scaling dimension of the operator with lowest dimension in the $\f \times \f$ OPE. We can relate \eqref{tch} and \eqref{sch} using crossing symmetry, and taking into account only terms of the type $t^{2\df -n/2}$ we get the new constraint
\be 
\mathcal{G}\left(\frac{i t+1 / 2}{i t-1 / 2}\right)=e^{-2\,i\pi\,\df}\,\left(\frac{1}{2}+i t\right)^{2 \Delta_{\varphi}}\left(1+\mathcal{O}\left(t^{-\D_{min}}\right)\right).
\ee
To sum up, by looking at the crossing equation in the three OPE limits for one-dimensional CFTs, we were able to derive two equations that the CFT data must satisfy. Since, as explained, our expansions only reproduce term of the type $t^{2\df -n/2}$, we can neglect $\mathcal{O}\left(t^{-\D_{min}}\right)$ and write the constraints as
\begin{align} \label{reggecross1}
&\left( \12-it \right)^{2\df}\,e^{2\,i\pi\,\df}\,\int_0^{\infty} [d\a]\,\hat{C}(\a\sqrt{t})\,e^{i\pi\,\g(\a\sqrt{t})}\,\cK(\a,t)\in \mathbb{R},\\ \label{reggecross2}
&\int_{0}^{\infty}[d \alpha]\, \hat{C}(\alpha \sqrt{t})\, \cK(\alpha, t)=e^{-2\,i\pi\,\df}\,\left(\frac{1}{2}+i t\right)^{2 \Delta_{\varphi}}.
\end{align}
At this point, one would like to solve these equations in terms of the CFT data. To this end, we consider $\hat{C}(\D)$ and $\g(\D)$ to be given as expansions in $1/\D$ with arbitrary coefficients, and solve for the coefficients. It turns out that \eqref{reggecross2} admits a rather simple solution, if we recall from \eqref{mftregge} that the MFT data $\hat{C}(\D)=1$, $\g(\D)=0$ are a solution of \eqref{reggecross2}. Since the r.h.s of \eqref{reggecross2} does not receive perturbative corrections, while the CFT data do, we conclude that $\hat{C}(\D)$ must completely cancel the contribution of the measure $[d\a]$, in such a way that the r.h.s. of \eqref{reggecross2} coincide with the r.h.s. of \eqref{mftregge} at every order, therefore giving the expected result. Therefore, the solution is
\be \label{derivativerule}
\hat{C}(\D)=\frac{1}{1-\g'(\D)}+\cO(\D^{-2\D_{min}}),
\ee
which generalises the derivative relation of \cite{PolchinskiPenedones} to every order in perturbation theory. Note, however, that there are corrections of $\cO(\D^{-2\D_{min}})$: while these are absent for tree-level contact diagrams, they are generically present. For instance, already in the case of tree-level exchange diagrams studied in Section \ref{sec:treeexchanges}, we have found that the derivative relation receives corrections when $\D_E<2\df$. In particular, we can see that eq. \eqref{derruleexch} agrees with the prediction of \eqref{derivativerule}: corrections to the derivative rule appear at $\cO(n^{-2\D_E})$, and when $\D_E<2\df$ we have that $\D_{min}=\D_E$. As we shall discuss in Section \ref{sec:transc - loop}, something similar happens for one-loop solutions corresponding to contact terms: the function expressing the correction to the derivative rule, that we call $\d C^{(2)}_n$, scales with $n^{-4\df}$ at large $n$, stemming from the fact that without exchanges $\D_{min}=2\df$. Finally, let us observe that a similar generalisation of the derivative relation in the Regge limit was already conjectured in \cite{Cornalba:2006xm}, and then proven in \cite{Cornalba:2007zb}, for CFTs in $d>1$.

Plugging this solution into \eqref{reggecross1}, we can simplify such equation and get the condition
\be  \label{reggecorssinggamma}
\left( \12-it \right)^{2\df}\,e^{2\,i\pi\,\df}\,\int_0^{\infty} d\a\,e^{i\pi\,\g(\a\sqrt{t})}\,\cK(\a,t)\in \mathbb{R},
\ee
which is a constraint for the anomalous dimensions only.

\subsection{Solutions to crossing symmetry in the Regge limit}\label{reggecrossing}

Summarising, we have derived a constraint, given by eq. \eqref{reggecorssinggamma}, which must be satisfied by the anomalous dimensions order by order in a $1/\D$ expansion. Once such an expansion is found, the OPE coefficients can be obtained using \eqref{derivativerule}. As explained, however, these expansions are only asymptotics, and receive corrections of order $\D^{-2\D_{min}}$.

We now want to solve for the anomalous dimensions, and to do so we must make an ansatz for their large $\D$ behaviour. We shall consider first Regge bounded solutions, {\it {\it i.e.},} anomalous dimensions that are {\it not} divergent for large $n$. We can therefore make an ansatz of the type
\be 
\g(\D)=p_0(\log\D)+\frac{p_1(\log\D)}{\D}+\frac{p_2(\log\D)}{\D^2}+...\,\,\, ,
\ee
where the $p_i$'s are polynomials in $\log\D$, and solve \eqref{reggecorssinggamma} for the coefficients of such polynomials. It turns out that the answer can be expressed in terms of the conformal Casimir $J^2=\D\,(\D-1)$, in agreement with the result found in \cite{Alday:2015eya} in higher dimension, which is often referred to as the reciprocity principle. The result reads
\be \label{reggesolbounded}
\begin{aligned} \gamma(\Delta)=& \lambda\left(\frac{1}{J^{2}}+\frac{2 \Delta_{\varphi}\left(\Delta_{\varphi}-1\right)}{J^{4}}\right)+f_{1}(\lambda) \frac{1}{J^{6}}+\cdots \\ &+\log J\left(2 \lambda^{2}\left(\frac{1}{J^{6}}+\frac{2\left(3 \Delta_{\varphi}^{2}\right)-3 \Delta_{\varphi}-2}{J^{8}}\right)+f_{2}(\lambda) \frac{1}{J^{10}}+\cdots\right) \\ &+\log ^{2} J\left(24 \lambda^{3}\left(\frac{1}{J^{10}}+\frac{10 \left(\df^2-\df-2\right)}{J^{12}}\right)+f_{3}(\lambda) \frac{1}{J^{14}}+\cdots\right)+\cdots \end{aligned}
\ee
where $\l$ is the coupling constant and $f_i(\l)$ are functions of $\l$ that cannot be fixed by our constraints\footnote{We are thinking in terms of a theory with only one coupling constant, and so where all possible exchanges are controlled by the same coupling. If one wants to think of this in terms of an EFT with arbitrary couplings for every interaction, then the functions $f_i$ will in general depend on all these couplings.}.

Several comments are in order. First, let us observe that we only assumed Regge boundedness and a gap in the anomalous dimensions from the identity operator. Hence, this solution describes, to all orders in perturbation theory, the anomalous dimensions coming from a $\f^4$ interaction plus an arbitrary (but {\it finite}) number of exchanges. Furthermore, as we shall comment later, the same equations also apply to fermions, so in principle it also encompasses fermionic interactions. This is the reason for the ambiguities $f_i(\l)$: there is no way to tell whether or not there is an exchange, or if the operators are bosons or fermions. Therefore, eq. \eqref{reggesolbounded} contains enough arbitrary parameters to describe all these cases. Indeed, we have checked it against our analytical results, and we can always find solutions for the $f_i(\l)$ such that \eqref{reggesolbounded} reproduces the correct answer. Despite all this freedom, we can make some interesting remarks about the powers of $\log J$ that appear in the expansion. First, we notice that $\log^k J$ appears for the first time at order $\l^{k+1}/J^{4k+2}$, and the coefficient of $(\log^k J)/J^{4k+2}$ is always fixed. Therefore, this describes a universal behaviour of the kind of theories that we are studying in the Regge limit, independent on the number and type of exchanged operators (provided it is finite). Furthermore, the fact that a new power of $\log J$ appears with a new power of the coupling can be interpreted in terms of the transcendentality principle discussed in the introduction, and that we shall exploit heavily in Section \ref{sec:transc - loop}: higher powers of $\l$ correspond to higher loop orders, and at every loops order the transcendentality is increased.

On a technical note, we can observe that the fact that $\g(\D)$ admits an expansion in powers of $J^2$ (rather than simply in powers of $\D$, or of $J$) is quite non-trivial. As we mentioned, this can be seen as a one-dimensional analogue of the reciprocity principle already observed in \cite{Dokshitzer:2005bf, Basso:2006nk, Alday:2015eya}. Given that such an expansion is possible, one could start from \eqref{intrep} and instead $\D=\a\sqrt{t}$, introduce a variable $j$ such that
\be 
J^2=j^2\,t=\D\,(\D-1)=\a\sqrt{t}\,(\a\sqrt{t}-1),
\ee
and write the kernel $\cK$ as a function of $j$ and $t$ rather than of $\a$ and $t$. In terms of $j$, we get an expansion of the type 
\be 
C^{(0)}_{\D}\,G_{\D}(1/2+i t)=e^{i\pi\,\D}\,\cK(j,\,t),
\ee
where
\be \label{kernelcasimir}
\cK(j,\,t)=t^{2\df}\,\left(\frac{8}{\sqrt{t}} \frac{\alpha^{4 \Delta_{\varphi}-1}}{\Gamma^{2}\left(2 \Delta_{\varphi}\right)} K_{0}\left(2 e^{i \pi / 4} j\right)+\cdots\right),
\ee
which is very similar to \eqref{kernel}. However, if we now choose as an ansatz for $\g(\D)$ an expansion in $1/J^2$ (rather than in $1/\D)$, then eq. \eqref{reggecross1} contains only {\it integer} powers of $1/t$, even before imposing any constraint on $\g$. This shows that the Kernel $\cK(j,t)$, together with an ansatz in terms of powers of $J^2$, is in some sense more natural. Indeed, the half-integer powers of $t$ have to cancel in any case to guarantee analyticity of the correlator, but while with $\cK(\a,t)$ and $\g(\D)$ this happens only ``on-shell'', {\it {\it i.e.},} after the solution is imposed, this is completely natural if we work with $\cK(j,t)$ and $\g(J)$.

Finally, let us comment on other possible applications of the same arguments. In this section we have been focused on Regge-bounded interactions, but one can generalize our equations to interactions that are {\it not} bounded in the Regge limit, such as derivative contact terms. It turns out that one can solve the crossing equation in the Regge limit also in this case, but only order by order in a perturbative expansion around MFT. In particular, at every order the anomalous dimensions will grow more and more as $n\ra \infty$, but we can imagine a very small coupling constant $\l$ such that $\l$ times any power of $n$ is still small, and solve the algebraic constraints coming from eq. \eqref{reggecorssinggamma} order by order in $\l$. Furthermore, although we did not discuss fermions, it turns out that very similar equations apply to the case of fields with Fermi statistics, and one gets similar constraints. Finally, the whole procedure can be generalized to models with $O(N)$ symmetry, both in the bosonic and in the fermionic case.

\section{Transcendentality ansatz - loop level}\label{sec:transc - loop}

We now turn to the study of loop level solutions for integer $\df$, using again a transcendentality ansatz and the constraints outlined at the beginning of the previous Section. The problem of loop-level AdS amplitudes was already considered in \cite{Aharony:2016dwx} from the point of view of the analytic bootstrap, where results where found in Mellin space, for $d=2$ and $d=4$. An interesting application to $\cN=4$ SYM was considered in \cite{Alday:2018pdi, Alday:2018kkw}, allowing to compute one-loop superstring amplitudes via the AdS/CFT correspondence. As it was observed in that paper, although the perturbative expansion of AdS amplitudes in terms of Witten diagrams is formally well-defined, already at one loop only a few results are available. Our construction of exact correlators in $d=1$ is then to be seen as a step forward in this direction, and it would be interesting to understand if it can be used as a constraint for higher dimensional theories, through the diagonal limit $z=\bar{z}$, on correlators in $d>1$. 

Our intuition at one loop is based on the result for $\df=1$ and $\f^4$ interaction given in \cite{mp}: in that case, the correlator has transcendentality four, and we will find that this remains valid for all contact term interactions at one loop, with or without derivatives. In the remainder of this section we shall discuss our results for contact terms with arbitrary number of derivatives at one loop for theories with a single field, and provide an example of similar solutions for theories with $O(N)$ symmetry.

\subsection{Contact terms, single field}

The conformal blocks expansion of a one-loop correlator $\cA^{(2)}(z)$ can be written as 
\begin{align} \label{CB1loop}
\begin{split}
\cA^{(2)}(z)\, =\sum_n &\Big( C^{(2)}_n
+\12 \,\left( C^{(0)}_n \,\g^{(2)}_n + C^{(1)}_n \,\g^{(1)}_n \,  \right)\frac{\pr}{\pr n} +\frac{1}{8} \,  C^{(0)}_n\,\left( \g^{(1)}_n \right)^2
 \frac{\pr^2}{\pr n^2} 
\Big)\, G_{2\,\D_{\f}+2n}(z)\\
=
 \sum_n &z^{2\df+2n}\, \Big[ 
 C^{(2)}_n+\12 \,\left( C^{(0)}_n \,\g^{(2)}_n + C^{(1)}_n \,\g^{(1)}_n \,  \right)\frac{\pr}{\pr n}+\frac{1}{8} \,  C^{(0)}_n\,\left( \g^{(1)}_n \right)^2
 \frac{\pr^2}{\pr n^2}\\&
 +\left( C^{(0)}_n \,\g^{(2)}_n + C^{(1)}_n \,\g^{(1)}_n +\12\, \,  C^{(0)}_n\,\left( \g^{(1)}_n \right)^2
 \frac{\pr}{\pr n}  \right)\,\log(z)
 \\&+\12\, \,  C^{(0)}_n\,\left( \g^{(1)}_n \right)^2\,\log^2(z)
\Big]\, F_{2\,\D_{\f}+2n}(z).
\end{split}
\end{align} 
and we recall that $F_{\b}(z)={}_2F_1(\b,\b;2\b;z)$. We now have to provide an ansatz for the correlator at one loop, in terms of a basis of transcendental functions up to some given transcendentality. To this end, we observe that for integer $\df$ all the tree-level solutions for contact terms (regardless the value of $q$) have anomalous dimensions that are rational functions of $n$. This implies that the sum 
\be 
\12\,\sum_n \,C^{(0)}_n\,\left( \g^{(1)}_n \right)^2\,G_{2\,\D_{\f}+2n}(z),
\ee
which determines the part proportional to $\log^2(z)$ of $\cA^{(2)}(z)$, has transcendentality two for every integer $\df>0$. This justifies our choice to follow \cite{mp} and assume a maximal transcendentality of four. Schematically, we write the correlator as
\be 
\cA^{(2)}(z)=\frac{1}{(1-z)^{2\,\Delta_{\f}}}\,\sum_i R_i(z)\,T_i(z),
\ee
where $R_i(z)$ are rational functions whose denominators only contain powers of $z$ and $1-z$, while $T_i(z)$ are chosen from a basis for transcendental function of transcendentality less than or equal to four. In particular, we are looking for a basis of functions which have discontinuities at either $z=0$ or $z=1$: this requirement selects those functions that can be built, using the Symbol map, from the ``letters'' $z$ and $1-z$. Hence, a basis of such transcendental functions with fixed transcendentality $t$ contains exactly $2^t$ functions, and a basis of functions with transcendentality {\textit{ up to}} four contains $\sum_{t=0}^4 2^t=31$ functions. However, as one can see from \eqref{CB1loop}, a one-loop correlator cannot contain terms with $\log^n(z)$ for $n\ge 3$ in its small $z$ expansion. Therefore, we have to remove from our basis of transcendental functions the ones that contain $\log^n(z)$ for $n\ge 3$ in the small $z$ expansion. With this caveat, we reduce our basis to the following 28 functions:
\begin{itemize}
\item Transc. 0: 1.
\item Transc. 1: $\log z$, $\log(1-z)$.
\item Transc. 2: $\Li_2(z)$, $\log^2(z)$, $\log(z)\,\log(1-z)$, $\log^2(1-z)$.
\item Transc. 3: $\Li_3(z)$, $\Li_3(1-z)$, $\Li_2(z)\,\log(z)$, $\Li_2(z)\,\log(1-z)$, $\log^2(z)\,\log(1-z)$,\\ $\log(z)\,\log^2(1-z)$, $\log^3(1-z)$.
\item Transc. 4: $\Li_4(z)$, $\Li_4(1-z)$, $\Li_4\left( \frac{z}{z-1} \right)$, $\Li_3(z)\,\log(z)$, $\Li_3(z)\,\log (1-z)$, $\Li_3(1-z)\,\log(z)$,\\ $\Li_3(1-z)\,\log (1-z)$, $\Li_2(z)^2$, $\Li_2(z)\,\log^2(z)$, $\Li_2(z)\,\log(z)\,\log(1-z)$, $\Li_2(z)\,\log^2(1-z)$,\\ $\log^2(z)\,\log^2(1-z)$, $\log(z)\,\log^3(1-z)$, $\log^4(1-z)$.
\end{itemize}
We can now apply our constraints to this ansatz and find solutions at one loop. As in the other cases, we find a finite number of ambiguities, that corresponds to the addition of tree-level solutions, possibly with derivatives. To fix this, first of all we require the mildest possible Regge behaviour, but this is not enough. According to the number of derivatives in the tree-level interaction, we still have some number $f$ of free parameters, which we conventionally fix by setting to zero the first anomalous dimensions:
\be 
\g^{(2)}_n=0 \q (0\le n \le f-1).
\ee

\subsubsection{Non-derivative $\f^4$ interaction}

We begin the analysis of loop-level solution with the case of a $f^4$ vertex with no derivatives, which we labelled $q=0$ in Section \ref{treesinglefield}. In this case, it is always possible to fix the freedom of adding tree level solutions in such a way that
\be 
\lim_{n\ra \infty}\g^{(2)}_n \sim \frac{1}{n^2},
\ee
and this leaves us with only one free parameter, which corresponds to a multiple of a tree level solution with $q=0$. This is nothing else but the necessity to fix a renormalization condition ({\it {\it i.e.}} to fix the coupling constant at one loop), and since the interaction that we are considering is renormalizable, it suffices to add one tree level diagram with the same kind of interaction. We fix this freedom with the following choice of the coupling constant:
\be 
\g^{(2)}_{n=0}=0.
\ee
With these caveats, we can completely fix all one-loop correlators for integer $\df$ and $q=0$. We can express our result for the correlation function as
\begin{align}
\begin{split} 
\mathcal{A}^{(2)}(z)|_{q=0}=&\frac{1}{(1-z)^{2\df}}
\Bigg\{
\frac{(z-2)z^{2\df+1}P_1^{6(\df-1)}}{(1-z)^{2(\df-1)}}\,\Li_4(1-z)
\\&+\frac{(z+1)(1-z)^{2\df+1}P_2^{6(\df-1)}}{z^{2(\df-1)}}\,\Li_4(z)
+\frac{(2z-1)P_3^{6(\df-1)}}{\left(z(1-z)\right)^{2(\df-1)}}\,\Li_4\left( \frac{z}{z-1} \right)\\&
+\frac{(z-2)z^{2\df+1}P_4^{6(\df-1)}}{(1-z)^{2(\df-1)}}\,\Li_3(z)\,\log(z)
+\frac{z^{2\df}P_5^{6\df-4}}{(1-z)^{2(\df-1)}}\,\Li_3(z)\,\log(1-z)\\&
+\frac{(z-2)P_6^{10(\df-1)}}{\left(z(1-z)\right)^{2\df-3}}\,\Li_3(z) 
+\frac{(1-z)^{2\df}P_7^{6\df-4}}{z^{2(\df-1)}}\,\Li_3(1-z)\,\log(z)\\&
+\frac{(1+z)(1-z)^{2\df+1}P_8^{6\df-6}}{z^{2(\df-1)}}\,\Li_3(1-z)\,\log(1-z) \\&
+\frac{(1+z)P_9^{10(\df-1)}}{\left(z(1-z)\right)^{2\df-3}}\,\Li_3(1-z) 
+\frac{(z-2)\,P_{10}^{10(\df-1)}}{\left(z(1-z)\right)^{2\df-3}}\,\Li_2(z)\,\log(z)\\&
+\frac{(z+1)\,P_{11}^{10(\df-1)}}{\left(z(1-z)\right)^{2\df-3}}\,\Li_2(z)\,\log(1-z)
+\frac{(2z-1)P_{12}^{6(\df-1)}}{\left(z(1-z)\right)^{2(\df-1)}}\,\log^4(1-z)\\&
+\frac{(2z-1)P_{13}^{6(\df-1)}}{\left(z(1-z)\right)^{2(\df-1)}}\,\log^3(1-z)\,\log(z)
+\frac{(1-z)^{2\df}P_{14}^{6\df-4}}{z^{2(\df-1)}}\,\log^2(1-z)\,\log^2(z) \\&
+\frac{(1+z)P_{15}^{10(\df-1)}}{\left(z(1-z)\right)^{2\df-3}}\,\log^2(1-z)\,\log(z)
+\frac{P_{16}^{8\df-6}}{\left(z(1-z)\right)^{2(\df-1)}}\,\log^2(1-z)
\nonumber
\end{split}
\end{align}
\begin{align}
\begin{split}
~~~~~~~~~~~~~~&
+\frac{(z-2)(1-z)^{2\df}P_{17}^{2(\df-2)}}{z^{2\df-3}}\,\log(1-z)\,\log^2(z)\\&
+\frac{(1-z)^2\,P_{18}^{10\df-8}}{\left(z(1-z)\right)^{2(\df-1)}}\,\log(1-z)\,\log(z)
+\frac{P_{19}^{10\df-7}}{\left(z(1-z)\right)^{2(\df-1)}}\,\log(1-z)\\&
+\frac{(1-z)^2\,P_{20}^{6\df-8}}{z^{2(\df-2)}}\,\log^2(z)
+\frac{P_{21}^{10\df-8}}{\left(z(1-z)\right)^{2(\df-1)}}\,\log(z)
+\frac{z\,P_{22}^{10\df-7}}{\left(z(1-z)\right)^{2(\df-1)}}
\Bigg\},
\end{split}
\end{align}
where $P_i^n(z)$ are polynomials of degree $n$ in $z$. The result for $\df=1$ was already given in \cite{mp}, and in appendix \ref{sec:1loop_phi^4} we provide for instance the result for $\df=2$. \\
Let us now discuss the one-loop CFT data corresponding to these solutions. First, we found useful to express our results in terms of harmonic sums, defined in Appendix \ref{sec:HarmSums}. In particular, we introduce the following combinations of harmonic sums:
\begin{align}
\begin{split}
\cS_3(n)=&S_{-3}(n)-2S_{-2,1}(n),\\
\s_2(n)=&2S_{-2}(n)+\zeta (2),\\
\s_3(n)=&S_{3}(n)-\zeta (3),\\
\s_4(n)=&8 \zeta (2) S_{-2}(n)-8 S_{4}(n)+16 S_{-2,-2}(n)+5 \zeta (4),
\end{split}
\end{align}
that will prove useful to have more compact expressions. We make the reciprocity principle of \cite{Dokshitzer:2005bf, Basso:2006nk, Alday:2015eya} manifest by expressing our results in terms of the conformal Casimir\footnote{This is actually the bare value of the conformal Casimir, {\it {\it i.e.}} the conformal Casimir for double trace operators computed with the MFT dimension of the double-trace operators. Appropriate combinations of CFT data will take into account the difference between the bare and the full Casimir $(2\df+2n+\g_n)\,(2\df+2n-1+\g_n)$.}
\be \label{ConfCas}
J^2=(2\df+2n)\,(2\df+2n-1).
\ee
Moreover, we write the one-loop OPE coefficients as
\begin{align}\label{derrel1loop}
\begin{split} 
C^{(2)}_n=&\12\frac{\pr}{\pr n}\left( C^{0}_n\,\g^{(2)}_n+C^{(1)}_n\,\g^{(1)}_n \right)-\frac{1}{8}\frac{\pr^2}{\pr n^2}\left( C^{(0)}_n\left( \g^{(1)}_n \right)^2 \right)+C^{(0)}_n\d C^{(2)}_n,
\end{split}
\end{align}
where $\d^{(2)}_n$ can be seen as a one-loop violation to the derivative relation. For general $\df$, we found that 
\begin{align}\label{loopgamma}
\begin{split} 
\g^{(2)}_n=&\12 \g^{(1)}_n\,\frac{\pr}{\pr n}\left( \g^{(1)}_n \right)+
P_1^{2(\df-1)}\, \cS_{3}(2 n+2\df-1)\\&
+\frac{1}{\prod_{r=0}^{\df-1}\left( J^2-2r(2r+1) \right)}
\Big[\frac{P_2^{2(\df-1)}}{\prod_{r=0}^{\df-1}\left( J^2-2r(2r-1) \right)} S_{-2}(2 n+2\df-1)\\&
+\frac{1}{\prod_{r=0}^{2(\df-1)}(J^2-r(r+1))}
\left(P_3^{5\df-4)}\,H_{2 n+2\df-1}+P_4^{5\df-6}\right)+ P_5^{3\df-2)}\,\zeta (3)\Big],
\end{split}
\end{align}
while
\begin{align} \label{loopdelta}
\begin{split} 
\d C^{(2)}_n=&\frac{1}{\prod_{r=0}^{2(\df-1)}(J^2-r(r+1))}
\Big(Q_1^{4\df-5}+Q_2^{4(\df-1)}\,\s_2(2n+2\df-1) \\&
+Q_3^{2(\df-1)}\,\s_3(2n+2\df-1)\Big)+Q_4^{2(\df-1)}\,\s_4(2n+2\df-1),
\end{split}
\end{align}
where the $P_i^n$ and $Q_i^n$ are polynomials of degree $n$ in the bare conformal Casimir $J^2$.

Our results extend those of \cite{mp} to higher (integer) values of $\df$, and in particular we were able to find results up to $\df=9$. However, since the polynomials appearing in the correlation functions and in the CFT data get more and more complicated when $\df$ increases, we will only give some explicit results in Appendix \ref{sec:1loop_phi^4}. Here, we limit ourselves to express the CFT data found in \cite{mp} for $\df=1$ into the form given in eqs. \eqref{loopgamma} and \eqref{loopdelta}:
\begin{align}
\begin{split}
\g^{(2)}_n=&\12 \g^{(1)}_n\,\frac{\pr}{\pr n}\left( \g^{(1)}_n \right)+4 \cS_{3}(2 n+1)\\&
+\frac{1}{J^2}\left(4 S_{-2}(2 n+1)-\frac{4 \left(J^2-1\right) H_{2 n+1}}{J^2}-2 \left(J^2-2\right) \zeta (3)+1\right),\\
\d C^{(2)}_n=&\s_4(2n+1)+\frac{4}{J^2}\Big( \s_2(2n+1)-2\,\s_3(2n+1)\Big).
\end{split}
\end{align} 
We can also compare these results to those of Section \ref{sec:Reggelimit}, and to this end let us compute the large $J$ expansion of these CFT data. Using the formulas given in Appendix \ref{sec:HarmSums}, we get
\begin{align} \label{gammad1loopexp}
\begin{split}
\g^{(2)}_n=&\12 \g^{(1)}_n\,\frac{\pr}{\pr n}\left( \g^{(1)}_n \right)+
\frac{12 \zeta (3)-\pi ^2+3}{3 J^2}+\frac{2 (4 \log (J)+4 \gamma_E -3)}{J^6}-\frac{4 (24 \log (J)+24 \gamma_E -31)}{3 J^8}
\\
&+\frac{4 (280 \log (J)+280 \gamma_E -447)}{5 J^{10}}-\frac{64 (11970 \log (J)+11970 \gamma_E -21767)}{315 J^{12}}\\
&+\frac{8 (1486800 \log (J)+1486800 \gamma_E -2967193)}{315 J^{14}}+\cO\left(\frac{1}{J^{16}}\right),
\end{split}
\end{align}
where $\g_E$ is the Euler-Mascheroni constant, and
\be 
\d C^{(2)}_n=\frac{2}{J^4}-\frac{2}{J^6}+\frac{10}{3 J^8}-\frac{28}{3 J^{10}}+\frac{632}{15 J^{12}}-\frac{880}{3 J^{14}}+\cO\left(\frac{1}{J^{16}}\right).
\ee
We can see that, as discussed in Section \ref{sec:Reggelimit} around eq. \eqref{derivativerule}, the violation to the derivative relation has an expansion in $J^2$ that begins with $1/J^{4\df}$\footnote{Here we showed it only for $\df=1$, but we found this to be true for every (integer) $\df$ that we have studied.}. As a final comment, let us note that one can find a coupling redefinition (corresponding to the addition of a $q=0$ tree-level solution with an appropriate normalization) such that the expansion of eq. \eqref{gammad1loopexp} begins with $\log(J)/J^6$. In the case at hand, corresponding to $\df=1$, this is particularly easy since there is no $J^{-4}$ term and $\g^{(1)}_n|_{q=0}=2/J^2$, but it turns out that this was possible for all the solutions we have found, and therefore we can conjecture that the same happens for any $\df$.

Finally, let us comment again on the transcendentality of the functions appearing in our solution. As we have discussed, all the one-loop correlators that we have found for integer $\df$ have transcendentality four, and the corresponding anomalous dimensions contain harmonic sums of weight three at most. The corrections to the derivative rule, instead, contain harmonic sums of weight four, and have therefore the same transcendentality as the correlator. These relations are completely analogous to the ones found for tree-level exchanges. Let us also observe that not all the reciprocity-respecting harmonic sums for a given weight are present in eqs. \eqref{loopgamma} and \eqref{loopdelta}. This might be due to the fact that not all weight four transcendental functions appear in $\cA^{(2)}(z)$, but only those that do not contain powers of $\log(z)$ higher than two.

\subsubsection{Loop level - derivative interactions}\label{sec:loopder}

We can now discuss loop-level interactions with $q\ge 1$. To study this kind of correlators, we found useful to observe that, for every $q\ge 1$ and integer $\df \ge 1$, there exists a number $\a(q,\df)$ such that
\be \label{loopderdiff}
\cG(z)=\cA^{(2)}_{q}(z)-\a(q,\df)\,\cA^{(2)}_{q=0}(z)
\ee
is actually of transcendentality two only (as opposed to the transcendentality four of the general one-loop correlator), and since all are constraints are satisfied both by $\cA^{(2)}_{q}(z)$ and by $\cA^{(2)}_{q=0}(z)$, then also $\cG(z)$ satisfies all the constraints\footnote{The fact that $\cG(z)$ has reduced transcendentality can be seen as consequence of the fact that, for every $q$ and $\df$, one can find a real number $\a(q,\df)$ such that $\left(\g^{(1)}_n|_q\right)^2-\a(q,\df) \,\left( \g^{(1)}_n|_{q=0}\right)^2$ is such that the sum $\sum_n C^{(0)}_n \left[\left( \g^{(1)}_n|_q\right)^2-\a(q,\df) \,\left( \g^{(1)}_n|_{q=0} \right)^2\right] G_{2\df+2n}(z)$ is a simple rational function, therefore giving transcendentality two for $\cG(z)$ when multiplied by $\log^2(z)$.}. The conformal blocks decomposition of $\cG(z)$ is easily read from the definition \eqref{loopderdiff} and the one-loops decomposition \eqref{CB1loop}. As anticipated, however, the constraints that we are using are not enough to fix all the possible ambiguities that correspond to the addition of tree level diagrams. In particular, as anticipated we shall require the mildest possible Regge behaviour, that turns out to be
\be \label{ReggeGammaLoopDer}
\lim_{n\ra \infty} \g_n^{(2)}\sim  n^{8q-6} \, \log n,
\ee
or equivalently
\be 
\lim_{t\ra \infty}\left(\frac{1}{2}+i\,t\right)^{-2\df}\,\cA^{(2)}\left(\frac{1}{2}+i\,t\right) \, \sim
\, t^{4q-2}. 
\ee
Even after fixing this behaviour, we are left with $2q$ free parameters. Again, this can be interpreted as a renormalization condition: since we are dealing with non-renormalizable interactions, one-loop renormalization requires the inclusion of tree-level interactions with more derivatives (higher value of $q$) than the one considered in the one-loop diagram. In particular, comparing eq. \eqref{ReggeGammaLoopDer} with eq. \eqref{ReggeGammaTreeDer} one can easily read that there are precisely $2q$ tree diagrams with milder (or equal) Regge behaviour than a one-loop diagram at level $q$. To fix these ambiguities, one can for instance require that
\be 
\g^{(2)}_n|_q=0 \qquad (0\le n \le 2q).
\ee
The general expression for the functions $\cG(z)$ that we have found is 
\begin{align}
\begin{split} 
\cG(z)=&\frac{1}{(1-z)^{2\df}}\Bigg\{\frac{z^{2\df}\,P_1^{4\df+8q-6}}{(1-z)^{2(2q-1)}}\,\log^2(z)+\frac{P_2^{4\df+12q-8}}{\left( z(1-z) \right)^{4q-3}}\,\log(z)\log(1-z) \\
&+\frac{(1-z)^{2\df}\,P_3^{4\df+8q-6}}{z^{2(2q-1)}}\,\log^2(1-z)+
\frac{P_4^{4\df+12q-10}}{(1-z)\,\left( z(1-z) \right)^{4(q-1)}}\,\log(z)\\ &
+
\frac{P_5^{4\df+12q-10}}{z\,\left( z(1-z) \right)^{4(q-1)}}\,\log(1-z)
+
\frac{P_6^{4\df+12q-12}}{\left( z(1-z) \right)^{4(q-1)}} \Bigg\},
\end{split}
\end{align}
where the $P_i^n$ are polynomials of degree $n$ in $z$. 

As for the anomalous dimensions, we can define
\be 
\G^{(2)}_n|_q=\g^{(2)}_n|_q-\a(q,\df) \, \g^{(2)}_n|_{q=0},
\ee
which turns out to have a simpler expression than the full $\g^{(2)}_n|_q$. For general (integer) $\df$ we found, in terms of the bare conformal Casimir defined in eq. \eqref{ConfCas},
\begin{align} \label{gammadiffloopder}
\begin{split}
\G^{(2)}_n|_q=&\12 \left( \g^{(1)}_n|_q \right)\frac{\pr}{\pr n} \left( \g^{(1)}_n|_q \right)-\12\,\a(q,\df)\,\left( \g^{(1)}_n|_{q=0}\right) \frac{\pr}{\pr n} \left(  \g^{(1)}_n|_{q=0} \right)\\ &
 +\frac{P_1^{4\df+8q-8}}{\prod_{r=0}^{\df-1}\left(J^2-r(r+1)\right)}H_{2n+2\df-1}+
 \frac{\left(J^2-2\df(2\df-1)\right)P_2^{2\df+8q-8}}{\prod_{r=0}^{\df-1}\left(J^2-2r(2r+1)\right)}\z(3)\\ &
 +\frac{P_3^{4\df+8q-8}}{\prod_{r=0}^{\df-1}\left(J^2-r(r+1)\right)},
\end{split}
\end{align}
where the $P_i^n$ are polynomials of degree $n$ in $J^2$. 

Similarly, for the OPE coefficients it is useful to define
\begin{align} \label{opediffloopder}
\begin{split}
C^{(0)}_n\,\D C^{(2)}_n =&\left[ C^{(2)}_n|_q-
\12\frac{\pr}{\pr n}\left( C^{0}_n\,\g^{(2)}_n|_q+C^{(1)}_n|_q\,\g^{(1)}_n|_q \right)-\frac{1}{8}\frac{\pr^2}{\pr n^2}\left( C^{(0)}_n\left( \g^{(1)}_n|_q \right)^2 \right)\right]\\ 
-\a(q,\df)\,&\left[ C^{(2)}_n|_{q=0}-
\12\frac{\pr}{\pr n}\left( C^{0}_n\,\g^{(2)}_n|_{q=0}+C^{(1)}_n|_{q=0}\,\g^{(1)}_n|_{q=0} \right)-\frac{1}{8}\frac{\pr^2}{\pr n^2}\left( C^{(0)}_n\left( \g^{(1)}_n|_{q=0} \right)^2 \right)\right],
\end{split}
\end{align}
which amounts to $C^{(0)}_n$. For general (integer) $\df$ we found, again in terms of the bare conformal Casimir,
\begin{align}
\begin{split}
\D C^{(2)}_n=&\frac{Q_1^{4\df+8q-8}}{\prod_{r=1}^{2(\df-1)}\left(J^2-r(r+1)\right)}\,\s_2(2n+2\df-1)\\&+\frac{Q_2^{4\df+16q-10}}{\prod_{r=0}^{2(\df-1)}(J^2-r(r+1)) \, \prod_{s=0}^{2(2q-1)}(J^2-s(s+1))},
\end{split}
\end{align}
which amounts to the quantity $C^{(0)}_n \left( \d C^{(2)}_n|_q-\d C^{(2)}_n|_{q=0} \right)$ where the $Q_i^n$ are polynomials of degree $n$ in $J^2$.

Interestingly, we have observed that the function $\d C^{(2)}_n|_q$ expressing the violation of the derivative rule for interactions with $4q$ derivatives is more suppressed than $\d C^{(2)}_n|_{q=0}$ for large $n$. In particular, as we also show with some examples in Appendix \ref{appendix:loopder}, we find
\be \label{violationderrulederivatives}
\d C^{(2)}_n|_q\sim n^{-4\df-4q}.
\ee
Finally, as we shall discuss in Appendix \ref{appendix:loopder}, our results are found to agree with the one found in Section \ref{sec:genericcontact} using the PM bootstrap.

\subsection{Contact terms, $O(N)$ global symmetry} \label{loop O(N)}

In this Section we consider one-loop solutions to models with $O(N)$ global symmetry. Essentially, we have to combine the machinery developed in the previous Section for loop-level solutions, using an ansatz of transcendentality four, with the observations of Section \ref{sec:O(N)tree} about crossing symmetry in the $O(N)$ model. The computations are very similar to those of the previous Section, but the results are more complicated due to the presence of three different representations. Therefore, for simplicity we shall limit to consider the case $\df=1$ and $p=0$ (in the notation of Section \ref{sec:O(N)tree}). 

In the case at hand, we find two ambiguities, which precisely correspond to the contact terms that one needs to add to the PM basis in the $O(N)$ case. In the language of Section \ref{sec:O(N)tree}, they correspond to the addition of a solution with $q=0$ and one with $p=0$. We fix these in such a way that
\be 
\g_n^{(2),T}=0 \qquad (n=0,1).
\ee
We give for instance the $T$-channel expression, from which the other two channels can be extracted using crossing symmetry:
\begin{align}\nonumber
\begin{split}
\cA^{(2),T}(z)=&\frac{1}{(1-z)^{2\df}}\Big\{ 
\frac{72}{25} (z-1)^2 \left(3 z^2-2\right)\Li_4(z)
+\frac{216}{25} (z-2) z^3\Li_4(1-z)\\
&
+\frac{72}{25} \left(z^2+4 z-2\right)\Li_4\left( \frac{z}{z-1} \right)-
\frac{108}{25}  z^2 \big( (z-2) z\log(z)\\
&+(z^2-2 z+3)\log(1-z) \big)\Li_3(z)
-\frac{36}{25}  (z-1)^2\big((3 z^2+4)\log(z)\\
&+(3 z^2-2)\log(1-z)\big)\Li_3(1-z)
+\frac{3}{25} \left(z^2+4 z-2\right) \big( \log(1-z)\\
&-4\log(z) \big)\log^3(1-z)+\frac{6}{25}\left( (z^2+4 z-2)\pi^2-3(z-1)^2 \left(z^2+2\right) \right)\log^2(1-z)\\
&+\frac{6}{25}\left( (z-1)^2 \left(3 z^2+4\right)\pi^2-6 z^2 (2 z-1) \right)\log(z)\log(1-z)\\
&+\frac{18}{25} z^2 \left(z^2+2 z-2\right)\log^2(z)+\frac{9}{50}\big((z-1) z \left(43 z^2+59 z-94\right)\\
&-8(3 z^4-6 z^3-z^2-4 z+2)\z(3)\big)\log(1-z)
-\frac{9}{50}\big(z^2 \left(43 z^2+24 z-24\right)\\
&-8(z-1)^2 \left(3 z^2+4\right)\z(3)\big)\log(z)\frac{1}{125} (-3) z^2 \left(z^2-2 z-6\right)\pi^4\\
& -\frac{108}{25}  (z-1) z^2\z(3)+\frac{387}{50} (z-1) z^2\\
&+N\Big[\frac{36}{25} (z-1)^2 z^2\Li_4(z)
+\frac{36}{25} (z-2) z^3\Li_4(1-z)+\frac{36 z^2}{25}\Li_4\left( \frac{z}{z-1} \right)\\
\end{split}
\end{align}
\begin{align}
\begin{split}
~~~~~~~~~~&
-\frac{18}{25}  (z-2) z^2 \left( z\log(z) +(z^2-2 z+3)\log(1-z)\right)\Li_3(z)\\
&-\frac{18}{25}(z-1)^2 z^2\left( \log(z)+\log(1-z) \right)\Li_3(1-z)\\
&+\frac{3}{50}z^2\left( \log(1-z)-4\log(z) \right)\log^3(1-z)-\frac{9}{25} (z-1)^2 z^2\log^2(1-z)\log^2(z)\\
&+\frac{3}{25}\left( z^2 \pi^2+6(z-1)^2 (z+1)^2 \right)\log^2(1-z)+\frac{3}{25}\big((z-1)^2 z^2\pi^2 \\
&-6(z-2) z \left(z^2+2 z-2\right)\big)\log(1-z)\log(z)-\frac{3}{50}\big((z-1) z \left(5 z^2+41 z-34\right)\\
&+12 z^2 \left(z^2-2 z-1\right)\z(3)\big)\log(1-z)+\frac{3}{50}\big(z^2 \left(5 z^2+48 z-48\right)\\
&+12(z-1)^2 z^2\z(3)\big)\log(1-z)-\frac{1}{250} z^2 \left(z^2-2 z-6\right)\pi^4\\
&-\frac{18}{25}  (z-1) z^2\z(3)
-\frac{3}{10}  (z-1) z^2
\Big]\Big\}.
\end{split}
\end{align}
The corresponding anomalous dimensions are
\begin{align}
\begin{split}
\g^{(2),T}(z)=&+\frac{1}{2}\g^{(1),T}_n\frac{\pr}{\pr n}\left( \g^{(1),T}_n \right)+\frac{36 (N+6)}{25}\cS_3(1+2n)+\frac{36 (N+6)}{25 J^2}S_{-2}(1+2n)\\
&-\frac{36 }{25 J^2}\left( (J^2-1)N-\frac{2 \left(J^4-5 J^2+2\right)}{J^2} \right)H_{1+2n}\\
&+\frac{1}{50J^2}\left( 87 \left(J^2-2\right) N -9 \left(59 J^2-102\right)\right),\\
\g^{(2),A}(z)=&+\frac{1}{2}\g^{(1),A}_n\frac{\pr}{\pr n}\left( \g^{(1),A}_n \right)+\frac{36 (N+2)}{25}\cS_3(2+2n)+\frac{36(N+2)}{25 J^2}S_{-2}(2+2n)\\
&-\frac{36 (N+4)}{25}\z(3)+\frac{36}{25 J^2}\left( (J^2-1)N-2(J^2+1) \right)H_{2+2n}\\
&-\frac{1}{50J^2}\left( (87J^2+36)N-531J^2 \right),\\
\g^{(2),S}(z)=&+\frac{1}{2}\g^{(1),S}_n\frac{\pr}{\pr n}\left( \g^{(1),S}_n \right)+\frac{108 (N+2)}{25}\cS_3(1+2n)+\frac{108 (N+2)}{25 J^2}S_{-2}(1+2n)\\
&+\frac{36}{25J^4}\left(\left(J^2+1\right)^2 N^2 -(3 J^4+J^2-4)N+2 \left(J^4-5 J^2+2\right)\right)H_{1+2n}\\
&-\frac{3}{50 J^2}\left( \left(29 J^2+35\right) N^2 -\left(206 J^2+119\right)N+3 (59 J^2-102) \right).
\end{split}
\end{align}
For the OPE coefficients, as in the previous Section we study functions $\d C^{(2)}_n$ expressing the violation to the derivative rule, and in the case we considered we found
\begin{align}\nonumber
\begin{split}
\d C^{(2),T}(z)=&\frac{9 (N+6)}{25}\s_4(1+2n)-\frac{72 (N+6)}{25 J^2}\s_3(1+2n)+\frac{36}{25 J^4}\left( (J^2+1)N-(J^2+2) \right)\\
& +\frac{36}{25 J^2}\left( (J^2-1)N-(J^2-6) \right)\s_2(1+2n),
\end{split}
\end{align}
\begin{align}
\begin{split}
\d C^{(2),A}(z)=&\frac{9 (N+2)}{25}\s_4(2+2n)-\frac{72 (N+2)}{25 J^2}\s_3(2+2n)+\frac{36}{25 J^2}\left( (J^2-1)N-(3J^2+2) \right)\\
&-\frac{36}{25 J^2}\left( (J^2-1)N-(3J^2+2) \right)\s_2(2+2n),\\
\d C^{(2),S}(z)=&\frac{27 (N+2)}{25}\s_4(1+2n)-\frac{216 (N+2)}{25 J^2}\s_3(1+2n)+\frac{36 \left(J^2+2\right) (N-1)}{25 J^4}\\
&+\frac{36}{25 J^2}\left( (J^2+3)N-(J^2-6) \right)\s_2(1+2n).
\end{split}
\end{align}
Notice that for the $T$ and $S$ representation we have $J^2=(2\df+2n)(2\df+2n-1)$, while for $A$ $J^2=(2\df+2n+1)(2\df+2n)$, due to the difference in the (bare) dimension of the corresponding double-trace operators.

In the Regge (large $J$) limit, we find that all anomalous dimensions scale with $\log(J)$: there is no power-law divergence in $j$ as $J\ra\infty$, and this stems from the fact that the interaction with $p=0$ is still renormalizable. The corrections to the derivative rule, instead, all scale with $J^{-4}$, in agreement with the fact that in this case the operator with lowest dimension in the OPE is the double trace $\f\,\f$, with dimension $\D_{min}=2\df=2$ at the MFT level. Finally, we can observe that the type of functions appearing in the correlators is the same as in the $N=1$ case, and therefore we find the same combinations of harmonic sums as for $N=1$ in the CFT data.

\section{Comments on higher dimensions}\label{sec:highdconclu}
In this section, we will briefly comment on what happens in higher dimensions. While the results so far are very encouraging for $1d$ PM bootstrap, at this stage it is not entirely clear what light they shed on the higher dimensional case. A proposal for a Mellin basis was recently put forward in \cite{Mazac:2019shk} but in the context of dispersion relations which are not crossing symmetric\footnote{Also see \cite{Paulos:2019gtx} for another interesting approach to find analytic functionals in higher dimensions.}.

Let us begin by considering the diagonal limit of the $2d$ Ising model. Can it be expanded in terms of the $1d$ PM blocks?
In the diagonal limit, the $2d$ ising model reduced correlation function takes the form 
\be
{\mathcal A}(z)=\sqrt[4]{\frac{1}{1-z}}\,,
\ee
which can be expanded in $1d$ conformal block where the OPE coefficients are given by 
\be
C_\D=\frac{\sqrt{\pi } (-1)^{\Delta } 2^{-2 (\Delta +1)} \Gamma \left(-\frac{1}{4}\right) \, _3F_2\left(\frac{3}{4},1-\Delta ,2-\Delta ;2,\frac{7}{4}-\Delta ;1\right)}{\Gamma \left(\frac{7}{4}-\Delta \right) \Gamma \left(\Delta -\frac{1}{2}\right)}\,,
\ee
and scaling dimensions are integers
\be
\D=1,~2,~3,\dots \,.
\ee

\begin{figure}[hbt!]
  \begin{subfigure}[b]{0.48\linewidth}
    \includegraphics[width=\linewidth]{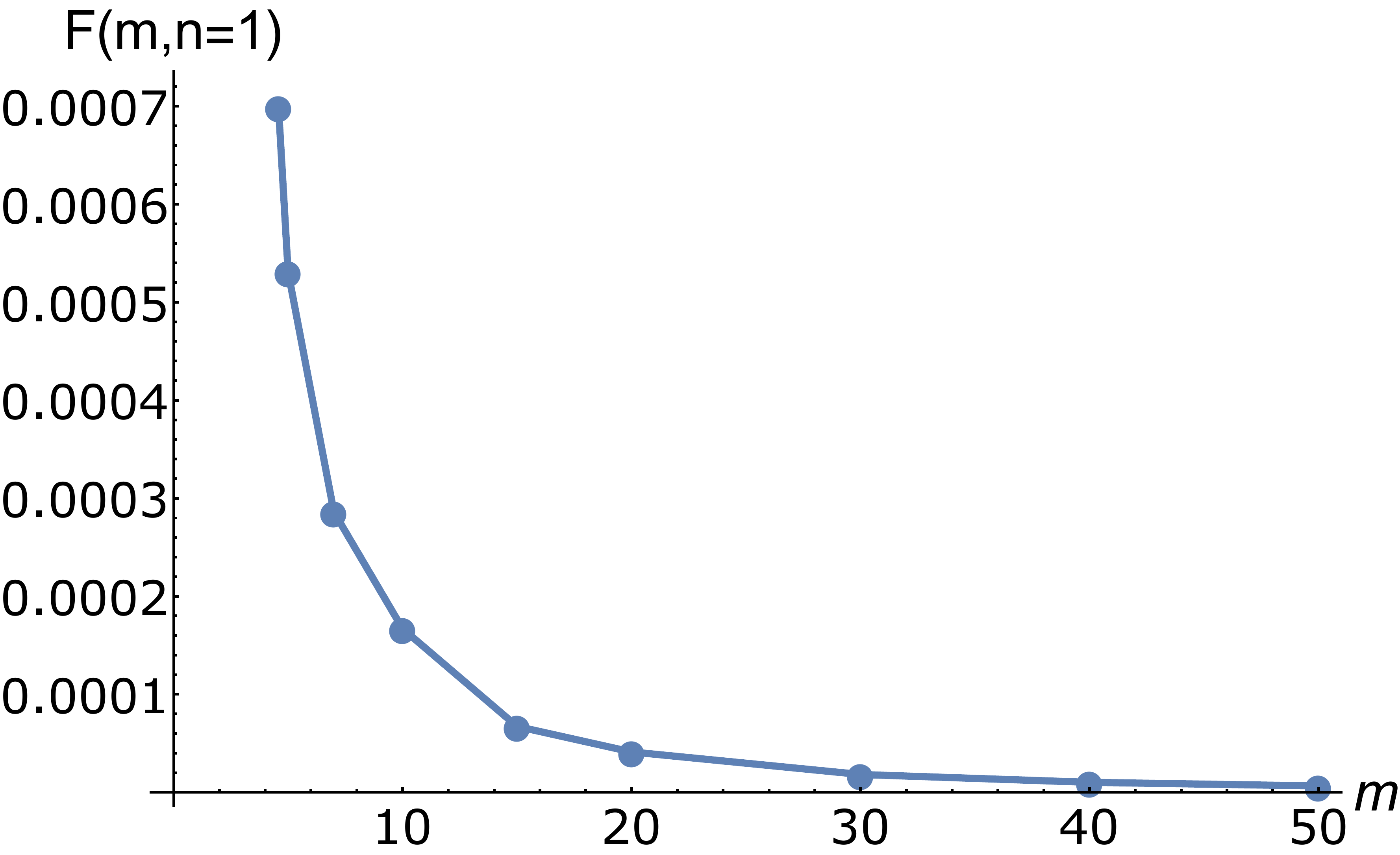}
    \caption{$F(m,n=1)~vs ~ m$}
  \end{subfigure}
  \begin{subfigure}[b]{0.48\linewidth}
     \includegraphics[width=\linewidth]{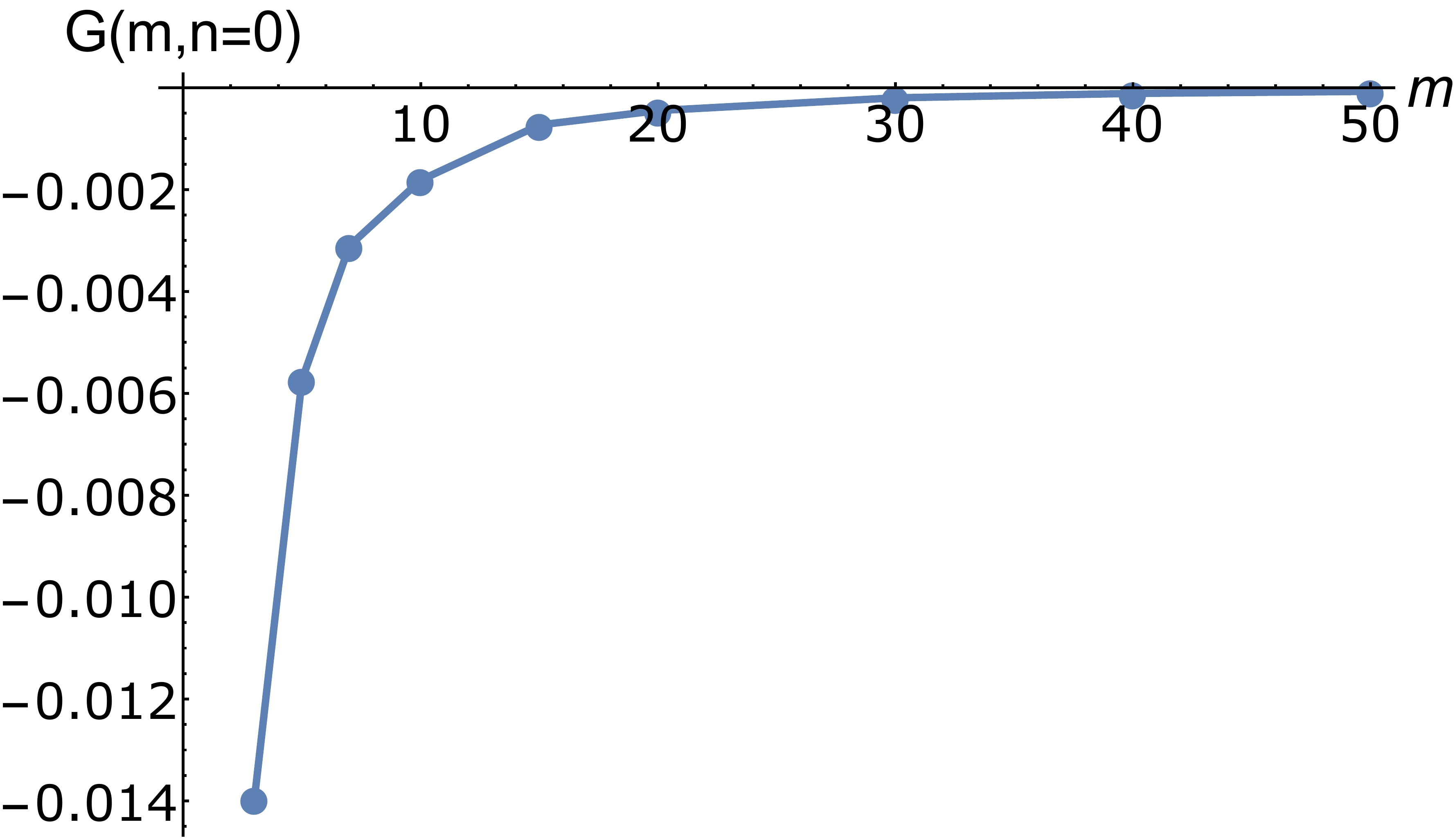}
    \caption{$G(m,n=0)~vs ~ m$}
   \end{subfigure}
   \caption{As $m$ increases $F(m,n=1)$ goes to $0$, also $G(m,n=0)$ goes to $0$ which is expected.  Note that  $q_{dis}'(\D_\phi)=2$ and $\Delta_{\phi}=\frac{1}{8}.$}
   \label{im:$1d$_2dising}
\end{figure}

To guarantee the expansion of diagonal $2d$ Ising amplitude in the $1d$ Poyakov-Mellin basis, it is necessary and sufficient to show that eq. (\ref{eq}, \ref{2eq}) holds for given $C_\D$ and $\D$. In order to show that, we truncate the sum over $\D$ as follows
\be
F(m,n)=\sum_{\D=1}^m C_{\D}N_{\D,0}f_{\D}(\D_\phi+n),~~
G(m,n)=\left(\sum_{\D=1}^m \left(C_{\D}N_{\D,0}f_{\D}'(\D_\phi+n)\right)\right)+q_{dis}'(\D_\phi+n)\,.
\ee
So as $m$ goes to $\infty$ we have $F(\infty,n)=0$ and $G(\infty,n)=0$. We illustrate this in the plots given below fig. (\ref{im:$1d$_2dising}). In the plots we show the $n=1$ of eq. (\ref{eq}) and $n=0$ of eq. (\ref{2eq}). Note that  $q_{dis}'(\D_\phi)=2$. One can easily see that as $m$ increases $F(m,n=1)$ goes to $0$, also $G(m,n=0)$ goes to $0$ which is expected. 

In light of this encouraging finding,  we ask what are the key considerations in our approach which can generalize to higher dimensions.

While \cite{mp} have argued the requirement for adding contact terms by demanding good Regge boundedness, this was not the argument we used in the PM bootstrap set up considered in this paper. The way we approached the contact term addition to the basis can be summarized as follows: we checked whether the sum over the spectrum was convergent or not and whenever we found that this was not the case we added appropriate contact terms to get rid of the divergence. In higher dimensions, there are two quantum numbers for each operators, spin and conformal dimension, unlike in $1d$. So, we have to make sure that for different limits our basis expansion has nice convergence properties. We will examine our basis in two extreme limits: a) fixed spin and large twist and b) fixed twist and large spin. For fixed spin and large twist (see also \cite{gs}), we find with MFT coefficients the blocks grow as $\frac{1}{\Delta^{2h-2+\ell}} $. Therefore, only spin zero will have a problem and it would be necessary to add a scalar contact term to the basis to cure it. For MFT, the fixed twist, large spin limit can be shown to be nicely convergent. If we were only considering CFTs with OPEs growing like MFT or slower, this then would be the punchline of our story--we just add the scalar contact term to fix the divergence problem. Now notice that since the fall off for fixed spin, large twist is $1/\D^{2h-2}$ for $\ell=0$, naively it would appear that for $d>3$, there would be no convergence issue. However, for the $\epsilon$-expansion, we need formulas that are analytic in $d$ and as such it is important to still add the scalar contact term. In \cite{gs}, this was shown to fix a mismatch with the $\phi^2$ anomalous dimension at $\epsilon^3$ order.  

As another extreme example, we take the $2d$ Ising model spectrum and study the $s-$channel behaviour in the fixed twist, large spin limit. As a concrete example, let us take twist zero operators. Our $s$-channel becomes (for $s=\Dphi$)
\begin{equation}
C_{\ell,\ell}\, N_{\Delta,\ell}\, q^{s}_{\Delta,\ell'|\ell}=C_{\ell,\ell} \frac{2^{-\ell -1} (\ell -1)^2 (2 \ell -1) \sin \left(\frac{\pi }{8}\right) \Gamma \left(\frac{5}{8}\right) \Gamma \left(\frac{\ell }{2}-\frac{1}{8}\right) \Gamma^2 \left(\frac{\ell -1}{2}\right) \Gamma \left(\ell -\frac{3}{4}\right) \Gamma^2 (2 \ell -1)}{\pi ^{3/2} \Gamma^2 \left(\frac{1}{8}\right) \Gamma \left(\frac{\ell }{2}+\frac{5}{8}\right) \Gamma \left(\ell -\frac{7}{8}\right) \Gamma \left(\ell -\frac{1}{2}\right) \Gamma ^4(\ell )},
\end{equation}  
and the OPE coefficients are given by
\begin{equation}
\begin{split}
C_{\ell,\ell} = &\frac{2^{1-2 \ell } \Gamma \left(\frac{7}{4}\right) \Gamma \left(\frac{\ell }{2}-\frac{1}{2}\right) \Gamma^2 (\ell ) \, _3F_2\left[\frac{1}{2},\frac{\ell }{2}-\frac{1}{2},\ell ;\frac{\ell }{2}+\frac{5}{4},\ell +\frac{1}{2};1\right]}{\pi  \Gamma \left(\frac{\ell }{2}+\frac{1}{4}\right) \Gamma \left(\ell -\frac{1}{2}\right) \Gamma \left(\ell +\frac{3}{2}\right)}\\
& -\frac{3\ 2^{-2 \ell -1} \Gamma \left(\frac{3}{4}\right) \Gamma \left(\frac{\ell }{2}\right) \Gamma (\ell ) \, _3F_2\left[\frac{1}{2},\frac{\ell }{2}-\frac{1}{4},\ell ;\frac{\ell }{2}+\frac{3}{2},\ell +\frac{1}{2};1\right)}{\pi  (\ell +1) \Gamma \left(\frac{\ell }{2}+\frac{1}{4}\right) \Gamma \left(\ell +\frac{1}{2}\right]}.
\end{split}
\end{equation}
Now in the large spin limit this behaves as
\begin{equation}
C_{\ell,\ell} N_{\ell,\ell} q^{s}_{\ell,0|\ell}(s=\Delta_{\phi})=\frac{3 \left(2-\sqrt{2}\right)^{3/2} \Gamma \left(\frac{5}{8}\right) \Gamma \left(\frac{3}{4}\right)}{16 \pi ^3 \ell^{7/8} \Gamma^2 \left(\frac{1}{8}\right)}
\end{equation}

Clearly, if we sum over spin this channel is going to diverge as it falls off as $\frac{1}{\ell^{7/8}}$.  So it seems we have to systematically study this limit and take linear combinations of equations to get rid of these divergences. Since this is a power law divergence, we believe appropriate linear combinations killing such divergences should exist. Thus we add another caveat to the list for the existence of the PM basis: divergences should be power law type for an appropriate linear combination (equivalent to adding a finite set of contact terms) of divergence free consistency equations to exist\footnote{
To make the case for the correctness of PM bootstrap in higher dimensions stronger, we note that
there are applications of PM bootstrap in higher dimensions, where it has been shown to work at higher orders in perturbation theory whenever we can make sure that the basis is free from any kind of divergences, e.g. if one considers $\lambda \phi^4$ theory in the AdS then adding this contact term to the basis we can work out the CFT data to one loop order of all double trace operators (including spin zero) in terms of one ambiguity which is equivalent to one renormalization condition in AdS \cite{KGloop}. Also, we can produce correct anomalous dimensions to first order in perturbation theory due to exchange of a singlet in the crossed channel and agrees with the answer found in \cite{Cardona:2018dov, Cardona:2018qrt,Liu:2018jhs}. As can be noticed, in none of the situations we have to do the spin sum, and we believe that this is the reason for the nice matches we have found so far. Recently there was also an application of Polyakov Mellin Bootstrap to show there are no perturbatively interacting CFTs with only fundamental scalars in  $d>6$ \cite{golden}. }. We hope to return with a systematic exploration of these issues in the future.

\section{Discussion}\label{sec:Discussion}
In this paper, we have developed the technology of the Polyakov-Mellin bootstrap in one dimension, where the contact term ambiguity can be completely fixed. As a proof that the machinery works, we have compared the results obtained using PM bootstrap with an independent method based on transcendentality -- we have found exact agreement in cases where the calculations can be performed in the latter approach. We were also able to reproduce effective field theory intuition by using both PM bootstrap and the transcendentality based approach.

The findings of this paper are quite encouraging from the perspective of fixing the crossing symmetric basis, including the contact terms, in higher dimensions. In particular, we found that the diagonal limit of the $2d$ Ising model can be expanded in terms of the $1d$ PM blocks. We proposed a potential strategy to extend this to fix the contact terms in the higher dimensional basis. 

In \cite{usprl, longpap, gs}, without adding contact terms, agreement was found to $O(\epsilon^2)$ order for $\phi^2$ and $O(\epsilon^3)$ for higher spins in the Wilson-Fisher epsilon expansion. In \cite{gs} it was realized that without adding contact terms, the higher order results would start disagreeing with the Feynman diagram results. With the strategy outlined above, it should be possible now to test the new basis with the contact terms to see if we can recover the higher order results. This itself does not appear straightforward since at higher orders, higher twist operators start contributing which will also lead to mixing with the twist-2 operators. Nonetheless, it appears possible that for the scalar $\phi^2$, the $O(\epsilon^3)$ term can be computed from this single correlator--for the epsilon expansion, it is very likely that only the constant contact term is sufficient to this order. Since for condensed matter applications, it is the lowest scalar that is of the most interest, taking this to fruition will be of utmost interest. 

As far as the transcendentality method is concerned, let us stress that all the techniques we have developed also apply to models with one-dimensional fermions, both with and without the $O(N)$ global symmetry. In that case, it is possible to find exact correlators in perturbation theory only for half-integer external dimensions (as opposed to integer, as in the bosonic case), and the arguments regarding transcendentality are essentially unaltered.  While this method of course has its limitations, both due to the requirement that the operators have integer (or half-integer) dimension, and because these ans\"atze become more and more complicated at higher orders in perturbation theory, this technique has proven to be quite powerful at loop level. A possible application of these ideas is to correlators of protected operators in defect CFT's arising as Wilson lines in supersymmetric gauge theories, such as the one considered in \cite{Giombi:2017cqn, Liendo:2018ukf, Gimenez-Grau:2019hez}. Interestingly enough, the tree-level CFT data for such model are particularly simple, and ignoring the mixing between double trace operators we were able to compute the correlators up to three loops. Therefore, as an extension of this work, we plan to attack the mixing problem and extend the results of \cite{Giombi:2017cqn, Liendo:2018ukf} to higher orders in perturbation theory. Similarly, one can consider $1d$ superconformal theories arising from Wilson lines in the ABJM theory (see \cite{Drukker:2019bev} for a recent review). 

As another possible application of these studies, one can consider the SYK model. This model can be described with a one-dimensional Hamiltonian for $N$ Majorana fermions and a $q$-fermions interaction (with $1/q$ being the scaling dimension of the fermions), and it can be studied in its nearly conformal \cite{Maldacena:2016hyu, Gross:2017hcz} or conformal \cite{Gross:2017vhb, Gross:2017aos} version, according to the choice of the kinetic term. Despite its simplicity, the SYK is an interesting toy model for holography: it is known that its bulk dual contains a tower of massive particles, which suggests a string-like formulation, but no concrete model has been proposed. One could then wonder how such a model fits into our study of $1d$ CFT's, and perhaps not surprisingly the answer is that it does not: a tower of massive particles in the bulk dual corresponds to an infinite tower of exchanges, which is a case we have not considered in detail, except to touch on it using the PM bootstrap formalism. It would be therefore very interesting to consider this model from the point of view of the conformal bootstrap, in order to understand a complicated setup with an infinite number of exchanged operators.

Finally, let us mention that in dimensions $d>1$ there are two possible limits that one can consider when taking the twist ($n$) to be large. One is the Regge limit, in which both the twist and the spin ($j$) are large, with fixed ratio $n/j$, and was considered in \cite{Cornalba:2006xm, Cornalba:2007zb, Li:2017lmh, Kulaxizi:2017ixa}. This led to very similar results to the ones that we found in one dimension. The other is the bulk-point limit, in which $n$ is taken to infinity at fixed $j$, which allows to extract local information about the bulk, such as flat space scattering amplitudes \cite{Gary:2009ae, Maldacena:2015iua, Alday:2017vkk} and the presence of extra dimensions \cite{Alday:2019qrf}. In our one-dimensional setting, it looks like only the former limit is possible, and the question remains open whether one can use our considerations about the Regge limit to compute two-dimensional flat-space amplitudes or to study the emergence of extra dimensions for AdS$_2$ String (or M) Theory compactifications. It would also be interesting to consider Froissart like bounds for the 1$d$ Mellin amplitudes \cite{hs}--in higher dimensions, the flat space limit led to nontrivial constraints on the number of subtractions needed to write a dispersion relation. The absence of spin in 1$d$ would make the analogous derivation very different.

\section*{Acknowledgments}
We especially thank Fernando Alday for initial collaboration and numerous useful discussions. 
We thank Apratim Kaviraj, Rajesh Gopakumar, Johan Henriksson for useful discussions and Dalimil Mazac and Xinan Zhou for correspondence.  A.S. thanks University of Oxford and CERN for hospitality during the completion of this work. A.S. acknowledges support from a DST Swarnajayanti Fellowship Award DST/SJF/PSA-01/2013-14 and from the Tata Trusts for a travel grant. We acknowledge partial support from F. Alday’s  European Research Council (ERC) grant under the European Union’s Horizon 2020 research and innovation programme (grant agreement No 787185) and a SPARC grant from MHRD, Govt of India.

\appendix

\section{Crossing kernel}
In this appendix, we explicitly compute the     analogue of the crossing kernel for PM bootstrap, in other words the decomposition into $s-$ channel partial waves of exchange Witten diagrams. First, we consider the $s-$ channel Witten diagram $W^{(s)}_{\Delta,\ell}(s,t)$ and expand the $t$ dependence in orthogonal Continuous Hahn polynomials $Q^{2s+\ell'}_{\ell',0}(t)$ \footnote{See \ref{ap:cHahn} for the definition of these polynomial and the orthogonality relations that they satisfy.},
\begin{equation}\nonumber
W^{(s)}_{\Delta,\ell}(s,t)=\sum_{\ell'} q^{(s)}_{\Delta,\ell'|\ell}(s) Q^{2s+\ell'}_{\ell',0}(t).
\end{equation}
Similarly, for the crossed channel we get
\begin{equation}\nonumber
W^{(t)}_{\Delta,\ell}(s,t)=\sum_{\ell'} q^{(t)}_{\Delta,\ell'|\ell}(s) Q^{2s+\ell'}_{\ell',0}(t).
\end{equation}

The explicit expressions for $q^{(s)}_{\Delta,\ell'|\ell}(s)$ and $q^{(t)}_{\Delta,\ell'|\ell}(s)$ \footnote{We call $q^{(t)}_{\Delta,\ell'|\ell}(s)$ the crossing kernel. To connect this to more familiar notion of crossing kernel for exchange Witten diagrams let us note here that knowledge of these $q^{(t)}_{\Delta,\ell'|\ell}(s)$ will enable us to find the $ \alpha_{n,J}$ and $\beta_{n,J}$ defined as $W^{(t)}_{\Delta,\ell}(u,v)=\sum_{J=0}^{\infty} \alpha_{n,J} g^{(s)}_{2\Delta_{\phi}+2n+J,J}(u,v)+\sum_{J=0}^{\infty} \beta_{n,J} \partial g^{(s)}_{2\Delta_{\phi}+2n+J,J}(u,v)$.} were derived in closed form in \cite{gs}, where it was shown how to express it as a sum of $_7F_6$ hypergeometric functions\footnote{See also \cite{Sleight:2018ryu,Sleight:2018epi,Liu:2018jhs}.}. However, as pointed out in \cite{gs}, these formulas are not complete and whenever the exchanged spin in the crossed channel is greater than spin exchanged in direct channel, there are a finite number of corrections which need to be added to the answer. Therefore, we first briefly revisit the derivation of the crossing kernel for this case and then diagnose the problem in the derivation. We will then propose a systematic prescription to cure it. Our formulas will be valid down to spin zero in the direct channel. In particular, this correction will be important in our discussion of $O(N)$ model and will have potential future applications as well.
\begin{figure}[hbt!]
\centering
    \includegraphics[width=0.7\linewidth]{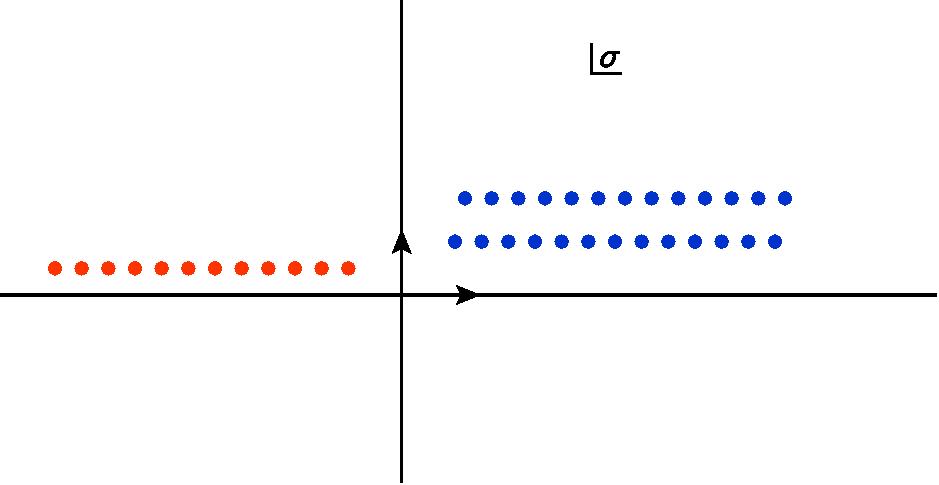}
    \caption{Mellin-Barnes type integration contour where the path of integration is parallel to imaginary vertical axis. The contour separates chains of poles which lie entirely on the right and the chains of poles that lie entirely on the left.  }
   \label{fn}
\end{figure}
The crossed channel, when decomposed in continuous Hahn basis, becomes \cite{gs}
\begin{eqnarray}\label{qtcoefftex}
q^{(t)}_{\D, \ell' |\ell}(s)  &=& \frac{2^{-\ell'}}{\ell'!}\frac{\G(2s+2\ell')}{\G^2(s+\ell')\G(a_\ell)} \frac{\Gamma \left(2\Delta_\phi +\ell -h\right)}{(a_\ell+\ell+2\Delta_{\phi}-h-1)} \\ \nonumber
&\times & \sum_{p=0}^{\ell'}\sum_{n=0}^{\ell}\sum_{m =0}^{\ell-n} \mu_{m,n}^{(\ell)}(\Dphi-s)_n \frac{\G^2(s+m+a_\ell-1)}{\G(2s+p+m+a_\ell-1)}\frac{(-\ell')_p(2s+\ell'-1)_p}{p!} \int_{0}^{1}  dy 
 y^{s-1}(1-y)^{a_\ell-1} \\ 
& \times & {}_2F_1[1, a_\ell; a_\ell+\ell+(2\Delta_{\phi}-h);y]{}_2F_1[s+p,s+m+a_\ell-1;2s+p+m+a_\ell-1,1-y]  \nonumber\, .
\end{eqnarray}
 Now we can use Mellin-Barnes representation for $_2F_1$ and $_3F_2$ and perform the $y$- integral to finally arrive at
  \begin{equation}
q^{(t)}_{\D, \ell' |\ell}(s)=  \frac{2^{-\ell'}}{\ell'!}\frac{\G(2s+2\ell')}{\G^2(s+\ell')\G(a_\ell)} \frac{\Gamma \left(2\Delta_\phi +\ell -h\right)}{(a_\ell+\ell+2\Delta_{\phi}-h-1)} \sum_{n=0}^{\ell}\sum_{m =0}^{\ell-n} \mu_{m,n}^{(\ell)}(\Dphi-s)_n \G^2(s+m+a_\ell-1) I^{m}_{\Delta,\ell'|\ell}(s)
 \end{equation}
 where $ I^{m}_{\Delta,\ell'|\ell}(s)$ is given by,
 \begin{equation}
 \begin{split}
 I^{m}_{\Delta,\ell'|\ell}(s)=&  \sum_{k=0}^{m} \binom{m}{k} \frac{\Gamma (a_\ell -h+\ell +2 \Delta_\phi )}{\Gamma^2(a_\ell +m+s-1) \Gamma (-h+\ell +2 \Delta_\phi )}\int [d\sigma]   \Gamma (-\sigma ) \Gamma (a_\ell +m+\sigma )\\
& \times \frac{\Gamma(a_{\ell}-k+m+\sigma)\Gamma(1-a_\ell+k-m-\sigma)}{\Gamma(a_\ell+m-\ell'+\sigma)} \Gamma^2 (a_{\ell} +m+s+\sigma -1) \\
& \times \frac{6 \Gamma^2(s+3) \Gamma \left(a_{\ell }\right) \Gamma \left(-h+\ell +2 \Delta _{\phi }-1\right) \Gamma \left(-h+\ell +a_{\ell }+2 \Delta _{\phi }\right)}{\Gamma (2 s+5) \Gamma^2 \left(s+a_{\ell }+3\right){} \Gamma \left(-h+\ell +2 \Delta _{\phi }\right) \Gamma \left(-h+\ell +a_{\ell }+2 \Delta _{\phi }-1\right)}
 \end{split}
 \end{equation}
 The contour of integration is shown in figure 1. 
 
  Now to write it in a form so that we can identify this as an integration which will give us a $_7F_6$  hypergeometric function, we use the reflection formula first to write
 \begin{equation} \label{reflectionform}
  \frac{\Gamma(a_{\ell}-k+m+\sigma)\Gamma(1-a_\ell+k-m-\sigma)}{\Gamma(a_\ell+m-\ell'+\sigma)} = (-1)^{k+\ell'} \Gamma(1-a_{\ell}-m+\ell'-\sigma)
  \end{equation}
 and then perform the $k$ sum to get the final form,
  \begin{equation} 
  \begin{split}
I^{m}_{\Delta,\ell'|\ell}(s)=& (-1)^{\ell'+m} \frac{(a_{\ell})_m \Gamma(2\Delta_{\phi}-h+\ell+a_{\ell})}{\Gamma^2(a_{\ell}+s+m-1) \Gamma(2\Delta_{\phi}-h+\ell)} \int_{-i\infty}^{i\infty} [d\sigma] \Gamma(-\sigma)\\
 &\times  \frac{\Gamma(1-a_{\ell}+\ell'-m-\sigma)\Gamma(a_{\ell}+m+\sigma)\Gamma(a_{\ell}+m+s-1+\sigma)^2 \Gamma(2\Delta_{\phi}-h+\ell-1+a_{\ell}+\sigma)} {\Gamma(a_{\ell}+\ell'+m+2s-1+\sigma)\Gamma(2a_{\ell}+\ell+m+2\Delta_{\phi}-h-1+\sigma)}\,.
 \end{split}
  \end{equation}

       \begin{figure}[hbt!]
\centering
    \includegraphics[width=0.7\linewidth]{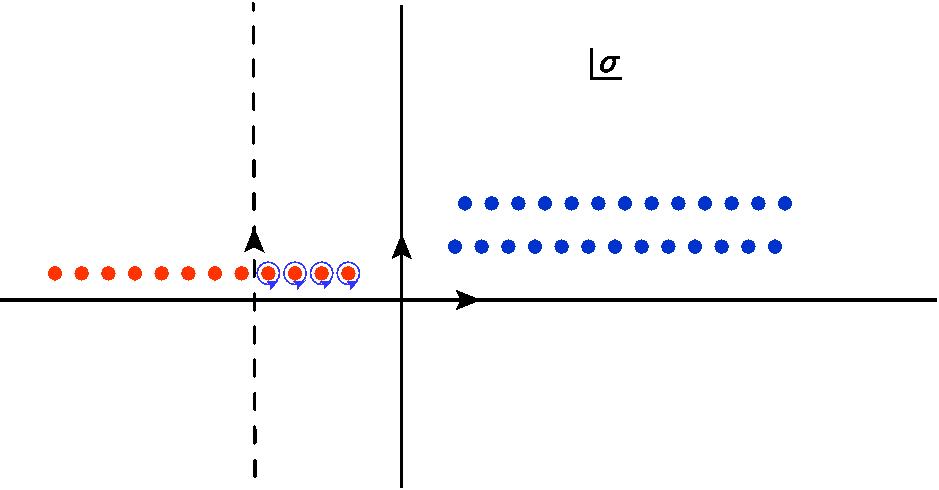}
    \caption{The contour gets pushed to left for $\ell>\ell'$ such that it can pick up the poles coming from left. Therefore we subtract this off from our final answer. }
   \label{fn}
\end{figure}

Using the integration given in section \ref{Wdef} we can write it as a $_7F_6$ Hypergeometric function. This formula does not give the correct decomposition for $\ell'<\ell$, because in such cases the contour gets shifted to the left for certain values of $m$ and if we look back at eq. (\ref{reflectionform}), we can see that for given $m$,  certain combinations of $k$ do not have these poles. Instead, the poles we considered came from the series of poles left to our actual contour, so for these cases we have to subtract off their contribution\footnote{In practice we have to add them back as we are closing the contour on right so the original answer was negative in sign.}. One can easily figure out the general pattern for $\ell> \ell'$. In general, we have to add the following quantity\footnote{In this formula we have put $m=\ell'+p$ and $p\geq 1$. If $p=0$ then there is no correction as discussed before. For a given $\ell$, $m\leq \ell$ and $p$'s highest value is $(\ell-\ell')$.}
\begin{equation}
\begin{split}
\mathcal{I}^{m}_{\Delta,\ell'|\ell}(s)= \sum_{k=1+\ell'}^{\ell'+p} \sum_{q=0}^{k-1-\ell'} & \frac{(-1)^{k+q+\ell '+1}  \binom{p+\ell '}{k} \Gamma \left(p-q+a_{\ell }-1\right)\Gamma \left(-h+\ell +a_{\ell }+2 \Delta _{\phi }\right)}{q! \Gamma \left(-h+\ell +2 \Delta _{\phi }\right) \Gamma \left(q+2 s+2 \ell '\right) \Gamma \left(p+s+a_{\ell }+\ell '-1\right){}^2 }\\
& \times \frac{\Gamma \left(q+\ell '+1\right) \Gamma \left(q+s+\ell '\right)^2 \Gamma \left(-h-k+q+\ell +2 \Delta _{\phi }+\ell '\right)}{\Gamma \left(-h-k+q+\ell +a_{\ell }+2 \Delta _{\phi }+\ell '\right)}\,.
\end{split}
\end{equation}

Let us give some explicit example for certain cases,
\begin{equation} 
\begin{split}
& \mathcal{I}^{1}_{\Delta,0|\ell}(s)=\frac{\Gamma (s)^2 \Gamma \left(a_{\ell }\right) \Gamma \left(-h+\ell +2 \Delta _{\phi }-1\right) \Gamma \left(-h+\ell +a_{\ell }+2 \Delta _{\phi }\right)}{\Gamma (2 s) \Gamma \left(s+a_{\ell }\right){}^2 \Gamma \left(-h+\ell +2 \Delta _{\phi }\right) \Gamma \left(-h+\ell +a_{\ell }+2 \Delta _{\phi }-1\right)}\\
& \mathcal{I}^{2}_{\Delta,0|\ell}(s)=\frac{s \Gamma^2 (s) \Gamma \left(a_{\ell }\right) \left(a_{\ell }+2 \Delta _{\phi }-h+\ell -1\right) \left(2 a_{\ell } \left(2 \Delta _{\phi }-h+\ell -2\right)-2 a_{\ell }^2+s \left(2 \Delta _{\phi }-h+\ell -2\right)\right)}{\Gamma (2 s+1) \Gamma^2 \left(s+a_{\ell }+1\right) \left(-2 \Delta _{\phi }+h-\ell +1\right) \left(-2 \Delta _{\phi }+h-\ell +2\right)}\,.
\end{split}
\end{equation}
So for $\ell'<\ell$, the correct  cross channel decomposition is given by, 
 \begin{equation}\nonumber
  \begin{split}
q^{(t)}_{\D, \ell' |\ell}(s)= &  \frac{2^{-\ell'}}{\ell'!}\frac{\G(2s+2\ell')}{\G^2(s+\ell')\G(a_\ell)} \frac{\Gamma \left(2\Delta_\phi +\ell -h\right)}{(a_\ell+\ell+2\Delta_{\phi}-h-1)} \sum_{n=0}^{\ell}\sum_{m =0}^{\ell-n} \mu_{m,n}^{(\ell)}(\Dphi-s)_n \G^2(s+m+a_\ell-1)\\
& ( I^{m}_{\Delta,\ell'|\ell}(s)+\mathcal{I}^{m}_{\Delta,\ell'|\ell}(s)\Theta(m-\ell'-1)),
\end{split}
 \end{equation}
where $\Theta(x)$is Heaviside step function.  $\mathcal{I}^{1}_{\Delta,0|1}(s)$ will be required in our discussion of the $O(N)$ model.

\section{Explicit Expressions for $q^{(s)}_{\D,\ell'|\ell}(s),q^{(t)}_{\D,\ell'|\ell}(s)$ and $q^{(u)}_{\D,\ell'|\ell}(s)$  }\label{allqs}
In this appendix we write down explicitly the expressions of the exchange scalar Witten diagrams in all three channels which we used in the discussion of Polyakov-Mellin bootstrap. We also defined other quantities such as Mack polynomial, Continiuous Hahn Polynomial and its orthogonality property which was important in our discussion.
\subsection{Mack Polynomials} \label{mackpol}
In our convention Mack Polynomial is given below,
\be
 \widehat P^{(s)}_{\D-h,\ell}(s,t) =\sum_{m,n} \mu_{m,n}^{(\ell)}(\frac{\D-\ell}{2}-s)_m (-t)_n=(-1)^\ell\sum_{m,n} \mu_{m,n}^{(\ell)}(\frac{\D-\ell}{2}-s)_m (s+t)_n\,,
 \ee
with
 \begin{eqnarray}\label{mudef}
 \mu_{m,n}^{(\ell)}&=&2^{-\ell} \frac{(-1)^{m+n}\ell!}{m! n! (\ell-m-n)!}(\l_1-m)_m (\l_2+n)_{\ell-n} (\l_2+m+n)_{\ell-m-n}(\ell+h-1)_{-m}(\ell+\D-1)_{n-\ell}  \nonumber \\
 &\times& {}_4F_3[-m,1-h+\l_2,1-h+\l_2,n-1+\D;2-2h+2\l_2,\l_1-m,\l_2+n;1]\,.\nonumber\\
 \end{eqnarray}
 Here $\lambda_1=(\D+\ell)/2, \lambda_2=(\D-\ell)/2$ and $h=d/2$ where $d$ is the number of spacetime dimensions. Further, the last ${}_4F_3$ is a well-balanced one. 
\subsection{Continuous Hahn Polynomials}\label{ap:cHahn}
Continuous Hahn polynomial $Q_{\ell, 0}^{2 s+\ell}(t)$ given by
\begin{equation} 
Q_{\ell, 0}^{2 s+\ell}(t)=\frac{2^{\ell}\left((s)_{\ell}\right)^{2}}{(2 s+\ell-1)_{\ell}}{}_3 F_{2} \left[ \begin{array}{c}{-\ell, 2 s+\ell-1, s+t} \\ {s, s}\end{array}\right],
\end{equation}
they satisfy the orthogonality relations 
\begin{equation} 
\frac{1}{2 \pi i} \int_{-i \infty}^{i \infty} \Gamma^{2}(s+t) \Gamma^{2}(-t) Q_{\ell, 0}^{2 s+\ell}(t) Q_{\ell^{\prime}, 0}^{2 s+\ell^{\prime}}(t)=\kappa_{\ell}(s) \delta_{\ell, \ell^{\prime}},
 \end{equation}
with
\begin{equation} \label{khahn}
\kappa_{\ell}(s)=\frac{4^{\ell}(-1)^{\ell} \ell ! \Gamma^{4}(\ell+s) \Gamma(2 s+\ell-1)}{\Gamma(2 s+2 \ell) \Gamma(2 s+2 \ell-1)}\,.
 \end{equation}
\subsection{${}_7F_6$ Integral}
We introduce the notation $W(a,b,c,d,e,f)$ and various parameters here which we referred in the main text and later part of appendix as well,
\begin{eqnarray}\label{Wdef}
&&W(a;b,c,d,e,f)\equiv \nonumber\\
&&{}_7F_6\bigg{(}\begin{matrix} a, & 1+\frac{1}{2}a, & b, & c, & d, & e, & f\\ ~&\frac{1}{2}a,& 1+a-b, & 1+a-c, & 1+a-d, & 1+a-e, & 1+a-f\end{matrix};1\bigg{)}\nonumber\\
&=&\frac{\G(1+a-b)\G(1+a-c)\G(1+a-d)\G(1+a-e)\G(1+a-f)}{\G(1+a)\G(b)\G(c)\G(d)\G(1+a-c-d)\G(1+a-b-d)\G(1+a-b-c)\G(1+a-e-f)}\nonumber\\
&\times&\frac{1}{2\pi i}\int_{-i\infty}^{i\infty}d\s\, \frac{\G(-\s)\G(1+a-b-c-d-\s)\G(b+\s)\G(c+\s)\G(d+\s)\G(1+a-e-f+\s)}{\G(1+a-e+\s)\G(1+a-f+\s)}\,.\nonumber\\
\end{eqnarray}
We use the $W$ notation of Bailey's \cite{bailey} where
\be\label{params}
a=\ell'+2(a_\ell+m+s-1)\,,\quad b=e=a_\ell+m,\quad c=d=a_\ell+m+s-1\,,\quad f=2(s-\Dphi)+h+m+\ell'-\ell\, ,
\ee
with $a_\ell =1+ \frac{\D-\ell}{2}-\Dphi$.

\subsection{s-channel coefficient}
In this section we write down the explicit form of a $s-$ channel exchange Witten diagram after expanding it in continuous Hahn basis,
\begin{equation}
W^{s}_{\Delta,\ell}(s,t)=\sum_{\ell'} q^{(s)}_{\D, \ell' |\ell}(s) Q_{\ell', 0}^{2 s+\ell'}(t)
\end{equation}
where \footnote{see \cite[eq 3.6]{gs}},
\begin{eqnarray}\label{qsgen}
q^{(s)}_{\D, \ell' |\ell}(s) &=& \sum_{m,n} \mu_{m,n}^{(\ell)}(\frac{\D-\ell}{2}-s)_m\chi^{(n)}_{\ell'}(s)
 \frac{\Gamma^2 \left(\frac{\Delta+\ell }{2}+\Delta_\phi -h\right)}{(\frac{\Delta-\ell}{2}-s) 
 \Gamma (\Delta-h +1)} \,\\
&\times & _3F_2\left[\frac{\Delta -\ell}{2}-s,1+\frac{\Delta-\ell }{2}-\Delta_\phi ,1+\frac{\Delta-\ell }{2}-\Delta_\phi ;1+\frac{\Delta-\ell }{2}-s,\Delta-h +1;1\right]\, .\nonumber 
\end{eqnarray}
where
\be\label{tpochexp}
(-t)_n=\sum_{\ell'=0}^n \c_{\ell'}^{(n)}(s) Q_{\ell',0}^{2s+\ell'}(t)\,,
\ee
with
\be\label{schi1}
\c_{\ell'}^{(n)}\!(s)=(-1)^{\ell'}2^{-\ell'}\frac{\G(2s+2\ell')\G^2(s+n)}{\ell'! \G^2(\ell'+s)\G(2s+n)}\,{}\frac{(-n)_{\ell'}}{(2s+n)_{\ell'}}\,.
\ee
Here we also give the expression of normalization which we have to multiply with the Witten block ,
\begin{equation}
N_{\Delta,\ell}=\frac{(-2)^{\ell } (\Delta -h) (\Delta +\ell -1) \Gamma (\Delta -h) \Gamma^2 (\ell +\Delta -1)}{\Gamma (\Delta -1) \Gamma^4 \left(\ell +\frac{\Delta -\ell }{2}\right) \Gamma^2 \left(\frac{1}{2} \left(\ell -\Delta +2 \Delta _{\phi }\right)\right) \Gamma^2 \left(\frac{1}{2} \left(-2 h+\ell +\Delta +2 \Delta _{\phi }\right)\right)}
\end{equation}
\subsection{t-channel coefficient}
The $t-$ channel exchange Witten diagram is given as,
\begin{equation}
W^{t}_{\Delta,\ell}(s,t)=\sum_{\ell'} q^{(t)}_{\D, \ell' |\ell}(s) Q_{\ell', 0}^{2 s+\ell'}(t)
\end{equation}
where \footnote{See \cite[eq D.22 of Appendix D]{gs} },
\be\label{eq:qt_noex}
\begin{split}
q^{(t)}_{\D,\ell'|\ell}(s)=& \sum_{m,n}^{\ell} (-1)^{\ell'+m}2^{-\ell'}\mu_{m,n}^{(\ell)} (\Dphi-s)_n (a_\ell)_m^2 \G(2s+2\ell')\G^2(d)\G(\frac{a}{2})\G(a+1)\G^2(1+a-f-b)\\
& \tilde W(a;b,c,d,e,f)\,,
\end{split}
\ee
where the parameters are given in eq. (\ref{params}) and the $\tilde W$ is the regularized version of $W$.


\subsection{u-channel coefficient}
Similarly the coefficient of  $u-$ channel exchange Witten diagram in continuous Hahn basis is given by,
\be
q^{(u)}_{\D,\ell'|\ell}(s)=(-1)^{\ell'+\ell}q^{(t)}_{\D,\ell'|\ell}(s).
\ee
\\
\textit{\textbf{Note  }}\\
One has to set $h=\frac{1}{2}$, $\ell=0$ in order to use the expressions in $1d$.

\section{Explicit formula for $d_{r,\ell}(s)$ } \label{ap:dderi}
In this appendix we essentially give details which we use to perform the $t-$ integral in the main text (\ref{gendPM}) and find the blocks in one dimension.
We write 
\be\label{e1}
(1-z)^{2t}=\sum_{r=0}^{\infty}(-1)^r ~{}^{2t}C_{r}z^r=\sum_{k_1,k_2=0}^{\infty}\frac{(-t)_{k_1}}{k_1!}\frac{(-t)_{k_2}}{k_2!}~z^{k_1+k_2}\,.
\ee
So we see that  
$(-1)^r~{}^{2t}C_r=\sum_{k_1=0}^{r}\frac{(-t)_{k_1}}{k_1!}\frac{(-t)_{r-k_1}}{(r-k_1)!}\,.
$
Now using
\be\label{e3}
(-t)_{m}(-t)_{n}=\sum_{\ell=0}^{m+n} \chi_{\ell}^{( m, n)}(s) Q_{\ell, 0}^{2 s+\ell}(t)\,,
\ee
with
\be
\begin{split}
\chi_{\ell}^{( m, n)}(s) \quad=&(-1)^{\ell} 2^{-\ell} \frac{\Gamma(2 s+2 \ell) \Gamma^2(m+s) \Gamma^2(n+s) }{\ell ! \Gamma^{2}(s) \Gamma^{2} (\ell+s)\Gamma(m+n+2 s)}\\
&\times_{4} F_{3}[-\ell, m+s, n+s, 2 s+\ell-1 ; s, s, 2 s+m+n ; 1]\,.
\end{split}
\ee
So plugging \ref{e3} in \ref{e1} and comparing with \ref{delldef} finally we have 
\be
\begin{split}\label{dell}
d_{r,\ell}(s)=&(-1)^{-r}\sum_{k_1=0}^{r}\frac{ \chi_{\ell}^{( k_1, r-k_1)}(s)}{k_1! (r-k_1)!}\,.
\end{split}
\ee

\section{Explicit calculations for $a_\ell(s)$}\label{ap:aell}
Here we want to write the contact terms and expand the $t$ dependence in continuous Hahn polynomial basis which we use in the main text,
\be
c(s,t)=\sum_{\ell=0}^{\infty} a_{\ell}(s) Q_{\ell, 0}^{2 s+\ell}(t)\,.
\ee
Using 
\be
(-t)_{n}(s+t)_{p}=\sum_{\ell=0}^{p+n} \Omega_{\ell}^{(n,p)}(s) Q_{\ell, 0}^{2 s+\ell}(t)\,,
\ee
with
\be\label{eq:Omega}
\begin{aligned} \Omega_{\ell}^{(n,p)}(s) &=(-1)^{\ell} 2^{-\ell} \frac{\Gamma(2 s+2 \ell) \Gamma(s) \Gamma(n+s) \Gamma(p+s) \Gamma(n+p+s)}{\ell ! \Gamma^{2}(s) \Gamma^{2}(\ell+s) \Gamma(n+p+2 s)} \\ & \times_{4} F_{3}[-\ell, s, n+s, 2 s+\ell-1 ; s, s, 2 s+n+p ; 1]\,. \end{aligned}
\ee
We begin by writing the contact term as,

\be
\begin{split}
c(s,t)&=\sum_{m+n=0}^{L}c_{mn}(-t)_m (s+t)_n+\sum_{m+n=0}^{L}c_{mn}(\Dphi-s)_m (s+t)_n\\
&+\sum_{m+n=0}^{L}c_{mn}(-t)_m (\Delta_{\phi}-s)_n\\
&=\sum_{\ell=0}^{\infty}\left[\sum_{m+n=0}^{L}\left(c_{m,n} \Omega_{\ell}^{(m,n)}(s)+c_{mn}(\Dphi-s)_m  \Omega_{\ell}^{(0,n)}(s)+c_{mn} (\Dphi-s)_n  \Omega_{\ell}^{(m,0)}(s)\right)\right] Q_{\ell, 0}^{2 s+\ell}(t)\,.
\end{split}
\ee

We immediately identify that 

\be
a_{\ell}(s)=\left[\sum_{m+n=0}^{L}\left(c_{m,n} \Omega_{\ell}^{(m,n)}(s)+c_{mn}(\Dphi-s)_m  \Omega_{\ell}^{(0,n)}(s)+c_{mn} (\Dphi-s)_n  \Omega_{\ell}^{(m,0)}(s)\right)\right]\,.
\ee

\subsection{Contact term in a theory with O(N) global symmetry} \label{ONcontact}

 We can write down any crossing symmetric contact term in a theory with O(N) global symmetry in the following way,

\be
\begin{split}
c(s,t)&=\delta_{ij}\delta_{k l}\sum_{m+n=0}^{L}c_{mn}(-t)_m (s+t)_n+\delta_{il}\delta_{j k}\sum_{m+n=0}^{L}c_{mn}(-s+\Delta_{\phi})_m (s+t)_n\\
&+\delta_{ik}\delta_{j l}\sum_{m+n=0}^{L}c_{mn}(-t)_m (\Delta_{\phi}-s)_n\,,
\end{split}
\ee
where $c_{mn}=c_{nm}$ are constant. One can write $c(s,t)$ as a expansion in Continuous Hahn polynomials as 
\be
c(s,t)=\sum_{\ell=0}^{\infty} a_{\ell}(s) Q_{\ell, 0}^{2 s+\ell}(t)\,.
\ee
where $a_\ell$ is given by (calculations are the same as in Appendix (\ref{ap:aell}))

\be
a_{\ell}(s)=\left[\sum_{m+n=0}^{L}\left(\delta_{ij}\delta_{k l}c_{m,n} \Omega_{\ell}^{(m,n)}(s)+\delta_{il}\delta_{j k}c_{mn}(\Delta_{\phi}-s)_m  \Omega_{\ell}^{(0,n)}(s)+\delta_{ik}\delta_{j l}c_{mn} (\Dphi-s)_n  \Omega_{\ell}^{(m,0)}(s)\right)\right]\,,
\ee
where $\Omega_{\ell}^{(m,n)}(s)$ given in eq. (\ref{eq:Omega}). 
We can break it into three irreducible sector,
\be
\begin{split}
a_{\ell}(s)&=\delta_{i j}\delta_{k l} a_{\ell}^{(S)}(s)+\left(\frac{(\delta_{ik}\delta_{j l}+\delta_{i l} \delta_{j k})}{2}-\frac{1}{N}\delta_{ij}\delta_{kl}\right)  a_{\ell}^{(T)}(s) +\frac{(\delta_{ik}\delta_{j l}-\delta_{i l}\delta_{j k})}{2}   a_{\ell}^{(A)}(s)\,,
\end{split}
\ee
where 
\be
\begin{split}
a_{\ell}^{(S)}(s)&=\sum_{m+n=0}^{L}c_{mn}\bigg(\Omega_{\ell}^{(m,n)}(s)+\frac{1}{N}((\Delta_{\phi}-s)_m  \Omega_{\ell}^{(0,n)}(s)+(\Dphi-s)_n  \Omega_{\ell}^{(m,0)}(s))\bigg)\,,\\
a_{\ell}^{(T)}(s)&=\sum_{m+n=0}^{L}c_{mn}\bigg((\Delta_{\phi}-s)_m  \Omega_{\ell}^{(0,n)}(s)+(\Dphi-s)_n  \Omega_{\ell}^{(m,0)}(s)\bigg)\,,\\
 a_{\ell}^{(A)}(s)&=\sum_{m+n=0}^{L}c_{mn}\bigg((\Dphi-s)_n  \Omega_{\ell}^{(m,0)}(s)-(\Delta_{\phi}-s)_m  \Omega_{\ell}^{(0,n)}(s)\bigg)\,.
\end{split}
\ee

\section{Identity contributions and determination of $q_{dis}(s)$}\label{identity}
In this appendix we elaborate on how to find the Polyakov block corresponding to contribution of identity operator. The amplitude corresponding to Identity contribution is given by,
\be
\mathcal{PB}_{\D=0}(z)=1+z^{2\Dphi}+\left(\frac{z}{1-z}\right)^{2\Dphi}\,.
\ee
For the last term we write 
\be
\begin{split}
\left(\frac{z}{1-z}\right)^{2\Dphi}&=\sum_m (-1)^m z^{2\Dphi+m}~~{}^{-2\Dphi}C_{m}\\
& =\sum_m z^{2\Dphi+m}\frac{(-1)^m \Gamma \left(1-2 \Delta _{\phi }\right)}{\Gamma (m+1) \Gamma \left(-m-2 \Delta _{\phi }+1\right)}\,.
\end{split}
\ee
There for we write the $PB_{\D=0}$ in Mellin space
\be
\begin{split}
\mathcal{PB}_{\D=0}(z)=-\int_{-i \infty}^{i\infty}[ds]z^{2s}\G^2(\Dphi-s)& \left(\frac{1}{s\G^2(\Dphi-s)}+\frac{1}{(s-\Dphi)\G^2(\Dphi-s)}\right.\\
&\left.-\frac{2\Gamma \left(1-2 \Delta _{\phi }\right) \Gamma \left(2 \Delta _{\phi }-2 s\right)}{\Gamma (1-2 s)\G^2(\Dphi-s)}\right).
\end{split}
\ee
Therefore we define the Mellin amplitude corresponding to the amplitude of  identity operator  to be $q_{dis}(s)$ and find,
\be\label{qdiss}
q_{dis}(s)=-\left(\frac{1}{s\G^2(\Dphi-s)}+\frac{1}{(s-\Dphi)\G^2(\Dphi-s)}-\frac{2\Gamma \left(1-2 \Delta _{\phi }\right) \Gamma \left(2 \Delta _{\phi }-2 s\right)}{\Gamma (1-2 s)\G^2(\Dphi-s)}\right)\,.
\ee

\subsection{Calculations for $q_{dis}^{(S)}(s)$, $q_{dis}^{(T)}(s)$ and $q_{dis}^{(A)}(s)$}\label{ap:ONqdis}
Now we define the mellin amplitude corresponding to the amplitude of  identity operator for O(N) theories.
\be\label{eq:ONqdis}
q_{dis}(s)=-\left(\delta_{ij}\delta_{k l}\frac{1}{s\G^2(\Dphi-s)}+\delta_{ik}\delta_{j l}\frac{1}{(s-\Dphi)\G^2(\Dphi-s)}-\delta_{il}\delta_{j k}\frac{2\Gamma \left(1-2 \Delta _{\phi }\right) \Gamma \left(2 \Delta _{\phi }-2 s\right)}{\Gamma (1-2 s)\G^2(\Dphi-s)}\right)\,.
\ee
We can write 
\be
\begin{split}
q_{dis}(s)&=\delta_{i j}\delta_{k l}q_{dis}^{(S)}(s) +(\frac{\delta_{ik}\delta_{j l}+\delta_{i l} \delta_{j k}}{2}-\frac{1}{N}\delta_{ij}\delta_{kl})q_{dis}^{(T)}(s)
 +(\frac{\delta_{ik}\delta_{j l}-\delta_{i l}\delta_{j k}}{2}) q_{dis}^{(A)}(s)\,,
\end{split}
\ee
where
\be
\begin{split}
q_{dis}^{(S)}(s)&=-\left(\frac{1}{s\G^2(\Dphi-s)}+\frac{1}{N}\left(\frac{1}{(s-\Dphi)\G^2(\Dphi-s)}-\frac{2\Gamma \left(1-2 \Delta _{\phi }\right) \Gamma \left(2 \Delta _{\phi }-2 s\right)}{\Gamma (1-2 s)\G^2(\Dphi-s)}\right)\right)\,,\\
q_{dis}^{(T)}(s)&=\left(- \frac{1}{(s-\Dphi)\G^2(\Dphi-s)}+\frac{2\Gamma \left(1-2 \Delta _{\phi }\right) \Gamma \left(2 \Delta _{\phi }-2 s\right)}{\Gamma (1-2 s)\G^2(\Dphi-s)}\right)\,,\\
q_{dis}^{(A)}(s)&=-\left( \frac{1}{(s-\Dphi)\G^2(\Dphi-s)}+\frac{2\Gamma \left(1-2 \Delta _{\phi }\right) \Gamma \left(2 \Delta _{\phi }-2 s\right)}{\Gamma (1-2 s)\G^2(\Dphi-s)}\right)\,.
\end{split}
\ee

\section{Anomalous dimensions for derivative contact terms}\label{sec:DerTree}

In this appendix we collect the explicit expressions for the polynomials $P_q(n,\,\df)$ that appear in the tree level anomalous dimensions for derivative contact terms, as in eq. \eqref{gammastree}. Recall that for $q=0$ we have $P_0(n,\,\df)=1$. For higher values of $q$:
\begin{align}
\begin{split} 
P_1(n,\,\df)=&(\df-1) \df^2 (4 \df+1) (4 \df+3)+\left(64 \df^4-28 \df^2-2 \df+2\right) n\\
&+2 (2 \df-1) (2 \df+1) (12 \df+1) n^2+8 (2 \df+1) (4 \df-1) n^3+8(2 \df+1) n^4,
\end{split}
\end{align}
\begin{align}
\begin{split} 
P_2(n,\,\df)=&
(\df-2) (\df-1) \df^2 (\df+1)^2 (4 \df+3) (4 \df+5) (4 \df+7) (4 \df+9)\\
&+4 (2 \df+3) (4 \df-1) \big(64 \df^7+208 \df^6-36 \df^5-605 \df^4-554 \df^3\\
&-30 \df^2+243 \df+90\big)n+4 (2 \df-1) (2 \df+3) \big(448 \df^6+1392 \df^5\\&+84 \df^4-2183 \df^3-2091 \df^2-1134 \df-105\big)n^2
+8 (2 \df+3) (4 \df-1)\\&
\times \left(224 \df^5+576 \df^4-158 \df^3-572 \df^2-243 \df-160\right)n^3\\
&+4 (2 \df+3) \left(2240 \df^5+4800 \df^4-2924 \df^3-2156 \df^2+246 \df-415\right)n^4\\
&+32 (2 \df+3) (2 \df+5) (4 \df-1) \left(28 \df^2-5 \df-5\right)n^5\\
&+ 32 (2 \df+3) (2 \df+5) \left(56 \df^2-22 \df-1\right)n^6\\
&+128 (2 \df+3) (2 \df+5) (4 \df-1)n^7+64 (2 \df+3) (2 \df+5)n^8,
\end{split}
\end{align}
\begin{align}\nonumber
\begin{split}
P_3(n,\,\df)=&
(\df-3) (\df-2) (\df-1) \df^2 (\df+1)^2 (\df+2)^2 (4 \df+5) (4 \df+7) (4 \df+9)\\
& (4 \df+11) (4 \df+13) (4 \df+15)\\
&+2 (2 \df+5) (4 \df-1) \big(3072 \df^{12}+36096 \df^{11}+132224 \df^{10}-3360 \df^9\\&-1214676 \df^8-2926395 \df^7-970776 \df^6+6196080 \df^5+10143424 \df^4\\&+5128059 \df^3-1542528 \df^2-3028860 \df-907200\big) n\\
&+ 6 (2 \df-1) (2 \df+5) \big(11264 \df^{11}+125184 \df^{10}+437120 \df^9+118880 \df^8\\&-2771604 \df^7-6808095 \df^6-4248981 \df^5+6860955 \df^4+13140919 \df^3\\&+9496058 \df^2+4002384 \df+360360\big) n^2\\ 
&+ 8 (2 \df+5) (4 \df-1) \big(14080 \df^{10}+142080 \df^9+423840 \df^8+8160 \df^7\\
&-2172753 \df^6-4187481 \df^5-1812050 \df^4+3606930 \df^3+3965596 \df^2\\
&+1661325 \df+791091\big) n^3\\
&+ 4 (2 \df+5) \big(253440 \df^{10}+2327040 \df^9+5816640 \df^8-1506240 \df^7\\&-22985970 \df^6-33151830 \df^5-9079800 \df^4\\
&+25792815 \df^3+10370477 \df^2-446534 \df+2131794\big) n^4\\
&+ 24 (2 \df+5) (2 \df+7) (4 \df-1) \big(8448 \df^7+45504 \df^6+22128 \df^5-79708 \df^4\\
&-143680 \df^3-114082 \df^2+52985 \df+27645\big) n^5\\
&+ 8 (2 \df+5) (2 \df+7) \big(118272 \df^7+556416 \df^6-8736 \df^5-656280 \df^4\\&-661308 \df^3-560400 \df^2+371392 \df+17415\big) n^6\\
&+ 96 (2 \df+5) (2 \df+7) (4 \df-1) \big(2112 \df^5+9792 \df^4+268 \df^3-5448 \df^2\\&-4628 \df-5493\big) n^7\\
&+ 96 (2 \df+5) (2 \df+7) \big(5280 \df^5+22080 \df^4-8610 \df^3-4790 \df^2\\&-798 \df-2931\big) n^8\\
&+ 640 (2 \df+5) (2 \df+7) (2 \df+9) (4 \df-1) \left(44 \df^2-7 \df-3\right) n^9\\
\end{split}
\end{align}
\begin{align}
\begin{split}
&+ 128 (2 \df+5) (2 \df+7) (2 \df+9) \left(264 \df^2-102 \df+5\right) n^{10}\\
&+ 1536 (2 \df+5) (2 \df+7) (2 \df+9) (4 \df-1) n^{11}\\
&+ 512 (2 \df+5) (2 \df+7) (2 \df+9) n^{12}.
\end{split}
\end{align}

\section{Harmonic sums}\label{sec:HarmSums}

For many of the computations involving harmonic sums that were presented in this paper, we used the Mathematica package ``HarmonicSums'' \cite{HarmonicSums}. 

The harmonic sums employed in the body of the present paper are defined as 
\be 
S_{k_1,\,k_2,\,k_3,\,...}(n)=
\sum_{m_1=1}^n 
\frac{(\text{sgn}(k_1))^{m_1}}{m_1^{|k_1|}}\,
\sum_{m_2=1}^{m_1} 
\frac{(\text{sgn}(k_2))^{m_2}}{m_2^{|k_2|}}\,
\sum_{m_3=1}^{m_2} 
\frac{(\text{sgn}(k_3))^{m_3}}{m_3^{|k_3|}}\,\,
.\,.\,. \quad.
\ee
For some values of $k_1,\,k_2,\,k_3,\,...$, the harmonic sums admit an expression in terms of generalised harmonic numbers
\be 
H_{n}^{(m)}=\sum_{k=1}^n \frac{1}{k^m},
\ee
such as 
\be 
S_{k,\,n}=H_{n}^{(m)},
\ee
or
\be 
S_{-k,\,n}=\frac{(-1)^n}{2^k} \left(H_{\frac{n}{2}}^{(k)}-H_{\frac{n-1}{2}}^{(k)}\right)- \left(1-2^{1-k}\right) \zeta (k),
\ee
which involve harmonic sums encountered in the present paper.

When studying the analytic expressions of one-loop CFT data, a part of the result was written in terms of a generalized derivative relation \eqref{derrel1loop}, involving the derivative of the one-loop anomalous dimensions $\g^{(2)}_n$. This involves taking the derivative of harmonic sums, and while most of the ones that we used can be written in terms of generalized harmonic numbers, we also exploited the following result:
\be 
\frac{\pr}{\pr n} S_{-2,1}(2n) =
-4 S_{-3,1}(2 n)-2 S_{-2,2}(2 n)+\12\z(2)\left( H_n^{(2)}-H_{n-\frac{1}{2}}^{(2)} \right)-\frac{37 \pi ^4}{720}.
\ee

In order to study the Regge limit of CFT data, one needs to expand the latter for large $n$. To do so, we have used the following expansions\footnote{One has two different analytic continuations of the harmonic sums according to whether $n$ is even or odd, due to the factors of $(-1)^n$. However, we do not have to worry about it because we are always interested in the case where the argument of the harmonic sums is $2\df+2n-1$, with integer $\df$, and so everything follows without arbitrary choices.}
\begin{align}
\begin{split}
S_{-2,1}(n)=&-\frac{5 }{8}\zeta (3) -(-1)^n\Big[(\log (n)+\gamma)\Big(\frac{38227}{2 n^{15}}-\frac{2073}{2 n^{13}}+\frac{155}{2 n^{11}}-\frac{17}{2 n^9}+\frac{3}{2 n^7}-\frac{1}{2 n^5}\\
&+\frac{1}{2 n^3}-\frac{1}{2 n^2}\Big)-\frac{19348413013}{480480 n^{15}}-\frac{220713001}{65520 n^{14}}+\frac{5001819}{2464 n^{13}}+\frac{28133}{132 n^{12}}-\frac{1393813}{10080 n^{11}}\\
&-\frac{9181}{480 n^{10}}+\frac{67379}{5040 n^9}+\frac{331}{126 n^8}-\frac{469}{240 n^7}-\frac{151}{240 n^6}+\frac{11}{24 n^5}+\frac{5}{12 n^4}-\frac{1}{2 n^3} \Big]+\cO\left( \frac{1}{n^{16}} \right),
\end{split}
\end{align}
\begin{align}
\begin{split}
S_{-2,-2}(n)=&+\frac{13 \pi ^4}{1440}+(-1)^n\pi^2\Big[
\frac{38227 }{24 n^{15}}-\frac{691 }{8 n^{13}}+\frac{155 }{24 n^{11}}-\frac{17 }{24 n^9}+\frac{1}{8 n^7}-\frac{1}{24 n^5}+\frac{1}{24 n^3}-\frac{1}{24 n^2}\Big]\\
&+\frac{23494}{45 n^{15}}+\frac{353}{8 n^{14}}-\frac{475}{12 n^{13}}-\frac{5}{n^{12}}+\frac{9}{2 n^{11}}+\frac{7}{8 n^{10}}-\frac{31}{36 n^9}-\frac{1}{4 n^8}+\frac{1}{3 n^7}+\frac{1}{8 n^6}\\
&-\frac{5}{12 n^5}+\frac{3}{8 n^4}-\frac{1}{6 n^3}+\cO\left( \frac{1}{n^{16}} \right),
\end{split}
\end{align}
which we derived following \cite{Albino:2005me, Albino:2009ci}.

Finally, to justify our claim that the CFT data we have obtained satisfy the reciprocity principle, they must admit in expansion in powers of $J^2$. While this is not true for generic harmonic sums, it is for specific combinations, defined in \cite{Beccaria:2009vt}\footnote{Relations between ordinary and reciprocity-respecting harmonic sums can be found at http://thd.pnpi.spb.ru/~velizh/5loop/.}. As an example, we note that the combination
\be 
\cS_3(2n+2\df-1)=S_{-3}(2n+2\df-1)-2S_{-2,1}(2n+2\df-1),
\ee
that appears in the one-loop anomalous dimensions, admits an expansion in $1/J^2$.

\section{One-loop results}\label{sec:1loop_phi^4}

\subsection{$\f^4$ interaction}

Here we collect some of the results for one-loop correlators and CFT data in the $\f^4$ case, {\it {\it i.e.}} the solution with $q=0$ in the notation of Section \ref{treesinglefield}. The expressions for the correlators are rather involved, therefore we shall limit to consider $\df=2$ ($\df=1$ was already discussed in \cite{mp}). In such case, we found:
\begin{align}\nonumber
&\mathcal{A}^{(2)}(z)=\frac{1}{(1-z)^{4}}\Bigg\{
\frac{144}{25}\Big[
\frac{ (z-2) z^5 \left(16 z^6-80 z^5+179 z^4-220 z^3+165 z^2-66 z+22\right)}{(z-1)^2}\Li_4(1-z)\\ \nonumber &
+\frac{ (z-1)^5 (z+1) \left(16 z^6-16 z^5+19 z^4-16 z^3+19 z^2-16 z+16\right)}{ z^2}\Li_4\left(\frac{z}{z-1}\right)\\ \nonumber &
+\frac{ (2 z-1) \left(22 z^6-66 z^5+165 z^4-220 z^3+179 z^2-80 z+16\right)}{ (z-1)^2 z^2}\Li_4(z)\Big]\\ \nonumber 
&
-\frac{72}{25}\Big[\Li_3(z)\Big(
\frac{(z-2) z^5 \left(16 z^6-80 z^5+179 z^4-220 z^3+165 z^2-66 z+22\right)}{ (z-1)^2}\log(z)\\ \nonumber &
+\frac{ z^4 \left(16 z^8-112 z^7+339 z^6-578 z^5+605 z^4-396 z^3+163 z^2-14 z+9\right)}{ (z-1)^2}\log(1-z)\\ \nonumber &+8\frac{ (z-2) \left(2 z^{10}-7 z^9+11 z^8-8 z^7+4 z^6-2 z^5+10 z^4-20 z^3+20 z^2-10 z+2\right)}{ (z-1) z}\Big)\\ \nonumber &
+\Li_3(1-z)\Big(
\frac{ (z-1)^4 \left(16 z^8-16 z^7+3 z^6+6 z^2-32 z+32\right)}{ z^2}\log(z)\\ \nonumber &
+\frac{ (z-1)^5 (z+1) \left(16 z^6-16 z^5+19 z^4-16 z^3+19 z^2-16 z+16\right)}{ z^2}\log(1-z)\\ \nonumber &
-8\frac{ (z+1) \left(2 z^{10}-13 z^9+38 z^8-68 z^7+88 z^6-92 z^5+88 z^4-68 z^3+38 z^2-13 z+2\right)}{ (z-1) z}\Big)\Big]\\ \nonumber
&
+\frac{576}{25 (z-1) z}\Li_2(z)\Big[
(z-2) \big(2 z^{10}-7 z^9+11 z^8-8 z^7+4 z^6-2 z^5+10 z^4-20 z^3+20 z^2\\ \nonumber &
-10 z+2\big)\log(z)
+(z+1) \big(2 z^{10}-13 z^9+38 z^8-68 z^7+88 z^6-92 z^5+88 z^4-68 z^3\\ \nonumber &+38 z^2-13 z+2\big)\log(1-z)\Big]+
\frac{6 (2 z-1) }{25 (z-1)^2 z^2}\log(1-z)^3\left( \log(1-z)-4\log(z) \right)\\ \nonumber &
\times \left(22 z^6-66 z^5+165 z^4-220 z^3+179 z^2-80 z+16\right)\\ \nonumber &
-\frac{36 (z-1)^4 \left(16 z^8-16 z^7+3 z^6+3 z^2-16 z+16\right)}{25 z^2}\log(1-z)^2\log(z)^2\\ \nonumber &
+\frac{576 (z+1)}{25 (z-1) z} \big(2 z^{10}-13 z^9+38 z^8-68 z^7+88 z^6-93 z^5+88 z^4-68 z^3\\ \nonumber &+38 z^2-13 z+2\big)\log(1-z)^2\log(z)
+\log(1-z)^2\Big[\frac{18}{25} z^2 \left(z^4-2 z^3+34 z^2-2 z+1\right)\\ \nonumber 
&
+\frac{12 (2 z-1) \left(22 z^6-66 z^5+165 z^4-220 z^3+179 z^2-80 z+16\right)}{25 (z-1)^2 z^2}\pi^2\Big]\\ \nonumber &
-\frac{576 (z-2) (z-1)^4}{25 z}\log(1-z)\log(z)^2
\end{align}
\begin{align}  \nonumber &
+\Big[\frac{12 (z-1)^4 \left(16 z^8-16 z^7+3 z^6+6 z^2-32 z+32\right)}{25 z^2}\pi^2+\\ 
 \nonumber &
+\frac{36}{25} \left(16 z^8-64 z^7+96 z^6-64 z^5-17 z^4+66 z^3-97 z^2+64 z-16\right)\Big]\log(1-z)\log(z)\\ \nonumber &
+\Big[\frac{1}{150} \big(1728 z^8-8374 z^7+15593 z^6-14870 z^5+8390 z^4-14870 z^3+15593 z^2\\ \nonumber &-8374 z+1728\big)
-\frac{96 (z+1)}{25 (z-1) z} \big(2 z^{10}-13 z^9+38 z^8-68 z^7+88 z^6-92 z^5+88 z^4-68 z^3\\ \nonumber &+38 z^2-13 z+2\big)\pi^2
+\frac{72}{125 (z-1)^2 z^2} \big(12 z^{11}-66 z^{10}+156 z^9-207 z^8+510 z^7-977 z^6\\ \nonumber &+2136 z^5-3091 z^4+2902 z^3-1695 z^2+560 z-80\big)\z(3)\Big]\log(1-z)\\ \nonumber &
+\frac{18}{25} (z-1)^2 \left(z^4-2 z^3+34 z^2-64 z+32\right)\log(z)^2\\ \nonumber
& +\Big[-\frac{72}{125 (z-1)^2 z^2} \big(12 z^{11}-66 z^{10}+156 z^9-207 z^8+510 z^7-1585 z^6+3960 z^5-6050 z^4 \\ \nonumber &+5780 z^3-3390 z^2+1120 z-160\big)\z(3)
+\frac{1}{150} \big(1728 z^8-5450 z^7+5359 z^6+398 z^5-4195 z^4\\ \nonumber &+14472 z^3-20952 z^2+13824 z-3456\big)\Big]\log(z)\\ \nonumber &
+-\frac{2 z^4 \left(16 z^8-112 z^7+339 z^6-578 z^5+605 z^4-396 z^3+136 z^2-104 z-18\right)}{125 (z-1)^2}\pi^4\\ \nonumber &
+\frac{288 \left(3 z^8-12 z^7+z^6+39 z^5-188 z^4+297 z^3-250 z^2+110 z-20\right)}{125 (z-1) z}\z(3)\\ &
+\frac{1}{75} \left(z^2-z+1\right) \left(864 z^6-2592 z^5+1213 z^4+1894 z^3+1213 z^2-2592 z+864\right)
\Bigg\}.
\end{align}
The corresponding anomalous dimensions are, in terms of harmonic sums,
\begin{align}
\begin{split} 
\g^{(2)}_n&=\12 \g^{(1)}_n\,\frac{\pr}{\pr n}\left( \g^{(1)}_n \right)+
\frac{72}{625}\Bigg[2  \left(J^4-7 J^2+15\right) \cS_3(2 n+3)\\&
+\frac{1}{J^2 \left(J^2-6\right)}\Bigg(\frac{30 \left(5 J^4-20 J^2+12\right)}{J^2-2}S_{-2}(2 n+3)\\&
-\frac{1}{J^2 \left(J^2-6\right) \left(J^2-2\right)}\Big(\frac{J^2}{24}\big(24 J^8+10475 J^6-105854 J^4+325940 J^2-265800\big)\\&+2\big( J^{12}-22 J^{10}+196 J^8-925 J^6+2226 J^4-4356 J^2+1800\big)H_{2 n+3}\Big)\\&
-\left(J^2-12\right) \left(J^6-J^4+45 J^2-90\right) \zeta (3)
\Bigg)\Bigg].
\end{split}
\end{align}
and the function expressing the violation to the derivative relation is
\begin{align}
\begin{split} 
\d C^{(2)}_n&=\frac{36}{625} \Bigg[-\left(J^4-7 J^2+15\right) \s_4(2n+3)\\&+\frac{1}{J^2 \left(J^2-6\right) \left(J^2-2\right)}\Bigg(-120 \left(5 J^4-20 J^2+12\right) \s_3(2 n+3)\\&+4 \left(J^8-16 J^6+100 J^4-100 J^2+276\right)  \s_2(2 n+3)+2 \left(J^6-17 J^4-32 J^2+660\right)\Bigg)\Bigg].
\end{split}
\end{align}
For large $J$, one gets the following expansions:
\begin{align}
\begin{split}
\g^{(2)}_n=&\12 \g^{(1)}_n\,\frac{\pr}{\pr n}\left( \g^{(1)}_n \right)
-\frac{3 \left(-2592 \zeta (3)+60 \pi ^2+2215\right)}{125 J^2}
-\frac{12 \left(-2592 \zeta (3)+60 \pi ^2+2215\right)}{125 J^4}\\
&+\frac{144 \left(90 \log (J)+1296 \zeta (3)-28 \pi ^2+90 \gamma_E -1295\right)}{125 J^6}\\
&
+\frac{288 \left(360 \log (J)+3888 \zeta (3)-82 \pi ^2+360 \gamma_E -4065\right)}{125 J^8}\\
&+\frac{144 \left(12600 \log (J)+46656 \zeta (3)-976 \pi ^2+12600 \gamma_E -62913\right)}{125 J^{10}}\\
&
+\frac{2304 \left(12600 \log (J)+122472 \zeta (3)-2555 \pi ^2+12600 \gamma_E -116140\right)}{875 J^{12}}+\cO\left( \frac{1}{J^{14}} \right),
\end{split}
\end{align}
and
\begin{align}
\begin{split} 
\d C^{(2)}_n&=\frac{72}{J^8}+\frac{432}{J^{10}}+\frac{394272}{125 J^{12}}+\frac{574573824}{4375 J^{16}}+\cO\left( \frac{1}{J^{18}} \right).
\end{split}
\end{align}
Note that the expansion of $\d C^{(2)}_n$ begins with $J^{-8}=J^{-2(2\df)}$, as discussed around eq. \eqref{derivativerule}. Let us also observe that one can perform a coupling redefinition by adding a tree-level $\f^4$ solution with an appropriate constant in such a way that the expansion starts with $J^{-6}$:
\begin{align}
\begin{split}
\g^{(2)}_n=&\12 \g^{(1)}_n\,\frac{\pr}{\pr n}\left( \g^{(1)}_n \right)
+\frac{72 \left(180 \log (J)+4 \pi ^2+180 \gamma_E -375\right)}{125 J^6}\\
&
+\frac{144 \left(720 \log (J)+16 \pi ^2+720 \gamma_E -1485\right)}{125 J^8}\\
&+\frac{144 \left(12600 \log (J)+104 \pi ^2+12600 \gamma_E -23043\right)}{125 J^{10}}\\
&+
\frac{576 \left(10080 \log (J)+224 \pi ^2+10080 \gamma_E -9185\right)}{175 J^{12}}+\cO\left( \frac{1}{J^{14}} \right).
\end{split}
\end{align}

For $\df=3$ we only give the CFT data:
\begin{align}
\begin{split} 
\g^{(2)}_n&=\12 \g^{(1)}_n\,\frac{\pr}{\pr n}\left( \g^{(1)}_n \right)+\frac{25}{43218}\Bigg[2\left( J^8-38 J^6+444 J^4-1752 J^2+3024 \right)\cS_3(2n+5)\\&+\frac{1}{J^2 \left(J^2-20\right) \left(J^2-6\right)}\Bigg(
\frac{10080 \left(7 J^8-196 J^6+1588 J^4-3744 J^2+1728\right) }{\left(J^2-12\right) \left(J^2-2\right)}S_{-2}(2 n+5)\\&
-\frac{1}{J^2 \left(J^2-20\right) \left(J^2-12\right) \left(J^2-6\right) \left(J^2-2\right)}\Big(\frac{J^2}{15}\big(15 J^{18}-1620 J^{16}+75870 J^{14}\\&+31287716 J^{12}-1764932904 J^{10}+36039913872 J^8-343873332736 J^6\\&+1575058355328 J^4-3201515606016 J^2+2303304837120\big)+2 \big(J^{22}-105 J^{20}+4726 J^{18}\\&-119588 J^{16}+1879728 J^{14}-19221824 J^{12}+130870656 J^{10}-603476736 J^8\\&+1873904256 J^6-3955889664 J^4+6135367680 J^2-2438553600\big) H_{2 n+5}\Big)\\&
-\left(J^2-30\right) \left(J^{12}-34 J^{10}+532 J^8-1896 J^6+44976 J^4-854784 J^2+1532160\right) \zeta (3)
\Bigg)\Bigg],
\end{split}
\end{align}
and the function expressing the violation to the derivative relation is
\begin{align}
\begin{split} 
\d C^{(2)}_n&=\frac{25}{86436}\Big[ -\left(J^8-38 J^6+444 J^4-1752 J^2+3024\right) \s_4(2n+5)\\&
+\frac{1}{J^2 \left(J^2-20\right) \left(J^2-12\right) \left(J^2-6\right) \left(J^2-2\right)} \Big(-40320 \big(7 J^8-196 J^6+1588 J^4-3744 J^2\\&+1728\big) \s_3(2n+5) +4 \big(J^{16}-79 J^{14}+2552 J^{12}-43756 J^{10}+435832 J^8-2463072 J^6\\&+7122144 J^4-6308352 J^2+15565824\big)\s_2(2n+5)+2 \big(J^{14}-80 J^{12}+2633 J^{10}\\&-46474 J^8+414820 J^6-824504 J^4-9362688 J^2+47859840\big)
\Big)\Big].
\end{split}
\end{align}
For large $J$, one gets the following expansions:
\begin{align}\nonumber
\begin{split}
\g^{(2)}_n=&\12 \g^{(1)}_n\,\frac{\pr}{\pr n}\left( \g^{(1)}_n \right)
-\frac{10 \left(-1026000 \zeta (3)+3150 \pi ^2+1190117\right)}{9261 J^2}\\
&
-\frac{40 \left(-1026000 \zeta (3)+3150 \pi ^2+1190117\right)}{3087 J^4}\\
&+\frac{40 \left(3500 \log (J)+2052000 \zeta (3)-5800 \pi ^2+3500 \gamma_E -2393359\right)}{343 J^6}\\
&+\frac{80 \left(56000 \log (J)+19836000 \zeta (3)-52900 \pi ^2+56000 \gamma_E -23225637\right)}{343 J^8}\\
&+\frac{80}{1029 J^{10}} \big(6951000 \log (J)+1177848000 \zeta (3)-3022800 \pi ^2\\
&+6951000 \gamma_E -1388318491\big)+\frac{320}{3087 J^{12}} \big(153594000 \log (J)
\end{split}
\end{align}
\begin{align}
\begin{split}
&+17612316000 \zeta (3)-44118900 \pi ^2+153594000 \gamma_E -20877432697\big)+\cO\left( \frac{1}{J^{14}} \right),
\end{split}
\end{align}
\begin{align}
\begin{split} 
\d C^{(2)}_n&=\frac{28800}{J^{12}}+\frac{14852498225}{14406 J^{14}}+\cO\left( \frac{1}{J^{16}} \right).
\end{split}
\end{align}
Note that the expansion of $\d C^{(2)}_n$ begins with $J^{-12}=J^{-2(2\df)}$, as discussed around eq. \eqref{derivativerule}. Let us also observe that one can perform a coupling redefinition by adding a tree-level $\f^4$ solution with an appropriate constant in such a way that the expansion starts with $J^{-6}$:
\begin{align}
\begin{split}
\g^{(2)}_n=&\12 \g^{(1)}_n\,\frac{\pr}{\pr n}\left( \g^{(1)}_n \right)
+\frac{5000 \left(28 \log (J)+4 \pi ^2+28 \gamma_E -105\right)}{343 J^6}\\
&
+\frac{10000 \left(1344 \log (J)+192 \pi ^2+1344 \gamma_E -5201\right)}{1029 J^8}\\
&
+\frac{2000 \left(92680 \log (J)+7912 \pi ^2+92680 \gamma_E -294189\right)}{343 J^{10}}\\
&+\frac{8000 \left(6143760 \log (J)+398160 \pi ^2+6143760 \gamma_E -17915371\right)}{3087 J^{12}}+\cO\left( \frac{1}{J^{14}} \right).
\end{split}
\end{align}

\subsection{Derivative interactions}\label{appendix:loopder}

Here we list some of our results for derivative interactions in the case of a single field, {\it {\it i.e.}} the solutions with $q>0$ in the notation of Section \ref{treesinglefield}. As discussed in Section \ref{sec:loopder}, we can express the difference from loop correlators with $q>0$ and that with $q=0$ in terms of a function $\cG(z)$ of reduced transcendentality. Here we provide results for that function, and the corresponding CFT data, for $\df=1$ and $1\le q \le 2$. All of our results are found to agree with those obtained using the PM bootstrap, as one can check for example comparing them to eqs. (\ref{gamma2d1q1_2}, \ref{gamma2d1q1_3}, \ref{gamma2d1q2_4}).

\begin{itemize}
\item $\df=1$, $q=1$. We have $\a(q=1,\df=1)=\frac{36}{1225}$, and the function $\cG(z)$ is
\begin{align}
\begin{split}
\cG(z)=&\frac{1}{(1-z)^{2\df}}\Big\{\frac{18 z^2 \left(9 z^6-44 z^5+85 z^4-80 z^3+35 z^2+6 z-2\right)}{1225 (z-1)^2}\log^2(z)\\
&-\frac{18 \left(18 z^8-72 z^7+107 z^6-69 z^5+17 z^4-3 z^3+9 z^2-7 z+2\right)}{1225 (z-1) z}\log(z)\log(1-z)\\
&+\frac{18 (z-1)^2 \left(9 z^6-10 z^5-10 z+9\right)}{1225 z^2}\log^2(1-z)\\
&+\frac{3}{42875 (1-z)}\Big[-120 z^4 \left(2 z^2-7 z+7\right)\z(3)-774 z^6+2499 z^5-3374 z^4+2170 z^3\\&-2135 z^2+1260 z-420\big)\Big]\log(z)+\frac{3}{42875 z}\Big[120 (z-1)^4 \left(2 z^2+3 z+2\right)\z(3)\\
&+774 z^6-2145 z^5+2489 z^4-1816 z^3+2489 z^2-2145 z+774\Big]\log(1-z)\\
&-\frac{3348 \left(z^2-z+1\right)^2}{42875}\Big\},
\end{split}
\end{align}
and the corresponding differences between CFT data given in \eqref{gammadiffloopder} and \eqref{opediffloopder} can be expressed as
\begin{align} \label{gq$1d$1comp}
\begin{split}
\G^{(2)}_n|_{q=1}=&\12 \g^{(1)}_n|_{q=1} \frac{\pr}{\pr n} \left( \12 \g^{(1)}_n|_{q=1} \right)-\12\,\frac{36}{1225}\, \g^{(1)}_n|_{q=0}\frac{\pr}{\pr n} \left( \12 \g^{(1)}_n|_{q=0} \right)\\ &
-\frac{18 J^2}{1225}H_{2n+1}+\frac{36 \left(J^2-2\right) \left(J^2+2\right)}{8575 J^2}\z(3)-\frac{3 \left(873 J^4-2722\right)}{85750 J^2},
\end{split}
\end{align} 
and
\be 
\D C^{(2)}_n|_{q=1}=\frac{9 \left(J^2-2\right) \left(J^2+4\right)}{1225}\s_2(2n+1)
+\frac{9 \left(J^{10}-7 J^8-4 J^6+72 J^4-32 J^2-96\right)}{1225 J^4 \left(J^2-6\right) \left(J^2-2\right)}.
\ee
This means that, using eqs. \eqref{gammadiffloopder} and \eqref{opediffloopder} to reconstruct the CFT data for $q=1$, we have the following large $J$ expansions:
\begin{align}
\begin{split}
\g^{(2)}_n|_{q=1}=&\12 \g^{(1)}_n|_{q=1} \frac{\pr}{\pr n} \left( \12 \g^{(1)}_n|_{q=1} \right)
-\frac{9 J^2 (140 \log (J)-40 \zeta (3)+140 \gamma_E +291)}{85750}-\frac{3}{1225}\\&-\frac{12 \left(-360 \zeta (3)+35 \pi ^2-447\right)}{42875 J^2}
-\frac{8}{42875 J^4}+\frac{6 (1680 \log (J)+1680 \gamma_E -1259)}{42875 J^6}\\&
-\frac{48 (9240 \log (J)+9240 \gamma_E -11933)}{471625 J^8}+\cO\left( \frac{1}{J^{10}} \right),
\end{split}
\end{align}
and
\be 
\d C^{(2)}_n|_{q=1}=\D C^{(2)}_n|_{q=1}+\frac{36}{1225}\d C^{(2)}_n|_{q=0}=\frac{24}{5 J^8}-\frac{48}{J^{10}}+\frac{1172832}{875 J^{12}}-\frac{898752}{25 J^{14}}+\cO\left( \frac{1}{J^{16}} \right).
\ee
The last equation agrees with the prediction of \eqref{violationderrulederivatives} for $\df=1$ and $q=1$.

\item $\df=1$, $q=2$. We have $\a(q=2,\df=1)=\frac{225}{1002001}$, and we shall give only the CFT data, still in terms of the quantities defined in eqs.  \eqref{gammadiffloopder} and \eqref{opediffloopder}:
\begin{align}
\begin{split}
\G^{(2)}_n|_{q=2}=&\12 \left( \g^{(1)}_n|_{q=2} \right)\frac{\pr}{\pr n} \left( \12 \g^{(1)}_n|_{q=2} \right)-\12\, \frac{225}{1002001}\,\left( \g^{(1)}_n|_{q=0}\right) \frac{\pr}{\pr n} \left( \12 \g^{(1)}_n|_{q=0} \right)\\ &
-\frac{25 J^2 \left(49 J^8+980 J^6+1652 J^4-5936 J^2+58848\right) }{13851661824}H_{1+2n}\\&
+\frac{\left(J^2-2\right)}{2423463666624 J^2} \big(13 J^{10}+286 J^8-72624 J^6-441360 J^4+90590400 J^2\\&+181180800\big) \zeta (3)+\frac{1}{241842286219742208000 J^2}\big(43266897946747 J^{12}\\
&+737011060234940 J^{10}+2955217554271876 J^8+5889841900120272 J^6\\
&-336513215167931520 J^4+1175663507488704000\big),
\end{split}
\end{align} 
and
\begin{align}
\begin{split}
\D C^{(2)}_n|_{q=2}=&\frac{25 \left(J^2-2\right)}{747989738496} \big(49 J^{10}+3136 J^8+40656 J^6+93240 J^4+1842912 J^2\\&+7153920\big)\s_2(1+2n)+
\frac{25}{747989738496 J^2 \prod_{s=0}^{6}\left( J^2-s(s+1) \right)} \big(49 J^{26}\\&-2499 J^{24}-75572 J^{22}+5979652 J^{20}-77502480 J^{18}-807810464 J^{16}\\&+25470218304 J^{14}-242090490816 J^{12}+1318392391680 J^{10}\\&-2730286994688 J^8-9721849319424 J^6+41023239782400 J^4\\&-7417184256000 J^2-51920289792000\big).
\end{split}
\end{align} 
This means that, using eqs. \eqref{gammadiffloopder} and \eqref{opediffloopder} to reconstruct the CFT data for $q=1$, we have the following large $J$ expansions:
\begin{align}
\begin{split}
&\g^{(2)}_n|_{q=2}-\12 \g^{(1)}_n|_{q=1} \frac{\pr}{\pr n} \left( \12 \g^{(1)}_n|_{q=1} \right)=\\&
\frac{J^{10} (-21387816450000 \log (J\,e^{\gamma_E})+1297296000 \zeta (3) +43266897946747)}{241842286219742208000}\\
&+\frac{J^8 (-21387816450000 \log (J\,e^{\gamma_E})+1297296000 \zeta (3) +36672321207997)}{12092114310987110400}\\
& +\frac{J^6 (-180268738650000 \log (J\,e^{\gamma_E})-1826093808000 \zeta (3) +721159439996719)}{60460571554935552000}\\
&+\frac{J^4 (5997641650000\log (J\,e^{\gamma_E})-68401872000 \zeta (3) +13388077995371)}{559820106990144000}\\
&+\frac{J^2 (-107026379460000 \log (J\,e^{\gamma_E})+38034523296000 \zeta (3) -1400260738625923)}{1007676192582259200}\\
&-\frac{6198755}{342828630144}+\frac{7544275200000 \zeta (3)-754247894400 \pi ^2+51285838980461}{10076761925822592 J^2}\\
&-\frac{11254325}{7799351335776 J^4}+\frac{25 (19054683648\log (J\,e^{\gamma_E}) -14279372279)}{265177945416384 J^6}\\
&-\frac{5 (5430584839680\log (J\,e^{\gamma_E}) -7013396580973)}{3778785722183472 J^8}
+\cO\left( \frac{1}{J^{10}} \right),
\end{split}
\end{align}
and
\be 
\d C^{(2)}_n|_{q=2}=\D C^{(2)}_n|_{q=2}+\frac{25}{1002001}\d C^{(2)}_n|_{q=0}=\frac{960}{7 J^{12}}-\frac{4800}{J^{14}}+\frac{3761376480}{3773 J^{16}}+\cO\left( \frac{1}{J^{18}} \right).
\ee
The last equation agrees with the prediction of \eqref{violationderrulederivatives} for $\df=1$ and $q=2$.

\end{itemize}

\end{document}